\documentclass[a4paper]{article}

\usepackage{graphicx}
\usepackage{epstopdf, epsfig}
\usepackage{amsmath}
\usepackage{dashrule,xcolor}
\usepackage{natbib}
\usepackage{array}
\usepackage{makecell}
\usepackage{latexsym}
\usepackage{lscape}
\usepackage{authblk}

\usepackage[left=2.5cm,right=2.5cm,top=2cm,bottom=2cm]{geometry}

\title{Machine-learning wall-model large-eddy simulation accounting for isotropic roughness under local equilibrium}

\author{Rong Ma$^*$} 
\author{Adri{\'a}n Lozano-Dur{\'a}n}
\affil{Department of Aeronautics and Astronautics, Massachusetts Institute of Technology, Cambridge, MA 02139, USA}
\affil{$^*$rongma@mit.edu}
\date{}

\begin{document}
\maketitle

\begin{abstract}
%
We introduce a wall model (WM) for large-eddy simulation (LES)
applicable to rough surfaces with Gaussian and non-Gaussian
distributions for both the transitionally- and fully-rough
regimes. The model is applicable to arbitrary complex geometries where
roughness elements are assumed to be underresolved, i.e.,
subgrid-scale roughness. The wall model is implemented using a
multi-hidden-layer feedforward neural network (FNN), with the mean
geometric properties of the roughness topology and near-wall flow
quantities serving as input. The optimal set of non-dimensional input
features is identified using information theory, selecting variables
that maximize information about the output while minimizing redundancy
among inputs.  The model also incorporates a confidence score based on
Gaussian process modeling, enabling the detection of potentially low
model performance for unseen rough surfaces.
The model is trained using a direct numerical simulation (DNS)
roughness database comprising approximately 200 cases. The roughness
geometries for the database are selected from a large repository
through active learning. This approach ensures that the rough surfaces
incorporated into the database are the most informative, achieving
higher model performance with fewer DNS cases compared to passive
learning techniques.
The performance of the model is evaluated both \emph{a priori} and
\emph{a posteriori} in WMLES of turbulent channel flows with rough
walls. Over 120 channel flow cases are considered, including untrained
roughness geometries, roughness Reynolds numbers, and grid resolutions
for both transitionally- and fully-rough regimes. The results show
that the rough-wall model typically predicts the wall shear stress
within a 1\% to 15\% accuracy range. The performance of the model is
also assessed on a high-pressure turbine blade with two different
rough surfaces.  The wall model predicts the skin
friction and the mean velocity deficit induced by the rough surface on
the blade within 1\% to 10\% accuracy except the region with shock waves.
This work extends the building-block flow wall model (BFWM) introduced
by \citet{lozano2023machine} for smooth walls, expanding the BFWM
framework to account for rough-wall scenarios.
\end{abstract}


\section{Introduction}

Turbulent boundary layers (TBL) over rough surfaces are prevalent in
engineering applications. Examples include the deposition of fuel and
airborne contaminants, as well as erosion in turbo-machinery
applications contributing to the formation of roughness on turbine
blades~\citep{bons2001many}. Another example is bio-fouling, resulting
from the accumulation of living organisms, which generates
multi-scaled roughness geometries on the immersed surfaces of marine
vessels~\citep{munk2009effects}. These rough surfaces significantly
increase hydrodynamic drag and reduce the overall efficiency of
engineering systems~\citep{bons2010review,
  kirschner2012bio}. Therefore, modeling the effects of roughness and
predicting drag in rough-wall flows are essential tasks in
engineering. Computational fluid dynamics (CFD) complements
experimental investigations by enabling more cost-effective
exploration under various operating conditions and reducing the need
for extensive physical testing. While roughness-resolved simulations
such as direct numerical simulation (DNS) and wall-resolved large-eddy
simulation (WRLES) provide valuable insights into the underlying
physics and aid in the development of models, their practical
applicability to high Reynolds number flows is limited due to their
high computational cost. Recently, wall-modeled large-eddy simulation
(WMLES) has emerged as a competitive approach for modeling the effects
of roughness on the outer flow without resolving the small-scale flow
and roughness details in the near-wall region. In this work, we
develop a rough-wall model for WMLES that is applicable to various
roughness geometries and flow conditions.

A surface is considered rough when its topographical features are
large enough to disrupt the near-wall eddies, resulting in increased
drag and momentum deficit across the TBL~\citep{raupach1991rough,
  jimenez2004turbulent, chung2021predicting}. In incompressible
zero-pressure-gradient TBLs (ZPGTBLs), the velocity deficit caused by
roughness is quantified by the roughness function $\Delta U^+= \Delta
U/u_{\tau}$, where $\Delta U$ represents the downward shift of the
mean velocity profiles in the logarithmic layer, and $u_{\tau}$
denotes the mean friction velocity. In the fully-rough regime, the
momentum deficit is primarily caused by form drag. This scenario is
typically easier to investigate as wall friction becomes independent
of Reynolds number. In contrast, both form drag and viscous drag
contribute to wall friction in the transitionally-rough regime. In
these cases, drag is highly sensitive to Reynolds number and roughness
topographies, making the search for universal scaling laws and models
for drag challenging tasks. For ZPGTBLs, the effect of roughness in
the fully-rough regime is generally characterized by the equivalent
sand-grain roughness height $k_s$. This hydraulic roughness scale,
proposed by \cite{nikuradse1933laws}, represents the size of uniformly
packed sand-grain roughness that produces the same frictional drag as
the actual roughness geometry. In the fully-rough regime, $k_s$
quantifies hydrodynamic drag through a logarithmic relationship with
the roughness function.

Models with varying fidelity have been devised to account for
wall-roughness effects, ranging from empirical correlations based on
Moody charts to wall functions for the Reynolds-averaged Navier-Stokes
equations (RANS), WRLES, and WMLES. Many of these rough-wall models
are formulated in terms of equivalent sand-grain roughness height,
making the prediction of $k_s$ the main goal. The first,
lower-fidelity family of models aims to establish correlations between
$k_s$ and other roughness geometrical parameters without explicitly
resolving the flow motion~\citep[e.g.,][]{bons2002st, flack2010review,
  forooghi2017toward, chung2021predicting}. Further details about this
topic can be found in the reviews by \cite{bons2002st},
\cite{flack2010review}, and \cite{forooghi2017toward}. More recently,
data-driven methods have also been leveraged to enhance the prediction
of $k_s$ without resolving the flow. \cite{jouybari2021data} used
machine-learning methods to predict $k_s$ based on a large set of
roughness parameters and demonstrated more accurate results than
previous empirical correlations. \cite{ma2023data} proposed a particle
swarm optimized backpropagation method to estimate $k_s$ and showed
better performance in evaluation metrics compared to both existing
roughness correlation formulas and the traditional backpropagation
model. \cite{yang2023prediction} utilized ensemble neural networks to
predict $k_s$ based on the roughness height probability density
function and power spectrum. Other methods do not rely on the use of
$k_s$. For example, \cite{yang2016exponential} proposed an analytical
roughness model based on the exponential velocity profile within the
roughness layer for rectangular-prism roughness elements and
demonstrated good predictions of mean velocity and drag forces for
this type of roughness where the flow separation point is easily
identified.

The second family of rough-wall models incorporates surface roughness
effects directly into RANS, WRLES, or WMLES. These methods usually
adopt one of the following approaches:
\begin{itemize}
  \setlength\itemsep{0em}
\item[(i)] In the first approach, wall roughness is represented using
  a closure model for RANS simulations. \cite{cebeci1978calculation}
  adapted the mixing-length formulation of eddy viscosity near rough
  walls by introducing an effective wall displacement as a function of
  $k_s$. \cite{feiereisen1986modeling} further refined the model
  proposed by \cite{cebeci1978calculation} by directly incorporating
  measurable roughness parameters instead of relying solely on
  $k_s$. \cite{durbin2001rough} extended the two-layer $k$--$\epsilon$
  model to rough walls by modifying the calculation of the eddy
  viscosity based on $k_s$. \cite{aupoix2003extensions} proposed two
  extensions of the Spalart–Allmaras model to account for roughness
  effects using the value of $k_s$ as key
  parameter. \cite{knopp2009new} presented an extension for
  $k$--$\omega$ type turbulence models to account for surface
  roughness based on $k_s$ and the rough-wall logarithmic law,
  demonstrating its capability of predicting the aerodynamic effects
  of surface roughness on the flow past an
  airfoil. \cite{brereton2018wall} proposed a model of equivalent
  shear force for the wall-roughness eddy viscosity, demonstrating
  good agreement with experimental data for zero and favorable
  pressure gradient turbulent boundary layers over fully-rough
  surfaces.
\item[(ii)] The second approach consists of imposing the fluxes as
  wall boundary condition obtained from analytical wall functions or
  rough-wall models that account for roughness
  effects. \cite{wilcox1998turbulence} incorporated roughness effects
  into the boundary condition for the $\omega$ equation of the
  $k$--$\omega$ turbulence closure model for RANS by introducing a
  functional dependence with $k_s$. \cite{suga2006analytical} derived
  an analytical wall function accounting for the effects of fine-grain
  surface roughness on turbulence and heat transfer in RANS
  simulations. In the context of WMLES, the logarithmic law for rough
  walls as a function of $k_s$ has been used to capture the downward
  shift of velocity profiles~\citep{yang2017large,
    li2021large}. \cite{li2022predictive} provided a systematic
  assessment of the predictive capability of the logarithmic law
  rough-wall model for WMLES and demonstrated good predictions of the
  mean velocity against DNS data. The logarithmic law rough-wall model
  has also been widely used to develop morphological models for flows
  over urban-like surfaces~\citep{theurer1993dispersion,
    macdonald1998improved, grimmond1999aerodynamic, hanna2010wind}.
\item[(iii)] The third approach involves introducing a body force term
  to the Navier–Stokes equations as a drag model representing
  roughness effects. The rationale for the forcing term was discussed
  by \cite{stripf2009extended}. The body-force drag model has been
  employed in both RANS equations~\citep{aupoix2016revisiting,
    chedevergne2020importance} and eddy-resolving
  simulations~\citep{shaw1992large,
    busse2012parametric}. \cite{busse2012parametric} used a body-force
  term model and showed good agreement for the mean flow and Reynolds
  stresses everywhere except in the immediate vicinity of the rough
  surface. However, they noticed that the model parameters need to be
  calibrated against experiments or DNS to be successfully applied in
  a simulation setting. \cite{anderson2011dynamic} developed a dynamic
  roughness model for LES applicable to multi-scale, fractal-like
  roughness by decomposing the surface into resolved and subgrid-scale
  height contributions. The unresolved height fluctuations were
  modeled using the equilibrium logarithmic law, and the subgrid-scale
  roughness parameter was dynamically estimated to achieve
  resolution-independent mean velocity profiles.
\end{itemize} 

The roughness modeling approaches presented above have greatly
facilitated the prediction of surface roughness effects on turbulent
flows. However, current rough-wall models still face important
limitations:
\begin{itemize}
  \setlength\itemsep{0em}
\item[1)] Many models lack true predictability, as they require the
  specification of the non-trivial roughness parameter $k_s$. This
  reliance on $k_s$ hinders their ability to provide predictions in
  the absence of accurate and universally applicable methods or
  correlations for $k_s$.
\item[2)] Although $k_s$ is effective for predicting drag in
  fully-rough flows, its utility diminishes in transitionally-rough
  flows. This limitation stems from the fact that the logarithmic
  relationship between $k_s$ and $\Delta U^+$ is valid only within the
  fully-rough regime.
\item[3)] The assumptions underlying many wall models are rooted in
  `equilibrium' turbulence, i.e., the presence of wall-attached,
  statistically steady turbulence under ZPG. Consequently, these wall
  models can accurately predict outcomes only for a limited number of
  cases and cannot be generalized to complex scenarios (e.g.,
  adverse/favorable mean pressure gradient and separated flows) which
  are of significant interest in practical applications.
\item[4)] Many models are tailored for specific roughness
  geometries. The challenge remains to develop a rough wall model that
  can accommodate a broad spectrum of surface topologies without
  sacrificing prediction accuracy.
\item[5)] On some occasions, the rough-wall models are, by
  construction, only applicable to simple flow configurations such as
  channel flows and flat plates. This limitation may be due to
  assumptions of flat walls, periodic boundary conditions, or the need
  for global flow quantities (e.g., turbulent channel height) that
  might not be well-defined in other scenarios. As a result, these
  models are unsuitable for the complex geometries typical in
  real-world engineering applications.
\end{itemize}
The reader is referred to the recent work by
\cite{durbin2023reflections} for a discussion on the strengths and
limitations of different approaches to formulate rough-wall models.

Addressing current limitations is crucial for developing accurate and
robust wall models that capture the effects of roughness on turbulent
flows across diverse conditions, geometries, and flow regimes. To
overcome these limitations, the present work aims to develop a wall
model by utilizing the flow over rough walls in minimal turbulent
channels. This effort builds upon the concept of a building-block-flow
wall model (BFWM) for WMLES of smooth-wall flows, previously
introduced by our group~\citep{lozano2023machine}. The core assumption
of the model is that a finite set of simple canonical flows contains
the essential physics necessary to predict wall stress in more complex
scenarios. \cite{lozano2023machine} demonstrated that the BFWM
approach successfully accounts for multiple flow regimes (e.g., zero,
adverse, favorable mean pressure gradients, and separation) within a
single unified model. In this work, we aim to extend this framework to
incorporate wall roughness, referred to as BFWM-rough. Our long-term
goal is to develop a rough-wall model for WMLES that accommodates
multiple roughness geometries and flow regimes, including both zero
and non-zero mean pressure gradient effects and separation. Toward
this goal, the primary objective of this work is to develop the
initial version of BFWM-rough for WMLES under near-wall equilibrium
assumptions applicable to transitionally- and fully-rough regimes with
Gaussian and non-Gaussian roughness geometries.

In this work, we build a database of rough-wall turbulent flows using
DNS cases selected through active learning. The resulting data are
utilized to train the machine-learning-based (ML-based) wall model,
which is implemented in both structured and unstructured WMLES
solvers. To evaluate the performance of the model, BFWM-rough is
assessed across a wide range of cases, spanning from canonical
turbulent channel flows to high-pressure turbine blade
configurations. The manuscript is organized as follows: The roughness
database is introduced in \S\ref{sec:database}. The formulation of the
newly proposed rough-wall model is discussed in
\S\ref{sec:formulation}. The model evaluation is presented in
\S\ref{sec:evaluation}. Finally, the conclusions are offered in
\S\ref{sec:conclusion}.
 
\section{Database of turbulence over rough surfaces}\label{sec:database}


We generate a database of turbulent channel flows over rough surfaces
to train the wall model. The streamwise, wall-normal, and spanwise
directions are denoted by $x$, $y$, and $z$, respectively, and
occasionally referred to as $x_1$, $x_2$, and $x_3$. The friction
Reynolds number is $Re_{\tau} = u_{\tau} \delta / \nu$, where
$u_{\tau}$ is the friction velocity, $\nu$ is the kinematic viscosity,
and $\delta$ is the channel half-height. The database is constructed
in three steps. First, we create a repository containing various
irregular rough surfaces. Second, we apply an active learning
framework to identify and select the most informative rough surfaces
from the repository. Third, we conduct DNS of turbulent channel flows
with the rough walls selected in the previous step. The resulting
database is used to train the wall model, as described in
\S\ref{sec:formulation}.

\subsection{Roughness repository}\label{sec:repository}

The roughness repository is a collection of rough surfaces designed
for generating the DNS database, which is utilized to train and
validate the wall model. The repository includes irregular rough
surfaces characterized by different probability density functions
(PDFs) and power spectra (PS), resembling the realistic roughness
encountered in engineering applications. These irregular rough
surfaces are created from the PDF and PS using a rough surface
generator~\citep{perez2019generating}. Two families of PDFs are
considered: Gaussian and Weibull.  The Gaussian distribution is chosen
due to its ubiquity in nature and engineering
applications~\citep{williamson1969shape,
  whitehouse2023handbook}. Examples of Gaussian roughness include
 turbine blades subject to erosion~\citep{bons2002st}, roughness on
painted high-pressure turbine vanes~\citep{bacci2021effect}, and
manufactured surfaces such as highly polished
steel~\citep{das2017evaluation}. The Weibull distribution is used to
represent roughness resulting from tribology and
wear~\citep{panda2015effects}, defect populations in materials due to
manufacturing processes, environmental factors, or operational
conditions~\citep{cook2019material}, as well as geophysical and
terrain roughness~\citep{barbosa2022structural}. The inclusion of
Weibull roughness in the repository enables the representation of a
broader spectrum of asymmetrical and non-Gaussian roughness features.
\begin{figure}
\centering
\includegraphics[width=65mm,trim={0.5cm 1.2cm 0.5cm 1.8cm},clip]{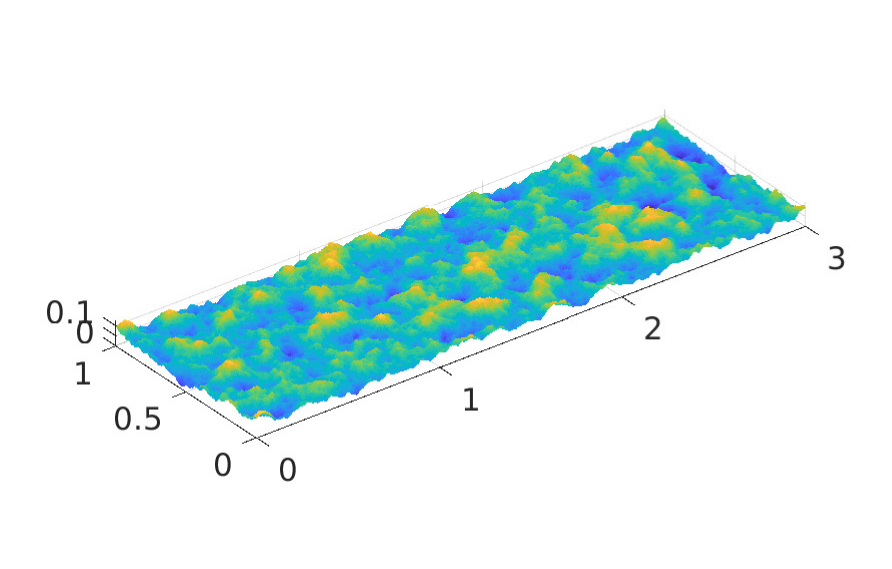}
\put(-72,20){$x/\delta$}
\put(-186,10){$z/\delta$}
\put(-183,70){(a)}
\includegraphics[width=65mm,trim={0.5cm 1.2cm 0.5cm 1.8cm},clip]{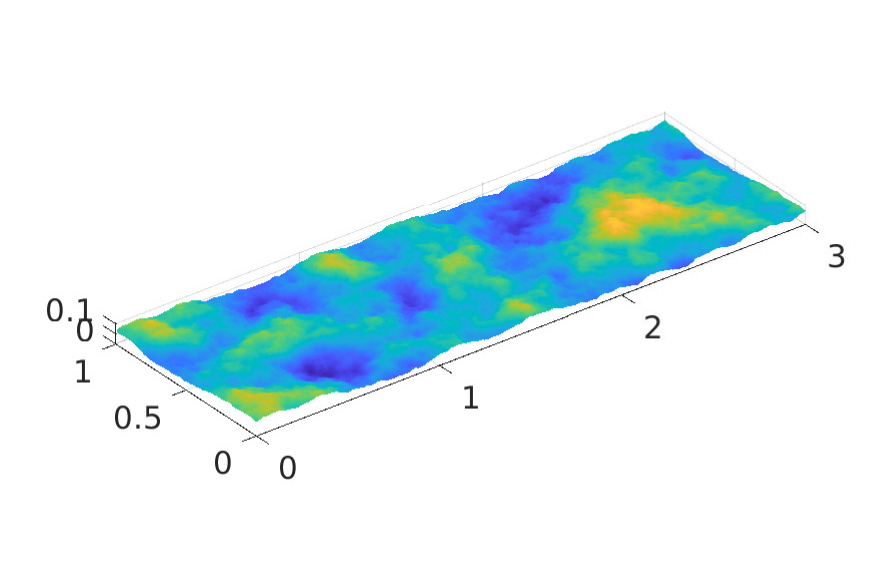}
\put(-72,20){$x/\delta$}
\put(-186,10){$z/\delta$}
\put(-183,70){(b)}
\hspace{1mm}
\includegraphics[width=65mm,trim={0.5cm 1.2cm 0.5cm 1.8cm},clip]{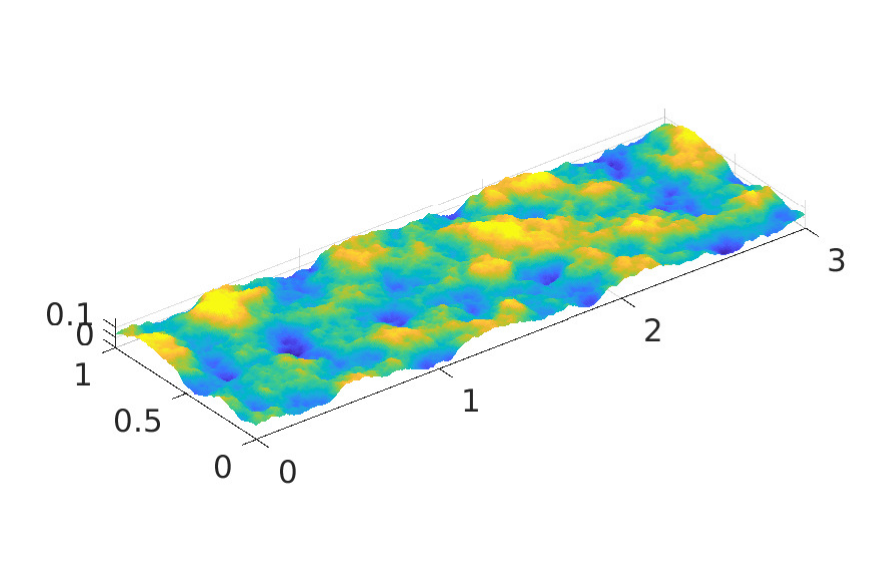}
\put(-72,20){$x/\delta$}
\put(-186,10){$z/\delta$}
\put(-183,70){(c)}
\includegraphics[width=65mm,trim={0.5cm 1.2cm 0.5cm 1.8cm},clip]{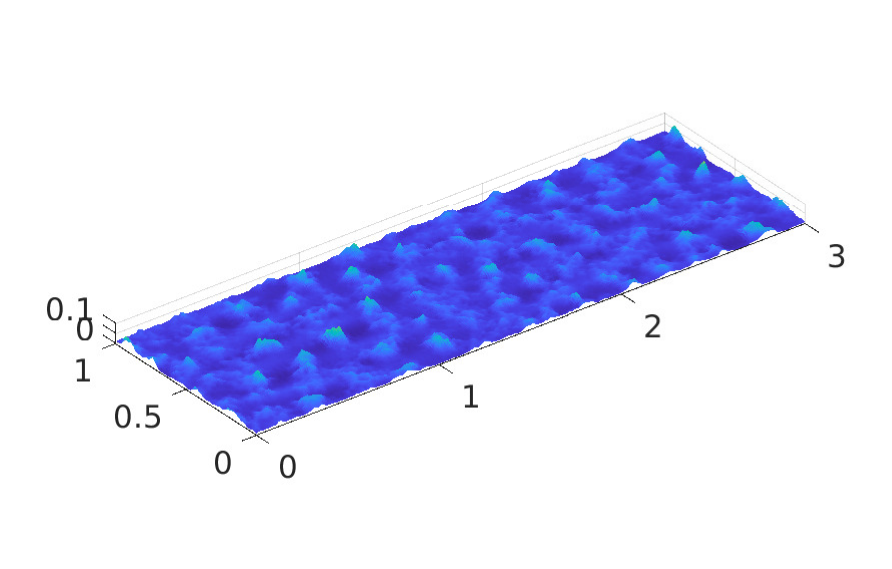}
\put(-72,20){$x/\delta$}
\put(-186,10){$z/\delta$}
\put(-183,70){(d)}
\hspace{1mm}
\includegraphics[width=65mm,trim={0.5cm 1.2cm 0.5cm 1.8cm},clip]{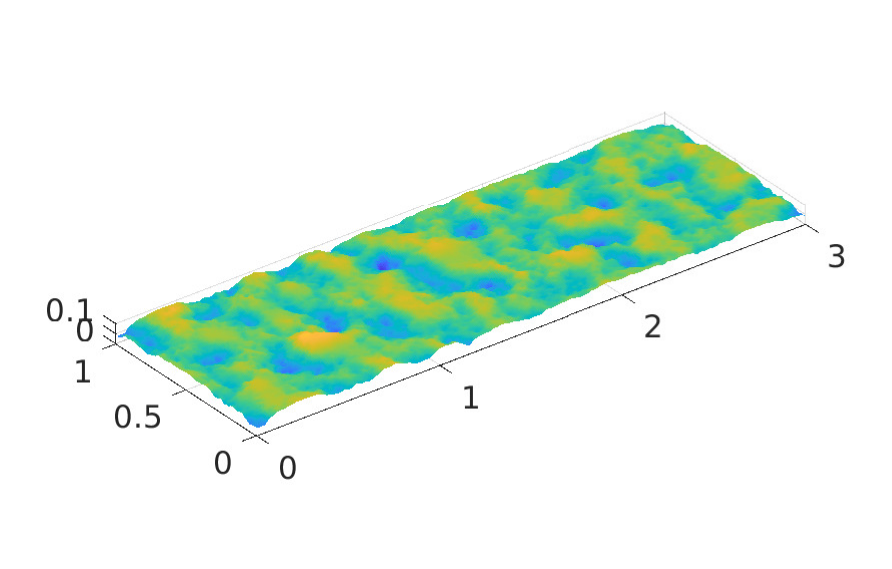}
\put(-72,20){$x/\delta$}
\put(-186,10){$z/\delta$}
\put(-183,70){(e)}
\includegraphics[width=65mm,trim={0.5cm 1.2cm 0.5cm 1.8cm},clip]{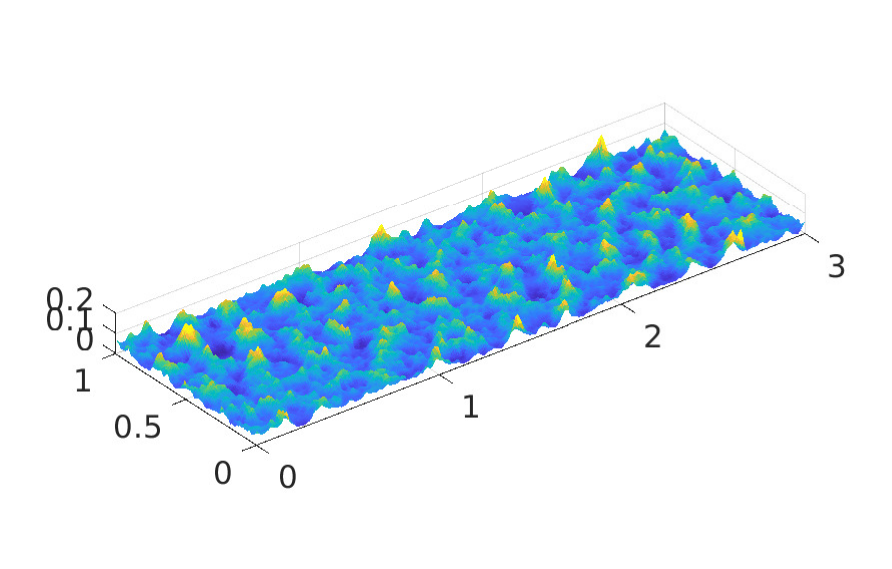}
\put(-72,20){$x/\delta$}
\put(-186,10){$z/\delta$}
\put(-183,70){(f)}
\caption{Visualization of roughness height for selected surface
  samples: (a)-(c) Gaussian roughness; (d)-(f) Weibull roughness. The
  colors represent the level of $k/\delta$ from 0.0 (blue) to 0.12
  (yellow), where $\delta$ is the channel half-height.}
\label{fig:rough_samples}
\end{figure}
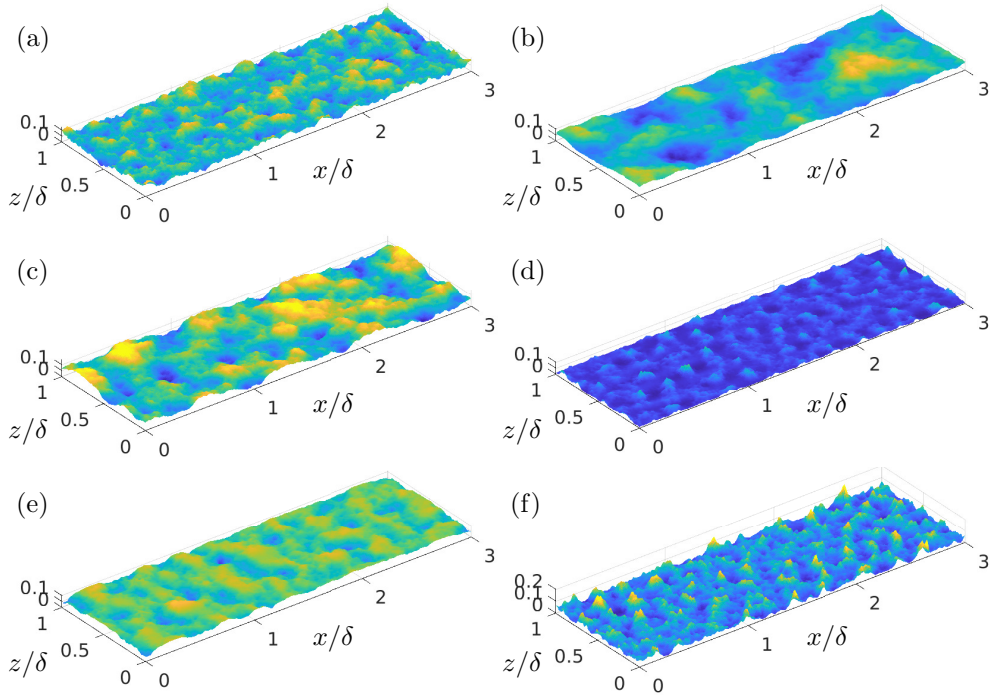

The Gaussian and Weibull rough surfaces considered are statistically
homogeneous along the wall-parallel directions. The Gaussian roughness
is generated based on the normal distribution of roughness height by
specifying the root-mean-square roughness height in a range from
0.005$\delta$ to 0.03$\delta$. The probability density function of the
Weibull distribution of the random variable $\Tilde{k}$ follows
\begin{equation}
    \text{PDF}_w(\Tilde{k}) = \frac{H}{\lambda} \biggl(\frac{\Tilde{k}}{\lambda}
    \biggr)^{H-1}e^{-(\frac{\Tilde{k}}{\lambda})^H}, \quad \Tilde{k} \ge 0
\end{equation}
where the shape parameter $H>0$ describes the shape of the probability
distribution and is randomly selected within [0.8, 2.3], and
$\lambda>0$ is the scale parameter. The PS describing the isotropic
self-affine fractal is:
\begin{equation}
    \begin{split}
      \text{PS}(\kappa) & = \kappa^{-2(1+H_f)}, \quad \kappa_0 \le \kappa \le \kappa_1 \\
      \text{PS}(\kappa) & = \kappa_0^{-2(1+H_f)}, \quad \kappa < \kappa_0
    \end{split}
\end{equation}
where $\kappa=\sqrt{\kappa_x^2+\kappa_z^2}$, and $\kappa_x$ and
$\kappa_z$ are the nondimensional wavenumbers in the streamwise $(x)$
and spanwise $(z)$ directions, respectively. The higher bound
wavenumber $\kappa_1=L_x/\lambda_1$ is set by giving the lower bound
of the roughness wavelength $\lambda_1 = 0.033\delta$ to ensure that
the smallest roughness length scales are resolved by adequate grid
points. The PS is controlled by two randomized parameters, the
roll-off (lower bound) wavenumber $\kappa_0$ and the Hurst exponent
$H_f$. The values of $\kappa_0$ are selected within the range [3,25],
and the values of $H_f$ are varied to obtain the power-law decline
rate $\theta$ within the range [-4,-3]. The resulting surface
generated based on the PDF and PS is then scaled from 0 to the
root-mean-square height $k_{rms}$ in a range from $0.005\delta$ to
$0.03\delta$. These values are determined to span the range of the
roughness parameters for the actual rough surfaces in engineering
applications \citep{bons2010review, kirschner2012bio}. The roughness
repository includes 50 Gaussian rough surfaces and 50 Weibull rough
surfaces. Six roughness samples are visualized in Figure
\ref{fig:rough_samples}.

The geometric properties of the roughness are characterized by
statistical quantities derived from the surface height
distribution. The definition of roughness parameters is given in table
\ref{tab:roughness_para}.  These include roughness height measures
such as mean height $k_{avg}$, first-order moment of height
fluctuations $R_a$, root-mean-square height $k_{rms}$, crest height
$k_c$, and mean peak-to-valley height $k_t$; high-order moments of
height fluctuations such as skewness $S_k$ and kurtosis $K_u$; height
gradients such as the effective slope $ES$, and inclination angle $I$;
surface porosity $P_o$; roughness density measures such as frontal
solidity $\lambda_f$; and the correlation length $L_{cor}$.  Note that
the rough surfaces considered are isotropic and the parameters $ES$,
$I$, and $L_{cor}$ are equivalent along any wall-parallel direction.
\begin{table}
\begin{center}
\def~{\hphantom{0}}
\renewcommand{\arraystretch}{2}
    \begin{tabular}{c|c}
     \hline
     Mean height & $k_{avg} = \frac{1}{A_t}\int_{x,z} k(x,z) dA$ \\
     Crest height & $k_c = \max\{k(x,z)\} -\min\{k(x,z)\}$ \\
     Mean peak-to-valley height & $k_t = \text{mean}\{\max|_{\delta \times \delta}\{k(x,z)\} - \min|_{\delta \times \delta}\{k(x,z)\}\}$ \\
     Root-mean-square height & $k_{rms} = \sqrt{\frac{1}{A_t}\int_{x,z} (k(x,z)-k_{avg})^2 dA}$ \\
     First-order moment of height fluctuations & $R_a = \frac{1}{A_t}\int_{x,z} |k(x,z)-k_{avg}| dA$ \\
     Skewness & $S_k = \frac{1}{A_tk_{rms}^3}\int_{x,z} (k(x,z)-k_{avg})^3 dA$ \\
     Kurtosis & $K_u = \frac{1}{A_tk_{rms}^4}\int_{x,z} (k(x,z)-k_{avg})^4 dA$ \\
     Effective slope & $ES = \frac{1}{A_t}\int_{x,z} |\frac{\partial k(x,z)}{\partial x}| dA$ \\
     Inclination angle & $I = \tan^{-1}\left(\frac{1}{2}S_k\{\frac{\partial k(x,z)}{\partial x}\}\right)$ \\
     Surface porosity & $P_o = \frac{1}{A_tk_c}\int_{0}^{k_c} A_f(y) dy$ \\
     Frontal solidity & $\lambda_f = \frac{A_p}{A_t}$ \\
     Correlation length &  $L_{cor}=\min_{\delta x }\{R_h(\delta x,0) \le 0.2 \}$\\
     \hline
    \end{tabular}
    \caption{\label{tab:roughness_para} Definitions of roughness
      geometrical parameters. $k(x,z)$ is the roughness height
      function, $A_f(y)$ is the fluid area at the $y$ location, $A_p$
      is the frontal projected area of the roughness elements, and
      $A_t$ is the total plan area. The correlation lengths are
      computed as the horizontal separation at which the roughness
      height autocorrelation function $R_h(\delta x,\delta
      z)=\frac{1}{k_{rms}^2}\langle k(x+\delta x, z+\delta z)k(x,z)
      \rangle_{xz}$ drops below 0.2, where $\langle \cdot
      \rangle_{xz}$ denotes average over $x$ and $z$. Given that the
      rough surfaces considered are isotropic, the parameters $ES$,
      $I$, and $L_{cor}$ are equivalent along any wall-parallel
      direction.  Similar definitions of roughness parameters can be
      found in \cite{thakkar2017surface, ma2021direct, jouybari2021data} and
      \cite{chung2021predicting}.  }
    \end{center}
\end{table}

\subsection{Active learning}\label{sec:AL}

We employ active learning (AL) to efficiently build the training
repository and minimize the computational expenses associated with
DNS. Ideally, DNS simulations of turbulent channel flow using all
surfaces from the roughness repository would be conducted to maximize
the amount of training data. However, in practice, the number of DNS
simulations we can perform is constrained by computational
resources. The AL approach iteratively selects the most valuable rough
surfaces from the repository for DNS simulations. AL focuses on
finding the most informative cases to enhance model performance while
reducing labeling costs~\citep{settles2009active}. This strategy
ensures effective exploration of the repository by incorporating DNS
data that is most useful for training robust, generalizable models.

The AL approach is implemented using a Gaussian process (GP) model to
predict the uncertainty from unseen roughness
surfaces~\citep{rasmussen2006gaussian}. The GP model is a
non-parametric method based on the assumption that the function to be
learned is drawn from a Gaussian process. This assumption enables the
model to make predictions with well-defined uncertainty.  The inputs
to the GP model include all roughness parameters described in
Table~\ref{tab:roughness_para}. The output is the non-dimensionalized
wall-shear stress $\langle\tau_w\rangle y_1/(\nu U_1)$, obtained from
DNS of turbulent channel flows. Here, $y_1$ is the wall-normal
distance, $\nu$ is the kinematic viscosity of the fluid,
$\langle\tau_w\rangle$ represents the mean wall shear stress (averaged
over the homogeneous directions and time), and $U_1 = U(y_1)$ is the
mean wall-parallel velocity magnitude at $y_1$. Detailed computations
of $\langle\tau_w\rangle$, $U(y_1)$, and the wall-normal locations
considered are presented in \S\ref{sec:DNS}. The GP model is defined
by a mean function and a covariance kernel. A zero mean function is
used as the prior mean function, and a squared exponential kernel
serves as the prior covariance function. The posterior distribution,
given the observed data, is obtained from the prior distribution and
is used to predict the uncertainty (namely, predictive variance
$\sigma^2$) of rough surfaces unseen by the GP model. During the
training process, the optimal hyperparameters for the GP model are
determined by minimizing the negative logarithm of the marginal
likelihood. Readers are referred to \cite{rasmussen2006gaussian} for
additional details about the GP model algorithm.

The steps for AL, summarized in Figure~\ref{fig:AL}, are as follows:
\begin{figure}
\centering
\includegraphics[width=130mm]{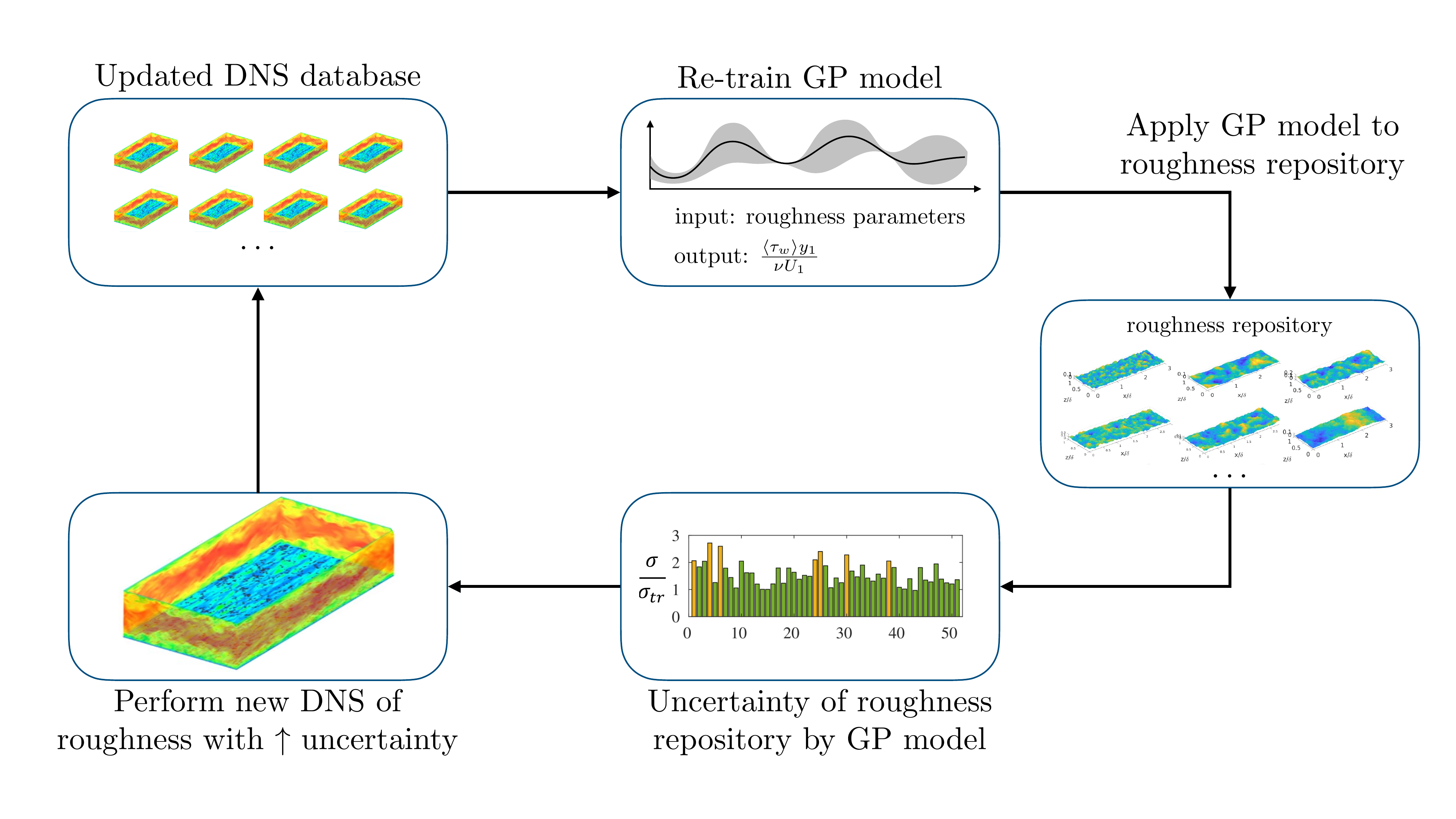}
\caption{Schematic of the
  active learning to select the rough surfaces for the DNS turbulent
  channel database.}
\label{fig:AL}
\end{figure}
%
\begin{enumerate}
  \item Initialization: A small set of DNS roughness data is used to
    initialize the process.
  \item GP model training: A GP model is trained using the initial
    labeled DNS data.
  \item Uncertainty sampling: The GP model is used to predict the
    uncertainty for all rough surfaces in the repository. Rough
    surfaces with the highest uncertainty are selected.
  \item New data generation: DNS of turbulent channel flows over the
    selected rough surfaces is performed at various Reynolds numbers.
  \item Model update: The newly labeled DNS data are added to the
    training set, and the GP model is retrained.
  \item Iteration: Steps (iii) through (v) are repeated until a
    stopping criterion is met.
\end{enumerate}
The rough surfaces selected through the AL framework are labeled as
GS$\#$ and WB$\#$, where GS and WB denote Gaussian and Weibull
roughness, respectively. The value of $\#$ is the ID number of the
surface used to locate the case in table \ref{tab:roughness_features},
where the properties of roughness topography are listed. At each
iteration, about $15\%$ of the total number of rough surfaces is
selected for performing DNS. This value is sufficient to reduce
uncertainty of the rough surfaces in the roughness repository at each
iteration, and was constrained by our computational resources to
conduct new DNS cases in each iteration. A total of four iterations of
the GP model are performed after which the value of
$\sigma^2/\sigma^2_{tr}$ is less than 2.5 for the whole roughness
repository, where $\sigma^2_{tr}$ is averaged variance of the last GP
model.
\begin{figure}
\centering
\vspace{2mm}
\includegraphics[width=125mm,trim={2.0cm 0.0cm 1.5cm 0.0cm},clip]{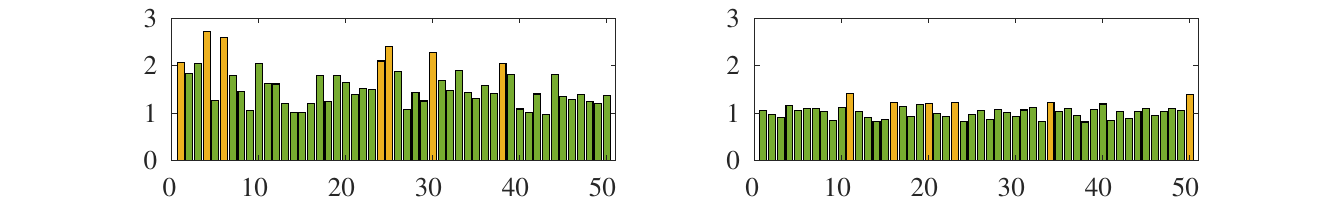}
\put(-325,-8){Index of Gaussian roughness}
\put(-363,23){\rotatebox{90}{$\sigma^2/\sigma^2_{tr}$}}
\put(-363,60){(a)}
\put(-330,62){1-st iteration using GP model-1}
\put(-140,-8){Index of Gaussian roughness}
\put(-179,23){\rotatebox{90}{$\sigma^2/\sigma^2_{tr}$}}
\put(-179,60){(b)}
\put(-146,62){2-nd iteration using GP model-2}
\vspace{3mm}
\hspace{3mm}
\includegraphics[width=125mm,trim={2.0cm 0.0cm 1.5cm 0.0cm},clip]{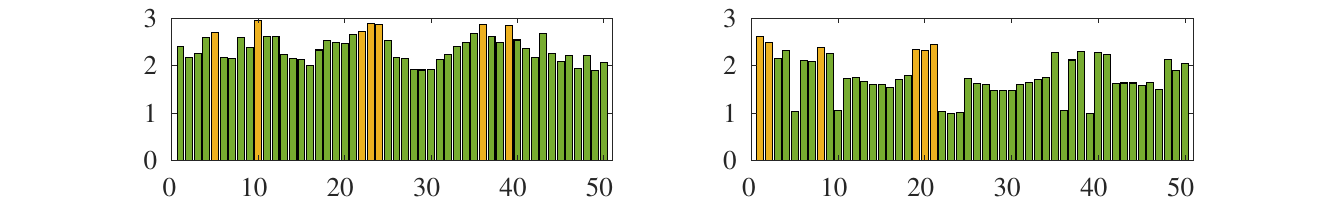}
\put(-322,-8){Index of Weibull roughness}
\put(-363,23){\rotatebox{90}{$\sigma^2/\sigma^2_{tr}$}}
\put(-363,60){(c)}
\put(-330,62){3-rd iteration using GP model-3}
\put(-137,-8){Index of Weibull roughness}
\put(-179,23){\rotatebox{90}{$\sigma^2/\sigma^2_{tr}$}}
\put(-179,60){(d)}
\put(-146,62){4-th iteration using GP model-4}
\vspace{2mm}
\caption{Uncertainty ($\sigma^2$) for the rough surfaces in the
  repository normalized by the mean uncertainty of the most updated GP
  model ($\sigma^2_{tr}$). In the first two iterations of the GP model,
  only Gaussian roughness is considered. In the third and fourth
  iterations, Weibull roughness is considered. (a) the 1-st iteration
  using GP model-1; (b) the 2-nd iteration using GP model-2; (c) the
  3-rd iteration using GP model-3; (d) the 4-th iteration using GP
  model-4. The surfaces with the highest prediction variance colored
  by yellow are selected for performing DNS.}
\label{fig:AL_variance}
\end{figure}

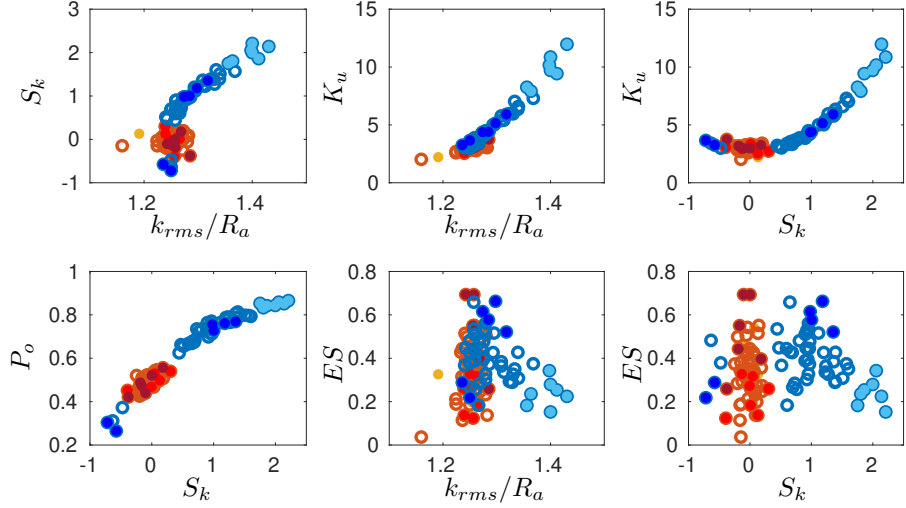
\begin{figure}
\centering
\includegraphics[width=140mm]{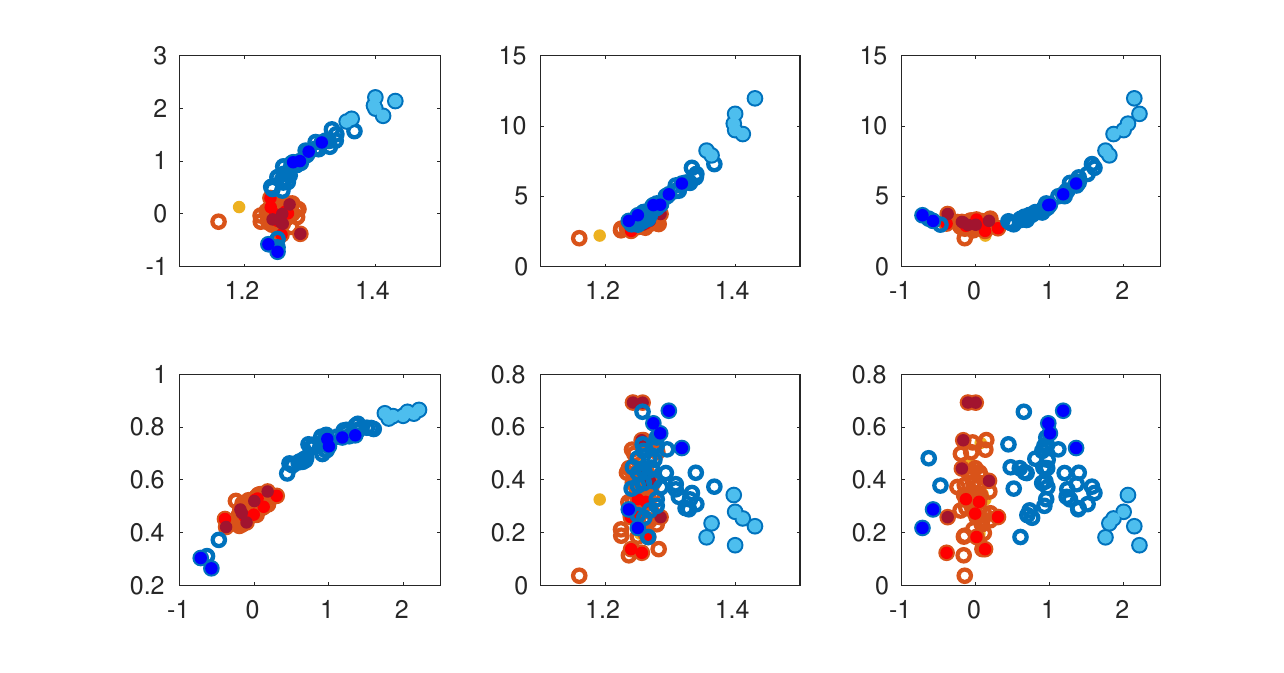}
\put(-320,107){$k_{rms}/R_a$}
\put(-368,155){\rotatebox{90}{$S_k$}}
\put(-210,107){$k_{rms}/R_a$}
\put(-255,155){\rotatebox{90}{$K_u$}}
\put(-85,107){$S_k$}
\put(-308,8){$S_k$}
\put(-372,55){\rotatebox{90}{$P_o$}}
\put(-143,155){\rotatebox{90}{$K_u$}}
\put(-210,8){$k_{rms}/R_a$}
\put(-255,50){\rotatebox{90}{$ES$}}
\put(-85,8){$S_k$}
\put(-143,50){\rotatebox{90}{$ES$}}
\caption{Scatter plots of roughness parameters for the sufaces
  selected for each iteration in AL. The roughness repository is
  circled and the roughness in the training set is filled. Gaussian
  roughness repository (red); Weibull roughness repository (blue);
  Initial GS01 to GS06 (yellow); GS07 to GS13 at the 1-st iteration
  (light red); GS14 to GS19 at the 2-nd iteration (dark red); WB01 to
  WB07 at the 3-rd iteration (light blue); WB08 to WB13 at the 4-th
  iteration (dark blue).  }
\label{fig:roughness_stats}
\end{figure}
%
As a starting point, six Gaussian rough surfaces not included in the
roughness repository, GS01 to GS06, are generated. These six rough
surfaces are created using the same method as the other Gaussian rough
surfaces in the roughness repository. DNS of turbulent channel flows
are performed for each roughness at six different $Re_{\tau}$ values:
180, 360, 540, 720, 900, and 1000. This data is used to initialize the
process and train the first GP model (GP model-1). For the Gaussian
rough surfaces, two iterations are conducted to improve the initial GP
model. Figure \ref{fig:AL_variance}(a) shows that seven new surfaces,
GS07 to GS13, with the highest uncertainty, are selected from the
roughness repository in the first iteration. Figure
\ref{fig:AL_variance}(b) shows that six new surfaces, GS14 to GS19,
are selected in the second iteration. The reduced uncertainty in the
second iteration, compared to the first, demonstrates that the current
strategy effectively explores the repository by adding new data in the
most uncertain regions of the parameter space. GP model-3 is used to
test the 50 Weibull rough surfaces from the repository. Figure
\ref{fig:AL_variance}(c) shows that seven Weibull rough surfaces, WB01
to WB07, with the highest prediction variance, are selected in the
third iteration. GP model-4 is then trained with the updated DNS data
and used to test the Weibull roughness in one more iteration. As shown
in Figure \ref{fig:AL_variance}(d), six additional Weibull rough
surfaces, WB08 to WB13, are selected in the fourth iteration.  In
summary, the resulting training database contains a total of 19
Gaussian and 13 Weibull rough surfaces, which are used to conduct DNS
of turbulent channel flows at six different $Re_{\tau}$ values. As a
result, the DNS roughness database includes 192 cases. The statistical
parameters of the rough surfaces in the DNS database are summarized in
Table \ref{tab:roughness_features}.
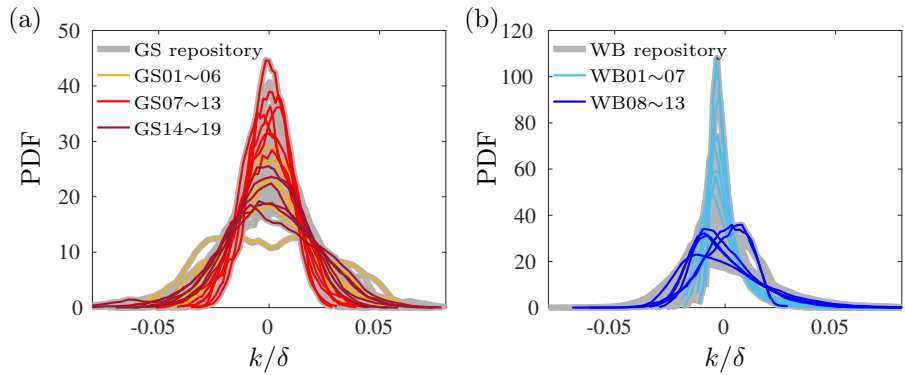
\begin{figure}
\centering
\includegraphics[width=60mm]{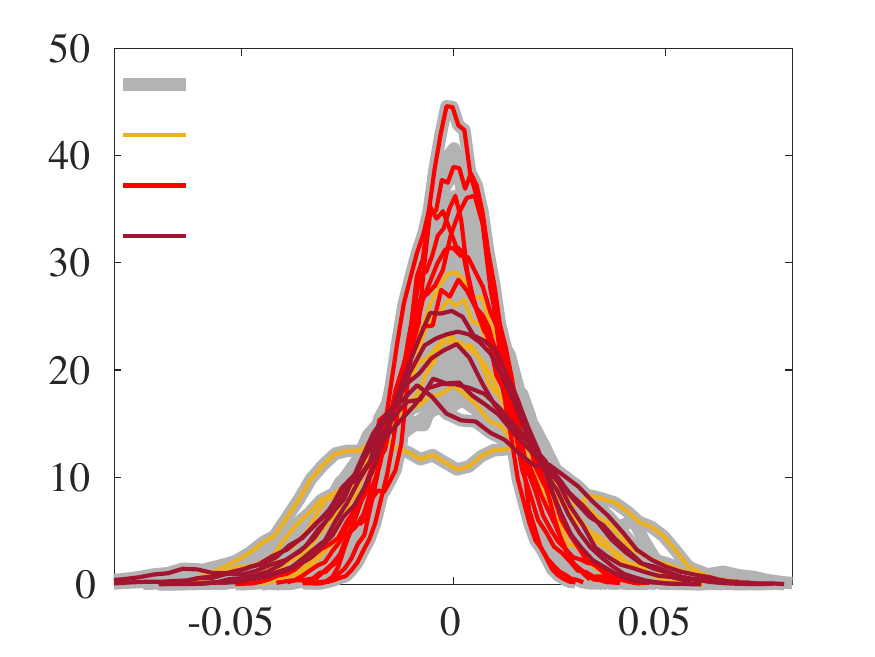}
\put(-180,120){(a)}
\put(-90,-8){$k/\delta$}
\put(-133,109){\scriptsize{GS repository}}
\put(-133,99){\scriptsize{GS01$\sim$06}}
\put(-133,89){\scriptsize{GS07$\sim$13}}
\put(-133,79){\scriptsize{GS14$\sim$19}}
\put(-175,60){\rotatebox{90}{PDF}}
\includegraphics[width=60mm]{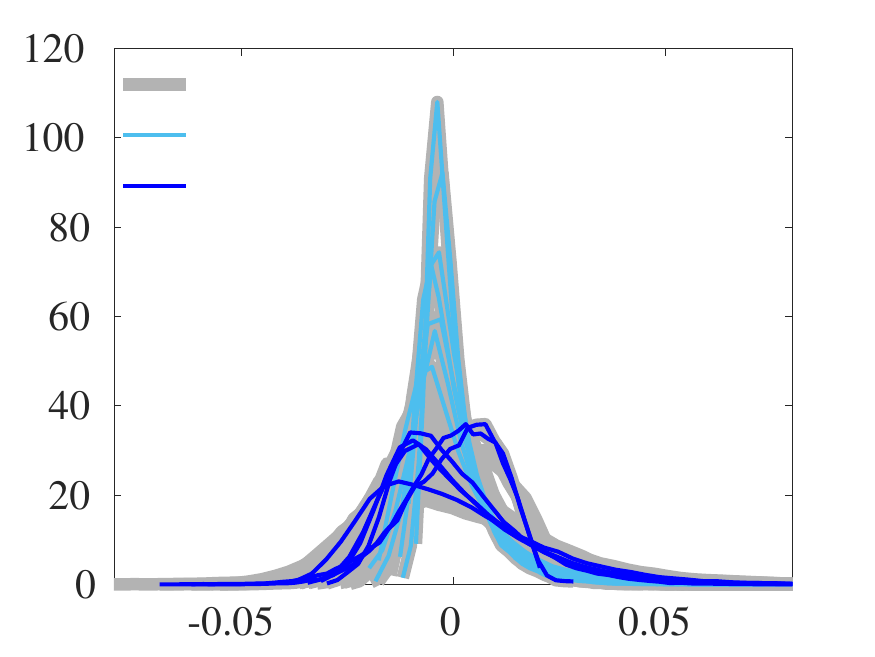}
\put(-180,120){(b)}
\put(-90,-8){$k/\delta$}
\put(-133,109){\scriptsize{WB repository}}
\put(-133,99){\scriptsize{WB01$\sim$07}}
\put(-133,89){\scriptsize{WB08$\sim$13}}
\put(-175,60){\rotatebox{90}{PDF}}
\caption{The PDF of roughness height for rough surfaces in the
  roughness repository and rough surfaces selected at different
  iterations in AL for (a) Gaussian roughness and (b) Weibull
  roughness.}
\label{fig:pdf_ps}
\end{figure}
%
\begin{table}
\begin{center}
\def~{\hphantom{0}}
\setlength\extrarowheight{-5pt}
\renewcommand{\arraystretch}{2}
    \begin{tabular}{cccccccccccccccc}
     \hline \\
      Case &  $k_{avg}/\delta$ &  $k_c/\delta$ &  $k_t/\delta$ &  $k_{rms}/\delta$ &  $R_a/\delta$ &  $S_k$ &  $K_u$ &  $ES$ &  $I$ &  $P_o$ &  $\lambda_f$ &  $L_{cor}/\delta$  \\[3pt] 
      GS01 &  0.062 &  0.125 &  0.120 &  0.018 &  0.014 &  0.068 &  2.902 &  0.540 &   -0.001 &   0.503 &  0.290 &  0.082  \\
      GS02 &  0.069 &  0.143 &  0.125 &  0.027 &  0.023 &  0.130 &  2.211 &  0.326 &   0.004  &   0.517 &  0.161 &  0.486  \\
      GS03 &  0.052 &  0.107 &  0.098 &  0.019 &  0.015 &  0.051 &  2.793 &  0.197 &   0.015  &   0.510 &  0.102 &  0.245  \\
      GS04 &  0.049 &  0.097 &  0.090 &  0.015 &  0.012 &  -0.001 &  2.752 &  0.290 &  -0.029 &   0.497 &  0.153 &  0.139  \\
      GS05 &  0.053 &  0.102 &  0.098 &  0.014 &  0.011 &  -0.095 &  2.974 &  0.463 &  -0.001 &   0.477 &  0.240 &  0.110  \\
      GS06 &  0.069 &  0.141 &  0.137 &  0.023 &  0.018 &  0.091 &  2.778 &  0.347 &   0.016  &   0.510 &  0.179 &  0.195  \\
      GS07 &  0.037 &  0.070 &  0.068 &  0.009 &  0.007 &  -0.007 &  3.046 &  0.272 &  -0.005 &   0.468 &  0.136 &  0.098  \\
      GS08 &  0.038 &  0.069 &  0.064 &  0.010 &  0.008 &  -0.392 &  3.065 &  0.124 &  0.025  &   0.452 &  0.056 &  0.234  \\
      GS09 &  0.035 &  0.069 &  0.066 &  0.012 &  0.010 &  0.127 &  2.543 &  0.137 &   0.006  &   0.498 &  0.065 &  0.257  \\
      GS10 &  0.042 &  0.089 &  0.084 &  0.013 &  0.010 &  0.045 &  3.078 &  0.317 &   -0.012 &   0.530 &  0.158 &  0.176  \\
      GS11 &  0.045 &  0.094 &  0.084 &  0.012 &  0.009 &  0.011 &  3.352 &  0.184 &   0.005  &   0.525 &  0.095 &  0.202  \\
      GS12 &  0.054 &  0.101 &  0.099 &  0.015 &  0.012 &  -0.132 &  2.908 &  0.327 &  0.001  &   0.463 &  0.160 &  0.250  \\
      GS13 &  0.036 &  0.077 &  0.070 &  0.011 &  0.009 &  0.301 &  2.733 &  0.259 &   -0.009 &   0.540 &  0.143 &  0.164  \\
      GS14 &  0.073 &  0.131 &  0.128 &  0.016 &  0.013 &  -0.104 &  2.967 &  0.694 &  -0.003 &   0.439 &  0.329 &  0.084  \\
      GS15 &  0.066 &  0.149 &  0.127 &  0.019 &  0.015 &  0.180 &  3.261 &  0.397 &   -0.008 &   0.556 &  0.213 &  0.148  \\
      GS16 &  0.100 &  0.173 &  0.170 &  0.025 &  0.019 &  -0.378 &  3.743 &  0.259 &  0.002  &   0.421 &  0.134 &  0.251  \\
      GS17 &  0.088 &  0.167 &  0.159 &  0.022 &  0.017 &  -0.167 &  3.146 &  0.551 &  0.009  &   0.475 &  0.277 &  0.119  \\
      GS18 &  0.066 &  0.128 &  0.118 &  0.016 &  0.012 &  -0.186 &  3.180 &  0.443 &  0.012  &   0.488 &  0.222 &  0.122  \\
      GS19 &  0.074 &  0.154 &  0.151 &  0.022 &  0.017 &  0.001 &  2.981 &  0.693 &   0.004  &   0.520 &  0.357 &  0.082  \\
      WB01 &  0.015 &  0.091 &  0.085 &  0.009 &  0.007 &  1.804 &  7.915 &  0.235 &   0.028  &   0.832 &  0.110 &  0.056  \\
      WB02 &  0.023 &  0.154 &  0.127 &  0.010 &  0.007 &  2.140 &  11.970 &  0.224 &  0.057  &   0.851 &  0.097 &  0.068  \\
      WB03 &  0.013 &  0.090 &  0.070 &  0.008 &  0.006 &  1.752 &  8.254 &  0.182 &   0.069  &   0.853 &  0.107 &  0.060  \\
      WB04 &  0.015 &  0.094 &  0.077 &  0.007 &  0.005 &  1.997 &  9.711 &  0.279 &   -0.028 &   0.843 &  0.129 &  0.033  \\
      WB05 &  0.022 &  0.157 &  0.139 &  0.013 &  0.009 &  2.055 &  10.171 &  0.343 &  0.072  &   0.859 &  0.196 &  0.053  \\
      WB06 &  0.011 &  0.084 &  0.074 &  0.007 &  0.005 &  2.210 &  10.867 &  0.152 &  0.078  &   0.866 &  0.064 &  0.094  \\
      WB07 &  0.025 &  0.091 &  0.085 &  0.009 &  0.007 &  1.804 &  7.915 &  0.235 &   0.028  &   0.832 &  0.146 &  0.066  \\
      WB08 &  0.072 &  0.104 &  0.098 &  0.012 &  0.010 &  -0.718 &  3.655 &  0.218 &  0.038  &   0.304 &  0.107 &  0.091  \\
      WB09 &  0.064 &  0.087 &  0.087 &  0.011 &  0.009 &  -0.575 &  3.254 &  0.288 &  -0.030 &   0.265 &  0.145 &  0.080  \\
      WB10 &  0.034 &  0.155 &  0.148 &  0.015 &  0.013 &  1.355 &  5.913 &  0.521 &   0.016  &   0.769 &  0.233 &  0.050  \\
      WB11 &  0.034 &  0.123 &  0.116 &  0.014 &  0.011 &  1.002 &  4.404 &  0.577 &   -0.001 &   0.728 &  0.294 &  0.035  \\
      WB12 &  0.042 &  0.171 &  0.161 &  0.020 &  0.016 &  0.980 &  4.375 &  0.615 &   -0.007 &   0.755 &  0.336 &  0.055  \\
      WB13 &  0.039 &  0.164 &  0.154 &  0.017 &  0.013 &  1.181 &  5.146 &  0.663 &   -0.032 &   0.761 &  0.408 &  0.037  \\
     \hline
    \end{tabular}
    \caption{\label{tab:roughness_features} Roughness parameters for
      rough surfaces in the DNS database. 
      }
    \end{center}
\end{table}

Scatter plots of roughness parameters and the PDFs of roughness height
are displayed in Figures \ref{fig:roughness_stats} and
\ref{fig:pdf_ps}, respectively. The results illustrate the
distribution of selected roughness at each iteration, demonstrating
how the AL framework assists in exploring uncertain regions within the
input roughness feature space. Figure \ref{fig:roughness_stats} also
reveals strong correlations between $k_{rms}/R_a$ and $S_k$,
$k_{rms}/R_a$ and $K_u$, $S_k$ and $K_u$, as well as $S_k$ and $P_o$.
The high correlation among different roughness parameters suggests
that only a reduced set of them might be required as input variables
for the wall model.

\subsection{DNS of rough-wall turbulent channel flows}\label{sec:DNS}


DNS of turbulent channel flows with the rough surfaces selected from
\S\ref{sec:AL} is performed to generate the training database. The
governing equations for momentum and continuity are given by the
incompressible Navier-Stokes equations:
\begin{equation}
\frac{\partial u_i}{\partial t} + \frac{\partial u_i u_j}{\partial x_j} = -\frac{1}{\rho}\frac{\partial p}{\partial x_i} + \nu \frac{\partial^2 u_i}{\partial x_j x_j} + F_i, \quad \frac{\partial u_i}{\partial x_i} = 0,
\label{eqn:nsme}
\end{equation}
where $u_i$ is the $i$-th component of the velocity (streamwise:
$i=1$, wall-normal: $i=2$, spanwise: $i=3$), $p$ denotes the pressure,
$\rho$ is the fluid density, and $\nu$ is the kinematic viscosity of
the fluid. An immersed boundary approach based on the volume-of-fluid
method is used, where the no-slip boundary condition on the rough
surface is enforced by the body force $F_i$ \citep{scotti2006direct,
  yuan2014numerical}. The solver utilizes second-order central finite
differences for spatial discretization, second-order Adams-Bashforth
semi-implicit time advancement, and is parallelized using a message
passing interface (MPI) method \citep{keating2004large}. The code has
been extensively validated in previous investigations of rough-wall
turbulence~\citep{yuan2014numerical, yuan2014roughness,
  yuan2018topographical, jouybari2021data}.
\begin{table}
  \begin{center}
    \def~{\hphantom{0}}
    \begin{tabular}{lccccccc}
    \hline
      $Re_{\tau}$ & $N_x\times N_y\times N_z$ & $L_x/\delta\times L_y/\delta\times L_z/\delta$ & $\Delta x^{+}$ & $\Delta z^{+}$ & $\Delta y^{+}_{\min}$ & $\Delta y^{+}_{\max}$\\[3pt] \hline
      180  & $400\times 300 \times 160$ & $3\times 1\times 1$ & $1.35$ & $1.13$ & $0.03$ & $1.66$\\
      360  & $400\times 300 \times 160$ & $3\times 1\times 1$ & $2.70$ & $2.25$ & $0.05$ & $3.33$\\
      540  & $400\times 300 \times 160$ & $3\times 1\times 1$ & $4.05$ & $3.38$ & $0.08$ & $4.99$\\
      720  & $400\times 300 \times 160$ & $3\times 1\times 1$ & $5.40$ & $4.50$ & $0.11$ & $6.65$\\
      900  & $400\times 300 \times 160$ & $3\times 1\times 1$ & $6.75$ & $5.63$ & $0.14$ & $8.32$\\
      1000 & $400\times 300 \times 160$ & $3\times 1\times 1$ & $7.50$ & $6.25$ & $0.15$ & $9.24$\\\hline
    \end{tabular}
    \caption{\label{tab:DNS} Simulation parameters for DNS of
      rough-wall channel flows at different $Re_{\tau}$. $N_x$, $N_y$,
      and $N_z$ are the number of grid points in the streamwise,
      wall-normal, and spanwise direction, respectively, $L_x$, $L_y$,
      and $L_z$ are the streamwise, wall-normal, and spanwise
      dimensions of the computational domain, $\Delta x^{+}$ and
      $\Delta z^{+}$ are the streamwise and spanwise grid resolutions,
      and $\Delta y^{+}_{\min}$ and $\Delta y^{+}_{\max}$ are the
      minimum and maximum wall-normal grid resolutions. Uniform grids
      are used in the streamwise and spanwise directions, and
      non-uniform grids with a hyperbolic tangent function are used in
      the wall-normal direction. The number of grid points is kept
      constant across $Re_{\tau}$ to resolve the roughness features
      and avoid interpolation between cases.}
  \end{center}
\end{table}
  
To cover both transitionally and fully rough regimes, turbulent
open-channel flows are simulated at six different frictional Reynolds
numbers: $Re_{\tau}=180, 360, 540, 720, 900, 1000$. A minimal-span
channel simulation approach is used to enhance computational
efficiency~\citep{jimenez1991minimal, chung2015fast,
  macdonald2017minimal}. \cite{chung2015fast} and
\cite{macdonald2017minimal} demonstrated that simulations in a
minimal-span domain can accurately capture the near-wall flow dynamics
by adhering to the domain constraints. The constraints proposed by
\cite{chung2015fast} and \cite{macdonald2017minimal} are:
\begin{gather}
    L_x \ge \max(3L_z, 1000\nu/u_{\tau}, \lambda_{r,x}), \\
    L_y \ge k_{ch}/0.15, \\
    L_z \ge \max(100\nu/u_{\tau}, k_{ch}/0.4, \lambda_{r,z}),
\end{gather}
where $L_x$, $L_y$, and $L_z$ are the domain lengths in the
streamwise, wall-normal, and spanwise directions, respectively;
$k_{ch}$ is the characteristic roughness height; and $\lambda_{r,x}$
and $\lambda_{r,z}$ are the streamwise and spanwise length scales of
the roughness elements. The crest roughness height $k_c$ is used as
the characteristic roughness height, and surface Taylor microscales
$\lambda_{T,x}$ and $\lambda_{T,z}$ are used as the streamwise and
spanwise roughness length scales for multiscale random roughness,
following \cite{jouybari2021data}. The simulation domain size is
$(L_x, L_y, L_z)=(3\delta, \delta, \delta)$, based on the criteria for
small-span channel simulations.  Periodic boundary conditions are used
in the streamwise and spanwise directions, and no-slip and symmetry
boundary conditions are imposed at the bottom surfaces and the top
boundary. The grid size is determined to ensure that roughness
elements are well resolved by at least four grid points per
$\lambda_{T,x}$ and $\lambda_{T,z}$, as recommended by
\cite{yuan2014estimation}. Note that although the simulations are
conducted in open channels, we still refer to $\delta$ as the channel
half-height.  The simulation details are shown in Table \ref{tab:DNS}.

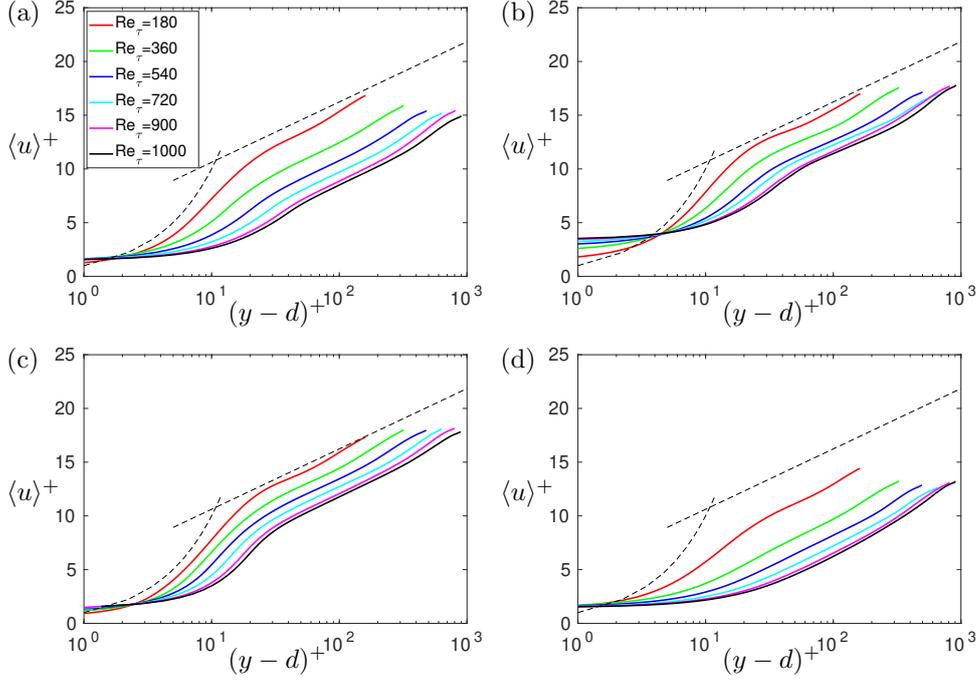
\begin{figure}
\centering
\includegraphics[width=65mm]{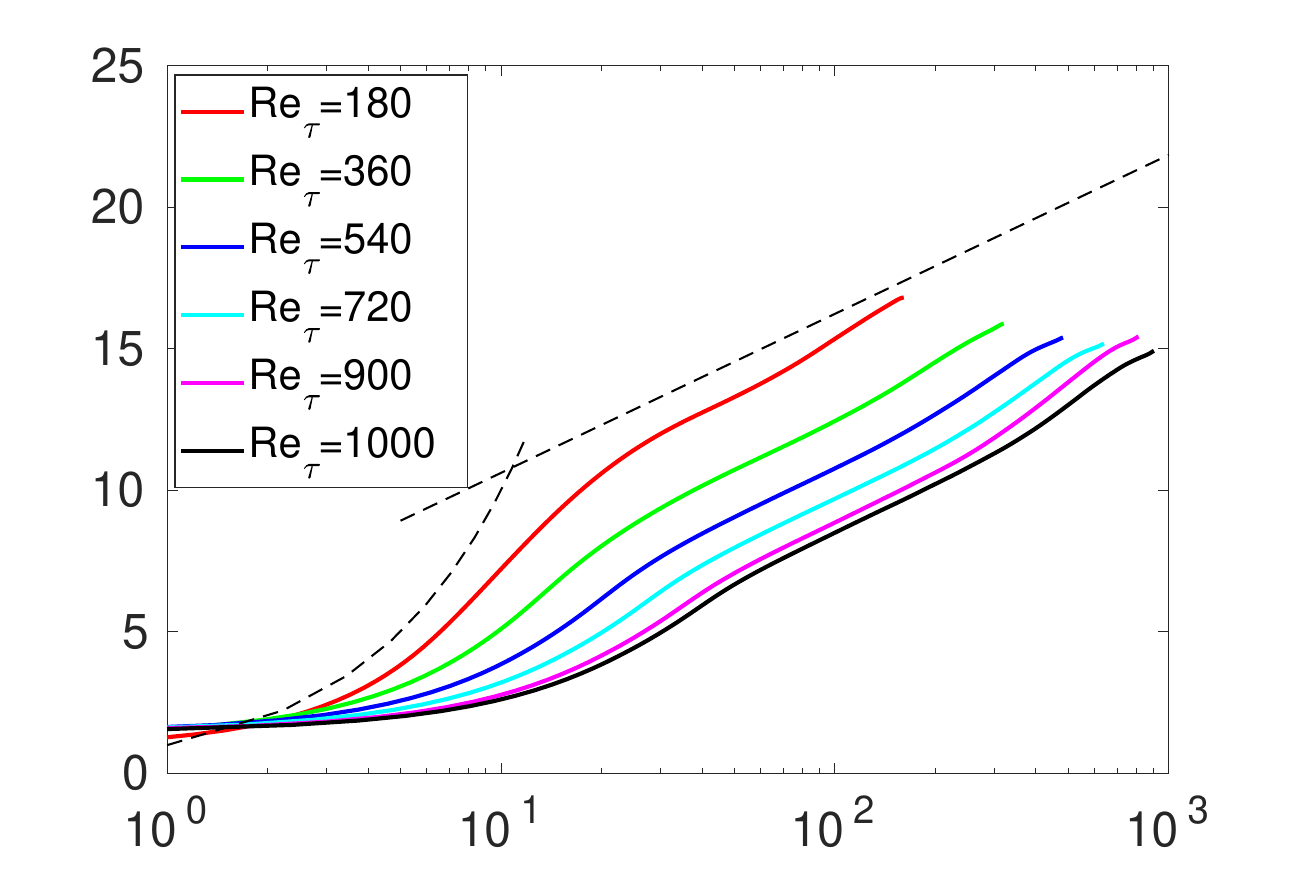}
\put(-107,-3){$(y-d)^+$}
\put(-190,60){$\langle u\rangle^+$}
\put(-190,110){(a)}
\includegraphics[width=65mm]{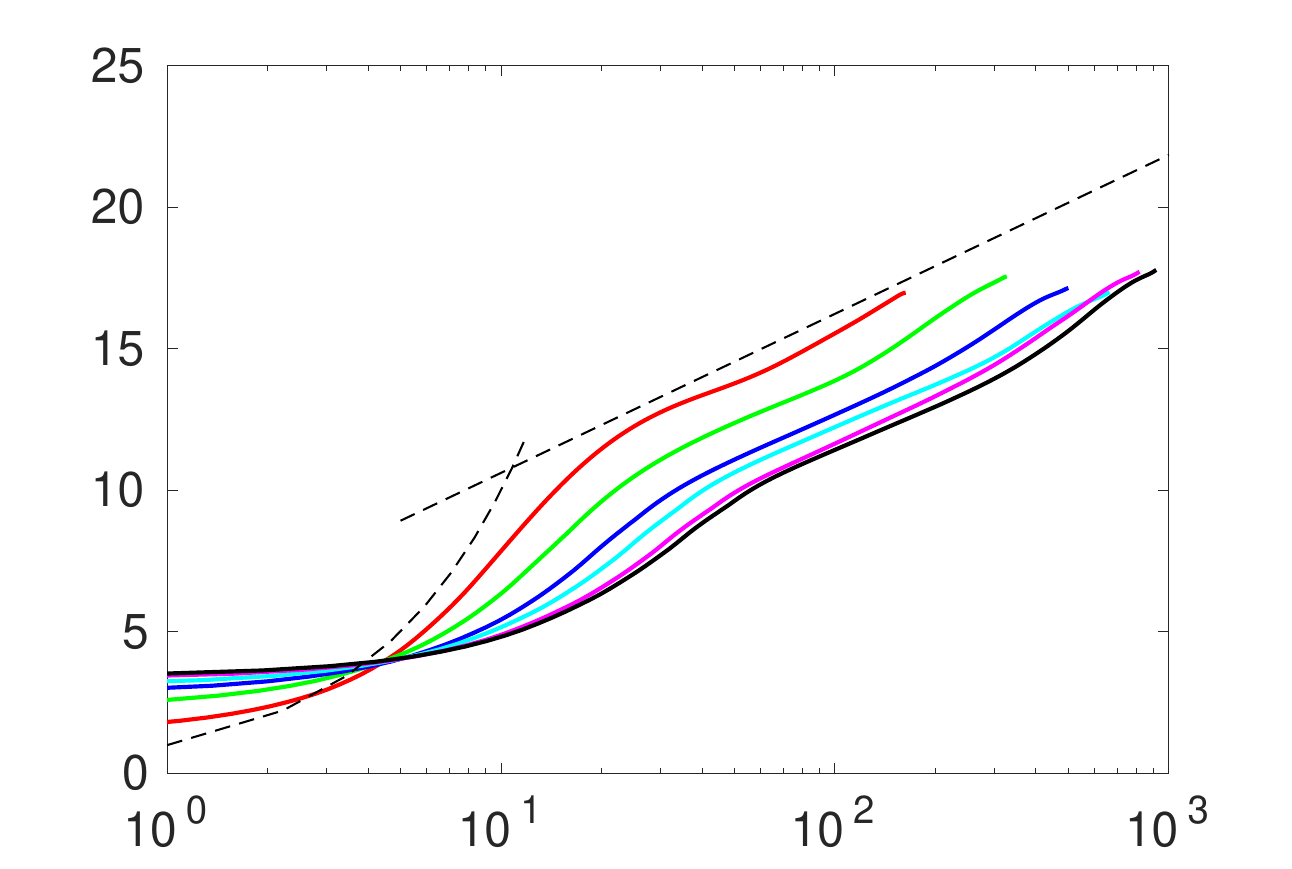}
\put(-107,-3){$(y-d)^+$}
\put(-190,60){$\langle u\rangle^+$}
\put(-190,110){(b)}
\hspace{1mm}
\includegraphics[width=65mm]{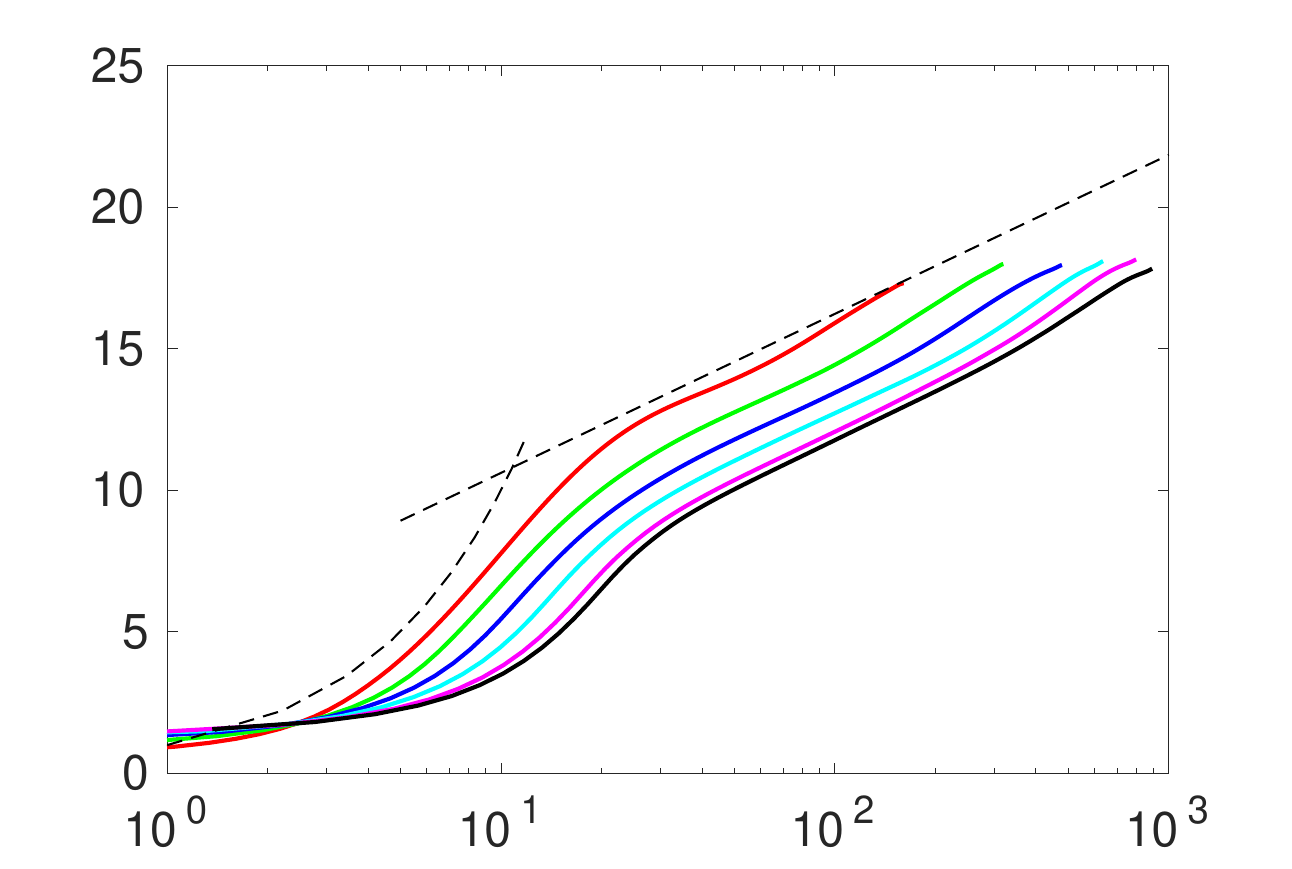}
\put(-107,-3){$(y-d)^+$}
\put(-190,60){$\langle u\rangle^+$}
\put(-190,110){(c)}
\includegraphics[width=65mm]{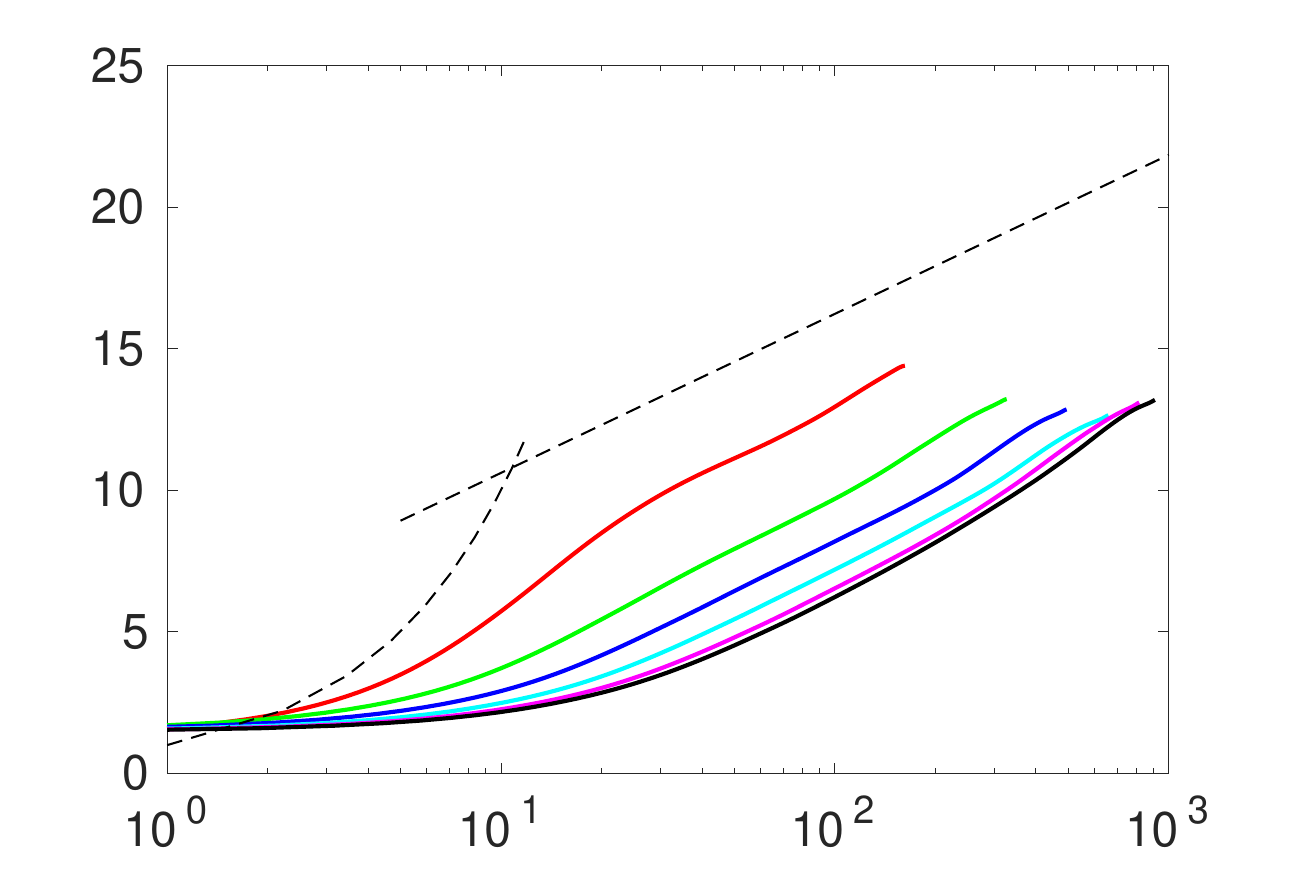}
\put(-107,-3){$(y-d)^+$}
\put(-190,60){$\langle u\rangle^+$}
\put(-190,110){(d)}
\caption{Streamwise mean velocity profiles of selected rough surfaces:
  (a,b) Gaussian roughness GS01 and GS03; (c,d) Weibull roughness WB08
  and WB10. The dashed lines are $U^+=y^+$ and
  $U_{s}^{+}=\frac{1}{\kappa} \ln{y^{+}}+ 5.0$.}
\label{fig:mean_velocity}
\end{figure}

As demonstrated by \cite{chung2015fast} and
\cite{macdonald2017minimal}, the minimal-span channel provides
reliable results for wall friction and turbulent statistics within the
region $y<0.3\delta$. In the context of wall model development, our
primary focus is on the flow close to the wall. Therefore, the
minimal-span channel approximation is sufficient to capture the
near-wall physics necessary for developing an accurate wall model,
provided that the near-wall grid size for WMLES is below $0.3\delta$.

\begin{table}
\begin{center}
\def~{\hphantom{0}}
\setlength\extrarowheight{-5pt}
\renewcommand{\arraystretch}{2}
    \begin{tabular}{ccccccc}
    \hline \\
     \multicolumn{1}{c}{ Case} & \multicolumn{6}{c}{ $\hat{k}_s^+$} \\
     - &  $Re_{\tau}=180$ &  $Re_{\tau}=360$ &  $Re_{\tau}=540$ &  $Re_{\tau}=720$ &  $Re_{\tau}=900$ &  $Re_{\tau}=1000$ \\
      GS01 &  7.7 &  20.1 &  39.5 &  61.1 &  85.8 &  99.5 \\
     GS02 &  6.5 &  11.6 &  18.7 &  23.4 &  28.5 &  31.6 \\
     GS03 &  5.6 &  10.2 &  17.6 &  21.5 &  27.6 &  30.8 \\
     GS04 &  5.7 &  11.9 &  20.4 &  28.7 &  36.4 &  40.8 \\
     GS05 &  5.7 &  11.9 &  21.8 &  33.4 &  45.1 &  52.9 \\
     GS06 &  8.0 &  21.5 &  40.1 &  60.3 &  74.9 &  87.6 \\
     GS07 &  5.1 &  6.4 &  9.8 &  13.6 &  17.7 &  20.0 \\
     GS08 &  4.7 &  4.8 &  6.4 &  7.3 &  8.3 &  8.8 \\
     GS09 &  6.5 &  11.6 &  18.7 &  23.4 &  28.5 &  31.6 \\
     GS10 &  5.0 &  7.5 &  12.6 &  17.5 &  22.8 &  25.7 \\
     GS11 &  4.4 &  7.1 &  11.1 &  14.2 &  16.4 &  18.1 \\
     GS12 &  4.5 &  8.7 &  13.8 &  19.6 &  25.4 &  29.6 \\
     GS13 &  4.8 &  8.0 &  12.4 &  17.2 &  21.8 &  25.0 \\
     GS14 &  5.3 &  16.7 &  33.9 &  52.7 &  71.3 &  82.6 \\
     GS15 &  7.3 &  18.6 &  32.5 &  49.0 &  65.0 &  76.3 \\
     GS16 &  10.4 &  17.2 &  29.4 &  41.2 &  53.7 &  55.6 \\
     GS17 &  7.9 &  27.0 &  53.2 &  80.8 &  107.7 &  122.0 \\
     GS18 &  4.7 &  13.3 &  24.4 &  36.1 &  48.6 &  55.8 \\
     GS19 &  7.6 &  31.9 &  61.6 &  92.6 &  125.4 &  144.9 \\
     WB01 &  9.1 &  26.0 &  46.0 &  66.0 &  82.8 &  93.9 \\
     WB02 &  4.6 &  10.2 &  16.5 &  23.3 &  29.0 &  31.3 \\
     WB03 &  6.0 &  14.1 &  28.0 &  42.7 &  58.4 &  65.1 \\
     WB04 &  13.8 &  47.2 &  88.0 &  121.1 &  161.7 &  181.8 \\
     WB05 &  10.9 &  32.7 &  59.0 &  83.5 &  108.8 &  122.7 \\
     WB06 &  6.6 &  15.5 &  28.5 &  42.8 &  54.0 &  60.9 \\
     WB07 &  7.1 &  20.2 &  38.8 &  56.7 &  76.4 &  84.1 \\
     WB08 &  5.6 &  8.7 &  13.1 &  17.7 &  23.1 &  26.0 \\
     WB09 &  15.0 &  54.3 &  101.5 &  142.1 &  192.3 &  216.1 \\
     WB10 &  17.8 &  63.1 &  114.3 &  169.6 &  218.1 &  242.1 \\
     WB11 &  4.9 &  7.4 &  11.5 &  16.1 &  20.9 &  23.6 \\
     WB12 &  9.5 &  35.6 &  67.7 &  103.6 &  131.7 &  152.1 \\
     WB13 &  17.6 &  61.0 &  110.3 &  157.3 &  204.4 &  236.5 \\   
     \hline
    \end{tabular}
    \caption{\label{tab:roughness_ks} The roughness parameter
      $\hat{k}_s^+$ at different $Re_{\tau}$ determined from the DNS
      results based on the rough-wall logarithmic law.  }
    \end{center}
\end{table}

In all simulations, mean quantities and statistics are averaged over a
time period $T \ge 20 \delta / u_{\tau}$ after transients to achieve
statistical convergence. The streamwise mean velocity $U(y)$ is
calculated as
\begin{equation}
 U(y) = \langle  u \rangle  = \frac{1}{A_f T} \int_T \iint_{A_f} u(x,y,z,t) \mathrm{d}x \mathrm{d}z  \mathrm{d}t,
\end{equation}
where $u$ is the instantaneous streamwise velocity, $A_f$ is the
fluid-occupied area at each $y$ location, $T$ is the total time
considered, and the angle brackets denote average over homogeneous
directions and time. The streamwise mean velocity profiles are shown
in Figure \ref{fig:mean_velocity} for four selected rough
surfaces. The mean wall shear stress $\langle \tau_w \rangle$ is
computed by integrating the time-averaged body force $F_1$ in the
streamwise direction~\citep{yuan2014numerical, yuan2014roughness}.

In the logarithmic region, the streamwise mean velocity profile in
smooth walls ($U_s$) can be approximated by
\begin{equation}
 U_{s}^{+} \approx \frac{1}{\kappa} \ln{y^{+}}+ 5.0,
 \label{eqn:low_smooth}
\end{equation}
where $\kappa\approx 0.41$ is the von K{\'a}rm{\'a}n
constant.  For rough-wall cases, the logarithmic velocity distribution
for the mean velocity profile ($U_r$) also holds in the fully-rough
regime:
\begin{equation}
 U_{r}^{+} \approx \frac{1}{\kappa} \ln(\frac{y-d}{\hat{k}_s})+ 8.5,
 \label{eqn:low_rough}
\end{equation}
where $\hat{k}_s^+=k_s^+$ for fully-rough cases (but not for
transitionally-rough cases), and $d$ is the zero-plane displacement,
computed based on the location of the centroid of the wall-normal
profile of the averaged drag force
\citep{jackson1981displacement}.  For small $d$, the roughness
function $\Delta U^+$ can be obtained by the difference of mean
velocities in wall units between smooth and rough walls within the
logarithmic layer:
\begin{equation}
 \Delta U^{+} \approx \frac{1}{\kappa} \ln{\hat{k}_{s}^{+}}-3.5.
 \label{eqn:low}
\end{equation}
To evaluate whether a rough-wall case is in the transitionally or
fully rough regime, $\hat{k}_s^+$ can be computed according to
equation (\ref{eqn:low}). The values of $\hat{k}_s^+$ for each flow
case in the current DNS database are presented in Table
\ref{tab:roughness_ks}. According to \cite{flack2010review}, a flow is
considered to be fully rough if $\hat{k}_s^+ \ge 70$. Using this
reference, it can be seen that the present DNS database spans a wide
range of both transitionally- and fully-rough regimes.
%

\section{Wall model formulation}\label{sec:formulation}


\subsection{Framework of WMLES}

The wall model is developed within the framework of WMLES, where only
the most energetic eddies in the outer layer of the flow are resolved
by the computational grid. The effects of the small scales far from
the wall are modeled by a subgrid-scale (SGS) model. Close to the
wall, the energy-containing eddies are under-resolved, and the
wall-shear stress is obtained using a wall model. The flow solver
integrates the coarse-grained incompressible Navier-Stokes equations
\begin{equation}
\frac{\partial \overline{u}_i}{\partial t} + \frac{\partial \overline{u}_i\overline{u}_j}{\partial x_j}=-\frac{1}{\rho}\frac{\partial \overline{p}}{\partial x_i} + \nu\frac{\partial^2 \overline{u}_i}{\partial x_k x_k}+\frac{\partial \tau_{ij}^{SGS}}{\partial x_j},~~\frac{\partial \overline{u}_i}{\partial x_i} = 0,
\label{eqn:nsme}
\end{equation}
where the overline denotes coarse-grained quantities and
$\tau_{ij}^{SGS}$ is the deviatoric part of the SGS stress tensor. At
the walls, the non-slip boundary condition is replaced by a shear
stress boundary condition, which is obtained from the wall model.

\begin{figure}
\centering
\includegraphics[width=130mm]{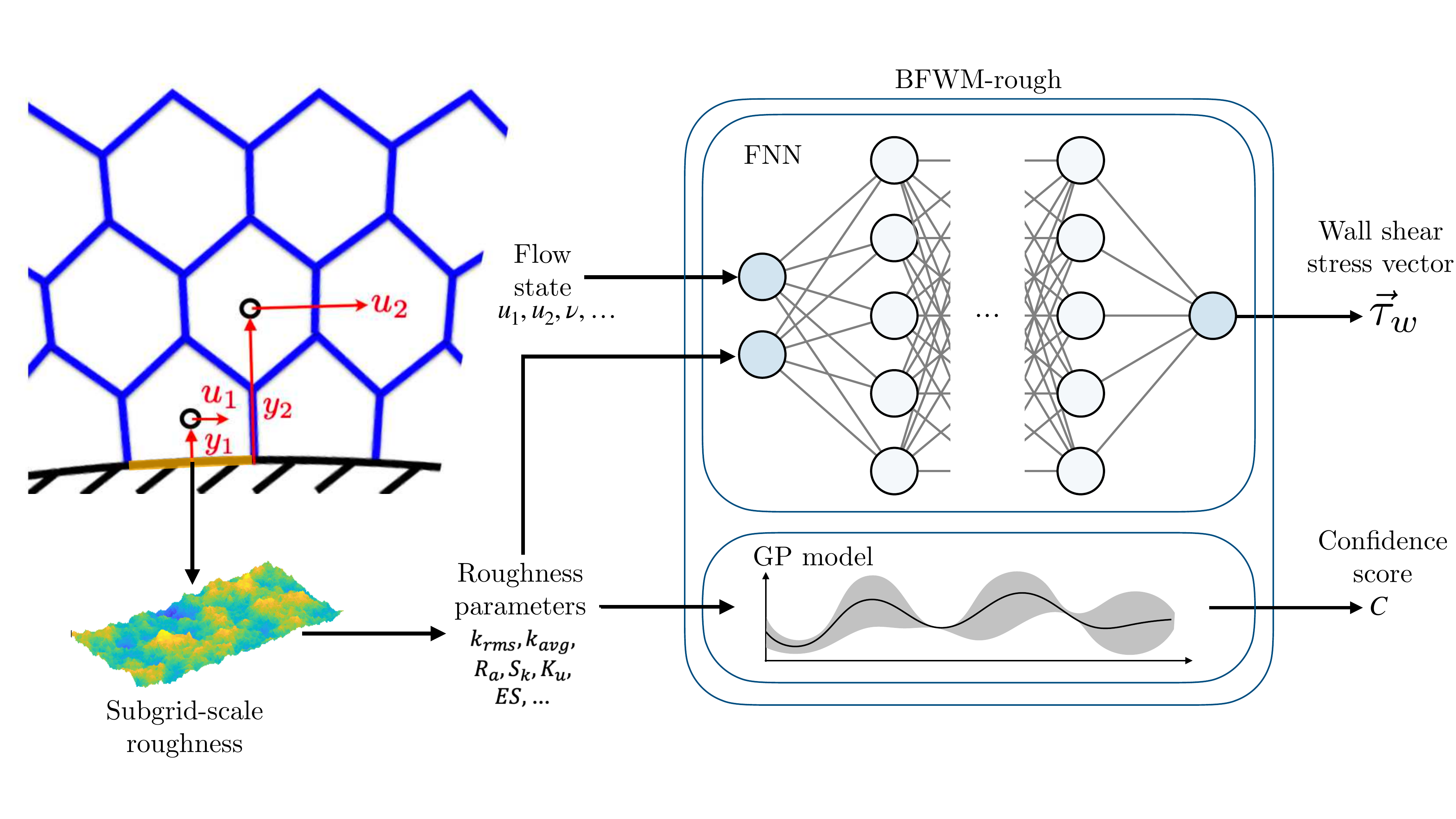}
\caption{Overview of BFWM-rough for WMLES: For each wall face, the
  inputs to the wall model include the local flow state and local
  roughness parameters. The flow state consists of the magnitudes of
  wall-parallel velocities ($u_1$ and $u_2$) at the center locations
  of the first and second off-wall control volumes ($y_1$ and $y_2$),
  along with the kinematic viscosity $\nu$. The roughness is assumed
  to be subgrid-scale and known for each control volume attached to
  the wall. The input roughness parameters are based on the
  statistical moments of the local roughness height distribution. The
  output of BFWM-rough is the wall-shear stress vector $\vec{\tau}_w$
  and the confidence score $C\in[0, 1]$. The wall-shear stress vector
  is predicted with a FNN, while the confidence score is determined
  using a GP model. The wall-shear stress vector from BFWM-rough is
  applied as the local boundary condition to the WMLES.}
\label{fig:BFM-rough}
\end{figure}

Figure \ref{fig:BFM-rough} depicts an overview of WMLES and the
BFWM-rough. The wall roughness is subgrid scale, i.e., the WMLES grid
does not resolve any of the geometric features of the roughness. The
BFWM-rough is implemented using FNNs, where the input comprises local
information from the flow in the WMLES grid and local roughness
features. The output of the BFWM-rough is the local wall-shear stress,
which is used as the boundary condition for WMLES at the wall.  In the
following sections, we discuss the model assumptions, the selection
of input and output variables, the architecture of FNNs, and the
details of wall model training.

\subsection{Wall model assumptions}\label{sec:assumption}

We summarize the main modeling assumptions of the BFWM-rough:
\begin{itemize}
\setlength\itemsep{0em}
\item[(i)] Building block flow assumption: A finite set of simple
  flows, based on minimal-span turbulent channel flows over rough surfaces,
  is sufficient to formulate a generalizable rough-wall model.
\item[(ii)] Quasi-equilibrium assumption: The near-wall region of
  complex cases is in a quasi-equilibrium state, i.e., statistically
  steady flow under mean zero-pressure-gradient effects.
\item[(iii)] Velocity/shear-stress alignment assumption: The direction
  of the wall-shear stress vector is aligned with the relative
  wall-parallel velocity at the first control volume attached to the
  wall.  
\item[(iv)] Statistical roughness description assumption: A collection
  of statistical parameters of the rough surface, such as the mean and
  root-mean-square roughness heights, along with high-order moments of
  height fluctuations (see Table~\ref{tab:roughness_para}), are
  sufficient to describe the geometrical effects of the surface
  topology on the wall shear stress.
\item[(v)] Space/time locality assumption: The relative wall-parallel
  velocity from the first two contiguous wall-normal control volumes
  above the wall, combined with (iii), provides enough information to
  predict the wall shear stress.
\item[(vi)] Subgrid-scale roughness assumption and outer-layer
  similarity: The roughness effects on the flow are assumed to be
  subgrid-scale and the only impact of roughness on the resolved
  scales is through the wall shear stress.
\item[(vii)] Viscous scaling assumption: The best-performing
  non-dimensional form of the velocity inputs and the model output is
  obtained by scaling the variables using the kinematic viscosity and
  wall-normal distance.
\item[(viii)] Mean-flow training data assumption: inputs and outputs
  samples based on mean velocity profiles and mean wall-shear stress
  are suitable for training accurate wall models.
\end{itemize}
Assumptions (i) and (ii) justify the use of the current DNS training
database, which targets equilibrium turbulent flows over rough
surfaces. Therefore, the application of the BFWM-rough to complex
geometries presupposes that the near-wall flow field maintains
quasi-equilibrium conditions. Assumptions (iii), (iv), and (v) are
adopted primarily for the sake of model simplicity. Incorporating flow
information from farther away from the wall could potentially enhance
the accuracy of the model, especially for rough surfaces with larger
roughness heights. However, this would also increase the complexity of
the model when dealing with unstructured grids in realistic
geometries. Similarly, the use of roughness statistics instead of
detailed local topography simplifies the model, akin to the use of
$k_s$ in other approaches. Assumption (vi) is supported by previous
observations from the literature~\citep{raupach1991rough} and by the
Townsend's outer-layer similarity
hypothesis~\citep{townsend1976structure}, which has been confirmed by
multiple studies~\citep{flack2005experimental, flores2006effect,
  leonardi2010channel, mizuno2013wall, chung2014idealised,
  chan2015systematic, lozano2019characteristic}.  The viscous scaling
from assumption (vii) is appropriate for the flow scenarios addressed
in this study but may become inaccurate under conditions with
significant pressure gradients or compressibility effects. The
rationale behind assumption (viii) is that the fluctuations of the
flow are less critical compared to the mean quantities for predicting
the mean wall shear stress. Consequently, the training dataset relies
on average flow values from DNS, with added Gaussian noise to improve
the robustness of the model.

\subsection{Input and output variables}\label{sec:inandout}

The goal of this section is to select the non-dimensional input
variables for the wall model that are most informative for predicting
the wall shear stress across the entire training dataset. The input
variables include both flow variables and roughness topography
parameters. The variables $y_1$ and $y_2$ denote the wall-normal
distances to the centers of the first and second control volumes off
the wall, respectively (as shown in Figure~\ref{fig:BFM-rough}), and
are related to the WMLES grid resolution by $y_1 \approx
\Delta/2$. The magnitudes of the corresponding wall-parallel
velocities relative to the wall at $y_1$ and $y_2$ are
$u_1=(u_s(y_1)^2+w_s(y_1)^2)^{1/2}$ and
$u_2=(u_s(y_2)^2+w_s(y_2)^2)^{1/2}$, where $u_s$ and $w_s$ are samples
generated as detailed below.  Three different grid resolutions,
$\Delta/\delta = 1/20, 1/10, 1/5$, are included in the generation of
data for the input and output variables. The candidate input variables
considered can be organized into five categories:
\begin{itemize}
\setlength\itemsep{0em}
\item[1)] Roughness parameters:
    \begin{align*}
        \frac{k_{avg}}{R_a}, \frac{k_{c}}{R_a}, \frac{k_{t}}{R_a}, \frac{k_{rms}}{R_a}, S_k, K_u, ES, P_o, I, \frac{L_{cor}}{R_a}, \lambda_f;
    \end{align*}
\item[2)] Products of roughness parameters:
    \begin{align*}
        ES^2, S_k^2,  ES S_k, ES K_u;
    \end{align*}        
\item[3)] Roughness height scaled by $y_1$:
    \begin{align*}
        \frac{k_{avg}}{y_1}, \frac{k_{c}}{y_1}, \frac{k_{t}}{y_1}, \frac{k_{rms}}{y_1}, \frac{y_1}{R_a};
    \end{align*}
\item[4)] Products of non-dimensional roughness parameters and
  roughness height scaled by $y_1$:
    \begin{align*}
        \frac{S_kk_{avg}}{y_1}, \frac{K_uk_{avg}}{y_1}, \frac{ES k_{avg}}{y_1}, \frac{P_ok_{avg}}{y_1}, \frac{I k_{avg}}{y_1},\\
        \frac{S_kk_{rms}}{y_1}, \frac{K_uk_{rms}}{y_1}, \frac{ES k_{rms}}{y_1}, \frac{P_ok_{rms}}{y_1}, \frac{I k_{rms}}{y_1};
    \end{align*}
\item[5)] Local Reynolds number at $y_1$ and $y_2$:
    \begin{align*}
        \frac{u_1y_1}{\nu}, \frac{u_2y_2}{\nu}.
    \end{align*}
\end{itemize}
The samples for $u(y_i)$ and $w(y_i)$ are generated based on a
Gaussian distribution $\mathcal{N}\left[ \mu, \sigma^2 \right]$ with
mean $\mu$ and variance $\sigma^2$ as
\begin{equation}
u_s(y_i) \sim \mathcal{N}\left[U(y_i),u_{rms}^2(y_i)\right], \quad  w_s(y_i) \sim \mathcal{N}\left[W(y_i),w_{rms}^2(y_i)\right],
\end{equation}
where $U(y_i)$ and $W(y_i)$ are the DNS streamwise and spanwise mean
velocities, and $u_{rms}(y_i)$ and $w_{rms}(y_i)$ are the DNS
root-mean-squared streamwise and spanwise velocity fluctuations at
$y_i$ (with $i=1,2$).

The non-dimensional output of the wall model is
\begin{equation}
  \Tilde{\tau}_w=\frac{\tau_{w,s}y_1}{\nu u_1},
\end{equation}
where the scaling factor $\nu u_1/y_1$ represents the naive estimation
of the wall shear stress using finite differences.  The values of
$\tau_{w,s}$ are generated based on a Gaussian distribution
$\tau_{w,s} \sim \mathcal{N}\left[ \langle \tau_w\rangle,
  \tau_{w,rms}^2 \right]$, where $\langle \tau_w\rangle$ is the mean
wall-shear stress calculated from DNS, and $\tau_{w,rms}$ is computed
based on the correlation $\tau_{w,rms}/\langle\tau_w\rangle = 0.298 +
0.018\ln Re_{\tau}$~\citep{orlu2011fluctuating}. As discussed in
\S\ref{sec:assumption}, the main goal of the current wall model is to
correctly predict mean flow quantities. It was found that this
approach facilitates the training of more robust models than the use
of actual instantaneous DNS values for velocities and wall shear
stress.

We aim to select the best-performing input variables from all the
candidates defined above. To that end, we use the Minimum Redundancy
Maximum Relevance (MRMR) algorithm to select an optimal set of the
candidate input features~\citep{ding2005minimum,peng2005feature}. The
MRMR algorithm is designed to rank input variables by considering both
their relevance to the output and their redundancy with respect to
each other.  This helps improve the accuracy of the model and reduce
overfitting by eliminating unnecessary inputs. The relevance and
redundancy are computed using mutual information, which measures the
amount of information shared between variables. The mutual information
$I$ between a pair of random variables $(\phi, \psi)$ is defined as:
\begin{equation}
    I(\phi,\psi) = \iint \text{PDF}(\phi,\psi)\log \Biggl(
    \frac{\text{PDF}(\phi,\psi)}{\text{PDF}(\phi)\text{PDF}(\psi)}
    \Biggr)\mathrm{d}\phi \mathrm{d}\psi,
\end{equation}
where $\text{PDF}(\phi,\psi)$ is the joint probability density
function of $\phi$ and $\psi$, and $\text{PDF}(\phi)$ and
$\text{PDF}(\psi)$ are the marginal probability density functions of
$\phi$ and $\psi$, respectively. The MRMR algorithm ranks features by
evaluating the importance score (a.k.a, mutual information quotient)
of an input $\phi_{in}$:
\begin{equation}
    \text{MIQ}_{\phi_{in}} = \frac{V_{\phi_{in}}}{W_{\phi_{in}}},
\end{equation}
where $V_{\phi_{in}}$ is the relevance of the input feature
$\phi_{in}$ with respect to the output variable $\phi_{out}$:
\begin{equation}
    V_{\phi_{in}} = I(\phi_{in},\phi_{out}),
\end{equation}
and $W_{\phi_{in}}$ is the redundancy of the input feature $\phi_{in}$
with respect to the rest of input features $\psi_{in}$:
\begin{equation}
    W_{\phi_{in}} = \frac{1}{|\textbf{S}|}\sum_{\psi_{in} \in
      \textbf{S}} I(\phi_{in},\psi_{in}),
\end{equation}
where $|\textbf{S}|$ is the number of features in the set of the input
variables $\textbf{S}$ maximizing $\text{MIQ}_{\phi_{in}}$.  Higher
relevance indicates a stronger association with the output, while
lower redundancy implies less similarity between inputs. 

\begin{figure}
\centering
\includegraphics[width=130mm,trim={0.0cm 5.5cm 0.0cm 4.0cm},clip]{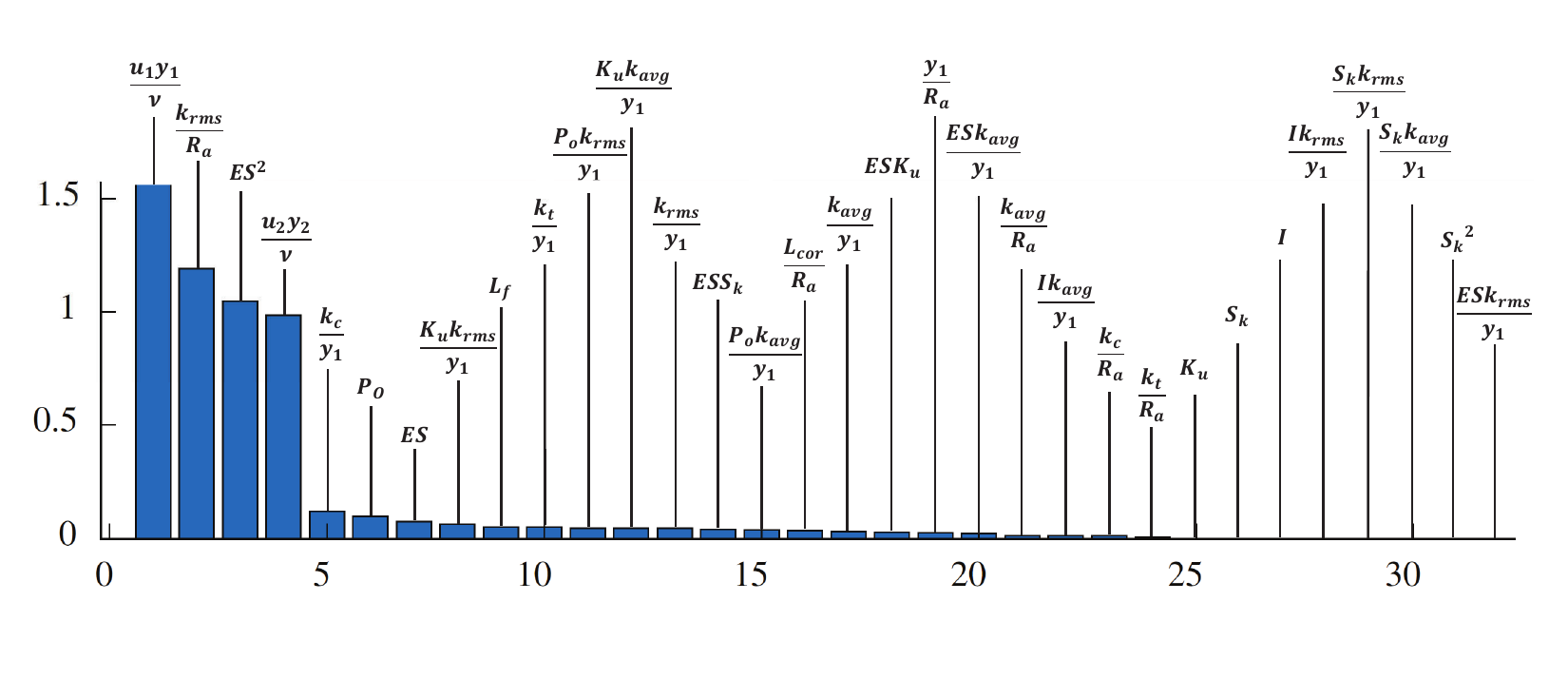}
\put(-376,36){\rotatebox{90}{Importance score}}
\put(-200,8){Input rank}
\caption{Ranking of candidate input variables for the wall model
  according to MRMR importance score in descending order.}
\label{fig:rank}
\end{figure}

We use MRMR to identify a subset of input variables that collectively
maximize relevance to the target while minimizing redundancy among the
selected inputs. Figure \ref{fig:rank} displays the MRMR importance
scores in descending order, with the top-ranked inputs being the most
informative for predicting the output. The results in Figure
\ref{fig:rank} reveal that the largest drop in importance score occurs
between the 4th and 5th inputs. This suggests that the local Reynolds
numbers, $u_1y_1/\nu$ and $u_2y_2/\nu$, along with the roughness
features $k_{rms}/R_a$ and $ES^2$, contain the most relevant
information for the wall model. Beyond the 5th input, the decrease in
importance score is relatively smaller, although it remains non-zero
up to the 24th input.

According to the Buckingham-$\pi$
theorem~\citep{buckingham1914physically}, the number of dimensionless
numbers required to uniquely determine the wall shear stress in the
rough-wall channel simulations equals the total number of parameters
needed to set up the cases minus the independent fundamental units. In
this study, there are seven parameters: $U_c$ (mean centerline
velocity), $\delta$, $\nu$, $k_{rms}$, $H$, $H_f$, and $\kappa_0$. The
number of fundamental units is two (length and time). Therefore, a
total of five dimensionless numbers is required to completely specify
the case, and hence the wall shear stress.  This implies that any
model aiming to accurately predict the wall shear stress is expected
to require five non-dimensional inputs. This explains why the model
error drops significantly from four to five inputs. In practice, the
non-dimensional inputs available to the wall model do not include
global parameters of the case (such as $U_c$ and $\delta$). Instead,
local quantities are used as proxies. For that reason, adding more
inputs can still inform the model predictions, and the importance
score remains non-zero when the number of inputs exceeds five. The
order of inputs provided by the ranking in Figure \ref{fig:rank} is
used in the following section to determine the number of model inputs.

\subsection{Wall model training}\label{sec:train}

An FNN is used to parameterize the relationship between inputs and
outputs. The layers are connected using hyperbolic tangent sigmoid
transfer functions as the activation function except for the last
layer, which is connected with rectified linear units (ReLUs).  The
training algorithm utilized is gradient descent with momentum and
adaptive learning rate
backpropagation~\citep{yu2002backpropagation}. A total of 192 cases of
turbulent channel flows are randomly divided into training dataset
(70\% of the total), validation dataset (15\% of the total), and
testing dataset (15\% of the total). The training dataset is used to
develop a candidate BFWM-rough model; the validation dataset, while
not used for parameter estimation, provides the stopping criterion
during training to prevent overfitting and enhance the
generalizability of the model. The model architecture --including the
number of hidden layers and neurons-- and hyperparameters are
optimized based on performance metrics from the validation
dataset. The testing dataset, not involved in the training process,
serves as an independent set of cases to assess the performance of
BFWM-rough on unseen scenarios.

To find the best-performing model, different numbers of input features
are tested according to the order presented in Figure
\ref{fig:rank}. Note that this ranking is key to limit the number of
possible input combinations.  For each number of inputs, 100 random
splits of the training, validation, and testing datasets are
conducted. For each split, the optimal number of hidden layers and
neurons per layer is determined through a grid search. The number of
hidden layers and neurons per layer considered in the grid search
ranges from 3 to 6 and from 5 to 20, respectively. The optimal model
for each combination of inputs is then determined based on the minimum
$L_2$-norm error over the entire training, validation and testing datasets from
the 100 random splits.

\begin{figure}
\centering
\includegraphics[width=130mm,trim={1.8cm 0.0cm 0.0cm 0.0cm},clip]{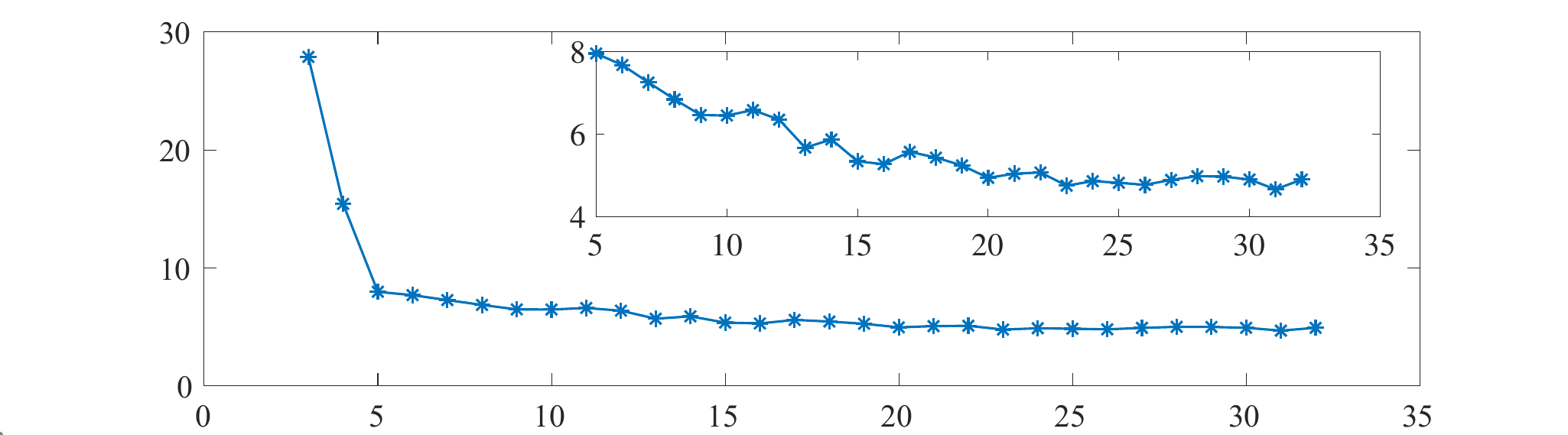}
\put(-127,70){\color{red}{$\downarrow$}}
\put(-127,78){\color{red}{No.=23}}
\put(-250,-10){No. of ranking input features}
\put(-370,33){\rotatebox{90}{\% Error}}
\caption{ \emph{A-priori} $L_2$-norm error in the prediction of
  $\Tilde{\tau}_w$ as function of the number of input
  features as ranked in Figure~\ref{fig:rank}.  The inset shows the
  errors when the number of inputs ranges from 5 to 32.}
\label{fig:error}
\end{figure}
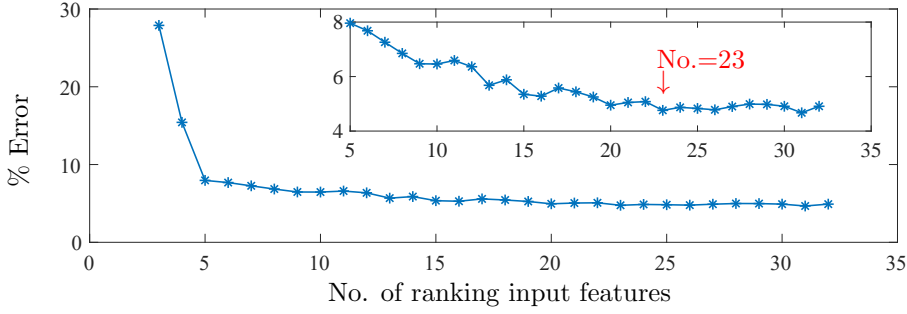

The errors for different candidate wall models are plotted against the
number of inputs in Figure \ref{fig:error}. Consistent with the
discussion in \S\ref{sec:inandout}, the wall model requires at least 5
non-dimensional inputs to achieve errors below 10\%. A less
significant reduction in error, from 8\% to 4.6\%, is observed with a
higher number of inputs, which aligns with the lower importance score
shown in Figure \ref{fig:rank} after the 5th ranked input. As an
interesting observation, it was not possible to train wall models with
errors below 10\% without including roughness parameters
non-dimensionalized by the grid size.

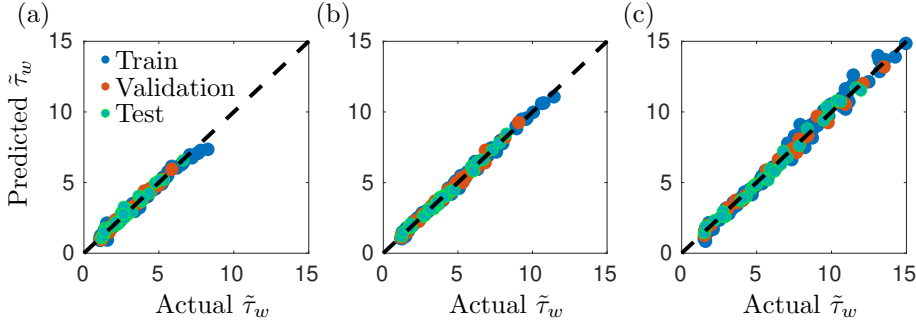
\begin{figure}
\centering
\includegraphics[width=140mm]{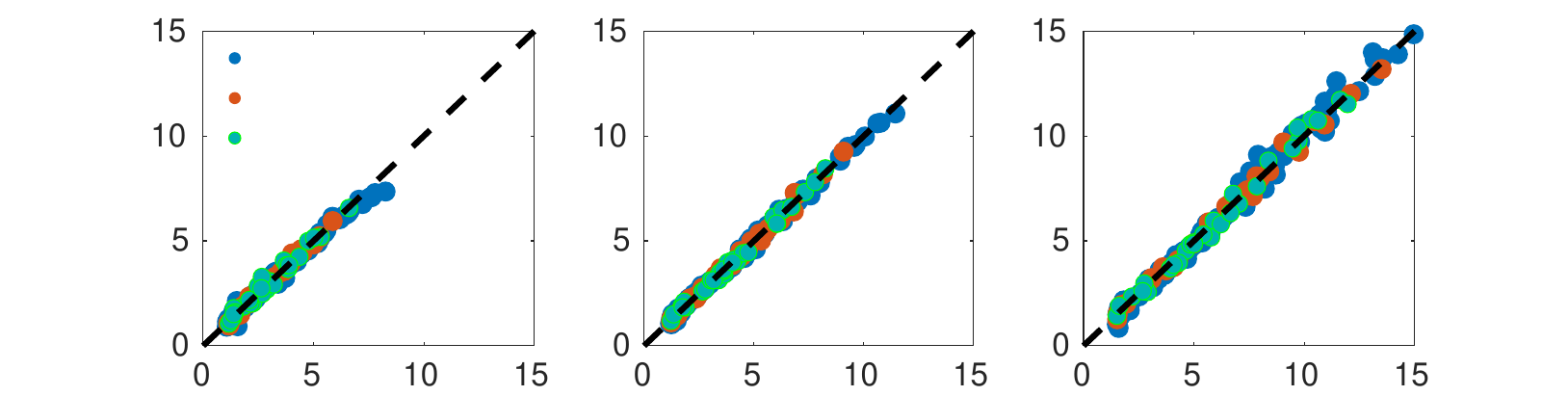}
\put(-372,100){(a)}
\put(-375,30){\rotatebox{90}{Predicted $\Tilde{\tau}_w$}}
\put(-335,82){Train}
\put(-335,72){Validation}
\put(-335,62){Test}
\put(-320,-10){Actual $\Tilde{\tau}_w$}
\put(-210,-10){Actual $\Tilde{\tau}_w$}
\put(-100,-10){Actual $\Tilde{\tau}_w$}
\put(-260,100){(b)}
\put(-145,100){(c)}
\caption{Scatter plot of regression results of actual $\Tilde{\tau}_w$
  and predicted $\Tilde{\tau}_w$ for BFWM-rough. The results are
  plotted for three grid resolutions: (a) $\Delta/\delta=1/20$, (b)
  $\Delta/\delta=1/10$, and (c) $\Delta/\delta=1/5$.}
\label{fig:regression}
\end{figure}

Based on the results shown in Figure \ref{fig:error}, the model with
23 inputs is selected as the BFWM-rough model. This model was chosen
because it has the minimum number of inputs while maintaining an
$L_2$-norm error below 5\%. The corresponding FNN comprises 5 hidden
layers with 15 neurons per layer. The $L_2$-norm error for BFWM-rough
is 4.76\% across all datasets.  As an example, regression results for
three different grid resolutions are illustrated in Figure
\ref{fig:regression}.  It is also useful to evaluate the range of
applicability of BFWM-rough in terms of typical roughness heights
relative to the WMLES grid resolution. The ratio of roughness height
parameters in the training database to the training grid size is
summarized in Table \ref{tab:applicability}. BFWM-rough covers a
peak-to-valley roughness height $k_c$ from 0.38$\Delta$ to
3.55$\Delta$, and a root-mean-square roughness height $k_{rms}$ from
0.04$\Delta$ to 0.56$\Delta$.  In WMLES scenarios characterized by
higher ratios (i.e., finer grid resolutions relative to typical
roughness element sizes) it may be more accurate to geometrically
resolve the roughness with the WMLES mesh and use a wall model for
smooth surfaces. The \emph{a-posteriori} performance of BFWM-rough is
assessed in actual WMLES in \S\ref{sec:evaluation}.
%
%
\begin{table}
\begin{center}
\def~{\hphantom{0}}
\setlength\extrarowheight{-5pt}
\renewcommand{\arraystretch}{2}
    \begin{tabular}{ccccc}
     \hline \\[0.5pt]
     $k_{avg}/\Delta$ & $k_c/\Delta$ & $k_t/\Delta$ & $k_{rms}/\Delta$ & $R_a/\Delta$ \\[3pt]
     0.06$\sim$2.1 & 0.38$\sim$3.55 & 0.35$\sim$3.49 & 0.04$\sim$0.56 & 0.03$\sim$0.47 \\
     \hline
    \end{tabular}
    \caption{\label{tab:applicability} Range of applicability for
      BFWM-rough for subgrid-scale roughness: ratio of roughness
      height parameters to the grid size of the training dataset.}
    \end{center}
\end{table}

\subsection{Confidence score}\label{sec:confidence}

A GP model, similar to the one introduced for AL in \S\ref{sec:AL}, is
used to calculate a confidence score for the roughness topology. The
main objective is to identify potential deficiencies in the BFWM-rough
when applied to surfaces with roughness features that differ
significantly from those in the training dataset. The GP model is
trained on all cases considered for BFWM-rough, but it only utilizes
the non-dimensional roughness parameters as inputs. The confidence
score is defined as $C = \min\{\sigma^2_{tr}/\sigma^2, 1\}$, where
$\sigma^2_{tr}$ is the mean prediction variance of the GP model over
the training dataset, and $\sigma^2$ is the predictive variance of the
case for which the confidence is computed. Note that the confidence
score provides an assessment only of the roughness geometry and does
not account for the flow conditions. When $\sigma^2_{tr} \approx
\sigma^2$, then $C \approx 1$ and confidence in the roughness geometry
is high. This indicates that the geometric properties of the roughness
considered are similar to those the BFWM-rough was trained
on. Conversely, large uncertainty in the roughness surface properties
will increase $\sigma^2$, thereby lowering the confidence score $C \ll
1$. In those cases, the prediction from BFWM-rough might be subject to
significant errors.

\section{Model evaluation}\label{sec:evaluation}




We evaluate the performance of the BFWM-rough across different
cases. The testing cases include 126 turbulent channel flows over
different rough surfaces. The model is also assessed in a turbulent
flow over a high-pressure turbine blade with two different surface
roughnesses.

Two sources of errors can be identified in the evaluation of a wall
model~\citep{lozano2022}: errors from the outer LES input data,
referred to as \emph{external wall-modeling errors}, and errors from
the wall model physical assumptions, referred to as \emph{internal
wall-modeling errors}. In the former, errors from the SGS model at the
matching locations propagate to the value of $\tau_w$ predicted by the
wall model. These errors can be labeled as external to the wall model
inasmuch as they are present even if the wall model provides an exact
physical representation of the near-wall region. The second source of
errors represents the intrinsic wall-model limitations: even in the
presence of exact values for the input data, the prediction might be
inaccurate when the physical assumptions the model is rooted in do not
hold.  In BFWM-rough, internal errors may come from the breakdown of
the assumptions discussed in \S\ref{sec:assumption} (e.g., lack of
quasi-equilibrium conditions, non-local effects, untrained roughness
topologies, etc.). The combined external plus internal error is
referred to as total error. In the following, we use \emph{a-priori}
testing to assess the internal errors of BFWM-rough and
\emph{a-posteriori} testing to evaluate the total errors.

\emph{A-priori} performance is assessed by the relative error in the
BFWM-rough model output when the input is generated from DNS data. The
model was implemented in actual WMLES to perform \emph{a-posteriori}
testing. For turbulent channel flows, WMLES with BFWM-rough was
conducted using our in-house code~\citep{bae2019dynamic,
  lozano2019characteristic, lozano2019error}. For high-pressure
turbine blade simulation with wall roughness, the BFWM-rough was
implemented into the high-fidelity solver charLES, developed by
Cascade Technologies, Inc~\citep{bres2018large, fu2021shock}.

\subsection{Rough-wall turbulent channel flow}

The performance of BFWM-rough is evaluated first in WMLES of turbulent
channel flows. The WMLES equations are solved using staggered
second-order finite differences and a fractional-step method with a
third-order Runge-Kutta time advancement scheme \citep{bae2019dynamic,
  lozano2019characteristic, lozano2019error}. The dynamic Smagorinsky
model~\citep{Germano1991, Lilly1992} is used as the SGS model. The
simulations are conducted by fixing the Reynolds number based on the
mean centerline velocity, $Re_c = U_c \delta/\nu$. The streamwise,
wall-normal, and spanwise lengths of the computational domain are
$2\pi\delta$, $2\delta$ and $\pi\delta$, respectively. Six grid
resolutions are considered, all of them with equal grid size in each
spatial direction: $\Delta/\delta = 1/5, 1/8, 1/10, 1/15, 1/20,$ and
$1/30$. The simulations were carried out for 30 eddy turnover times
after transients.

Hereafter, we use the term `unseen' (or testing) to indicate that the
model was never trained for that particular rough surface, rough
Reynolds number ($k_s^+$), and/or grid resolution.  The term `seen'
(or trained) is used when the model was explicitly trained for that
condition. The four categories of cases examined are listed below:
\begin{enumerate}
\item Seen Gaussian/Weibull rough surfaces;
\item Unseen Gaussian/Weibull rough surfaces;
\item Unseen bimodal Gaussian rough surfaces;
\item Unseen ellipsoidal, sinusoidal, Fourier-mode, and sandgrain
  rough surfaces from \cite{jouybari2021data}.
\end{enumerate}
Cases in (i), (ii) and (iii) are evaluated at unseen grid resolutions
and unseen $k_s^+$, whereas cases in (iv) are only conducted for
unseen $k_s^+$.

\subsubsection{Seen Gaussian/Weibull rough surfaces}\label{sec:validation-1}
\begin{table}
\begin{center}
\scriptsize
\def~{\hphantom{0}}
\setlength\extrarowheight{-5pt}
\renewcommand{\arraystretch}{2}
    \begin{tabular}{ccccccccccccccc}
    \hline \\
     \multicolumn{1}{c}{Case} & \multicolumn{1}{c}{$Re_{\tau}$} & \multicolumn{1}{c}{$\hat{k}_s^+$} & \multicolumn{6}{c}{\emph{A-priori} error of ${\tau}_w$} & \multicolumn{6}{c}{\emph{A-posteriori} error of $\tau_w$} \\
       &   &   & \multicolumn{3}{c}{Training grid $\Delta/\delta$} & \multicolumn{3}{c}{Testing grid $\Delta/\delta$} & \multicolumn{3}{c}{Training grid $\Delta/\delta$} & \multicolumn{3}{c}{Testing grid $\Delta/\delta$}\\
       &   &   & 1/20 & 1/10 & 1/5 & 1/15 & 1/8 & 1/30 & 1/20 & 1/10 & 1/5 & 1/15 & 1/8 & 1/30 \\
     \hline \\
     WB13 & 720 & 142.1 & -2.42 & 5.48 & 4.52 & 1.50 & 2.41 & 34.41 & -17.05 & -7.38 & -7.38 & -12.5 & -6.76 & -23.56 \\
     WB06 & 360 & 10.2 & -0.51 & -2.99 & -11.76 & -9.53 & -0.03 & -26.45 & 2.55 & 7.64 & 17.10 & 1.41 & 11.47 & -10.06 \\
     GS03 & 720 & 21.5 & 0.75 & 4.56 & 1.17 & 0.06 & 4.51 & 6.84 & -4.14 & 4.68 & 9.77 & -1.02 & 6.72 & -11.00\\
     WB05 & 360 & 32.7 & 1.60 & -2.40 & -5.92 & -3.41 & -3.82 & 13.49 & -8.33 & 2.64 & 11.99 & -4.67 & 8.33 & -16.46 \\
     GS14 & 540 & 33.9 & 1.97 & -1.85 & -2.31 & 8.27 & 5.53 & -5.65 & -0.77 & 10.95 & 16.11 & 5.16 & 13.84 & -9.09 \\
     GS18 & 1000 & 55.8 & 4.95 & 2.61 & 7.23 & -0.11 & -6.34 & 10.42 & 2.34 & 13.04 & 12.42 & 7.04 & 15.11 & -5.59 \\
     WB14 & 1000 & 131.4 & 11.80 & 5.68 & 1.27 & 0.61 & -1.02 & 22.73 & -5.53 & 4.44 & 3.64 & -2.83 & 5.66 & -12.73\\
     WB15 & 1000 & 139.0 & 4.95 & -7.00 & -10.39 & -10.28 & -11.54 & 22.13 & -11.64 & -3.41 & -6.61 & -8.22 & -3.01 & -19.24\\
     WB16 & 1000 & 198.5 & 2.19 & 0.90 & -4.68 & -11.78 & -11.86 & 25.32 & -12.36 & -5.07 & -5.27 & -7.71 & 0.20 & -19.67\\
     WB17 & 1000 & 80.8 & 8.25 & 0.48 & 1.40 & -5.76 & -6.22 & 1.43 & 2.42 & 6.48 & 7.08 & 1.42 & 8.30 & -9.31\\
     BM01 & 1000 & 230.9 & -18.06 & -7.17 & -7.05 & -12.64 & -8.27 & 9.02 & -12.16 & -2.43 & -5.88 & -8.32 & -1.82 & -19.67\\
     BM02 & 1000 & 186.3 & -10.27 & -5.93 & -6.37 & -6.11 & -7.97 & 15.06 & -12.56 & -2.87 & -6.35 & -9.63 & -2.87 & -20.90\\
     BM03 & 1000 & 132.6 & 0.31 & 8.89 & 13.19 & 3.83 & 10.83 & -15.08 & -8.80 & 2.84 & 0.20 & -3.24 & 2.64 & -15.62\\
     BM04 & 1000 & 53.0 & 35.26 & 29.20 & 27.10 & 30.92 & 25.14 & -12.05 & 8.14 & 20.96 & 17.82 & 14.05 & 22.43 & -0.63\\
     \hline
    \end{tabular}
    \caption{\label{tab:test_datasets_error} 
      Relative errors (in \%) of $\tau_w$ for the testing cases at
      various grid resolutions and Reynolds numbers.  The relative
      error is computed based on the predicted value from the
      BFWM-rough and the actual value from DNS in turbulent channel
      flows. The table shows case name, $Re_{\tau}$, and $\hat{k}_s^+$. }
    \end{center}
\end{table}
%
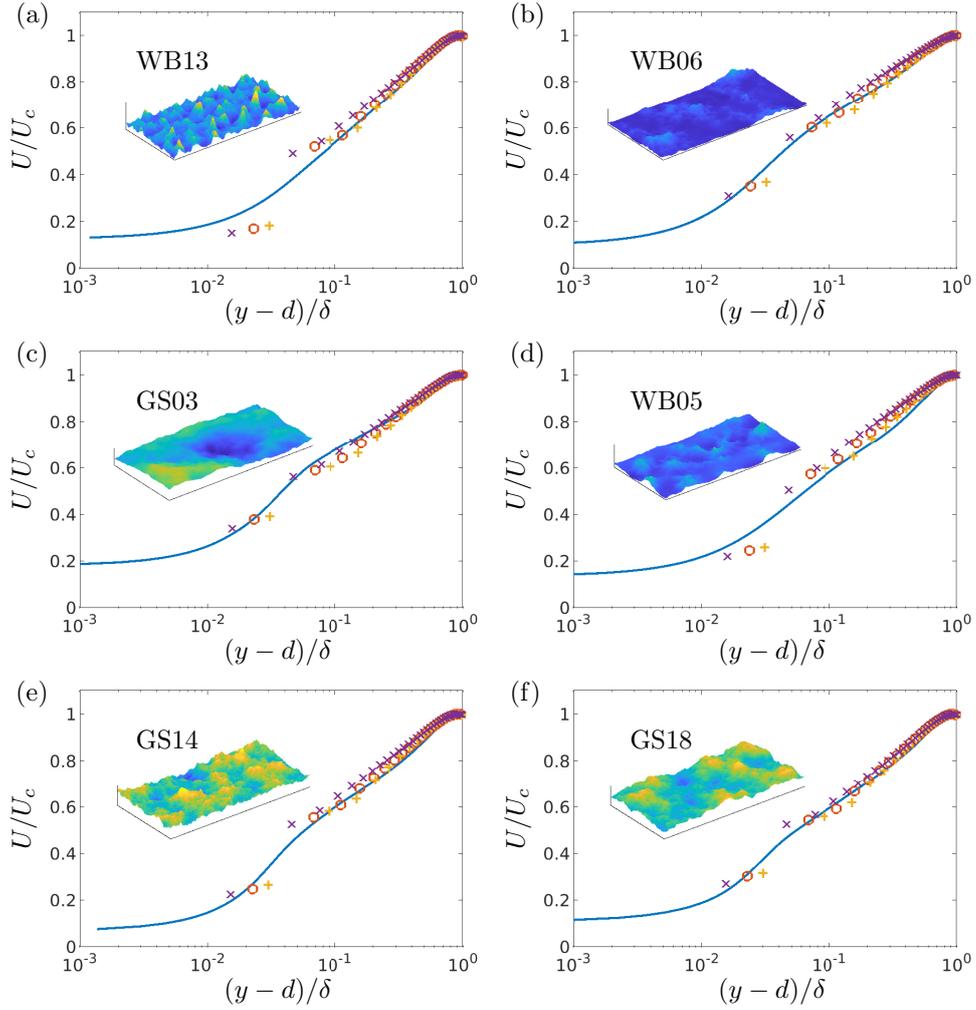
\begin{figure}
\centering
\includegraphics[width=65mm]{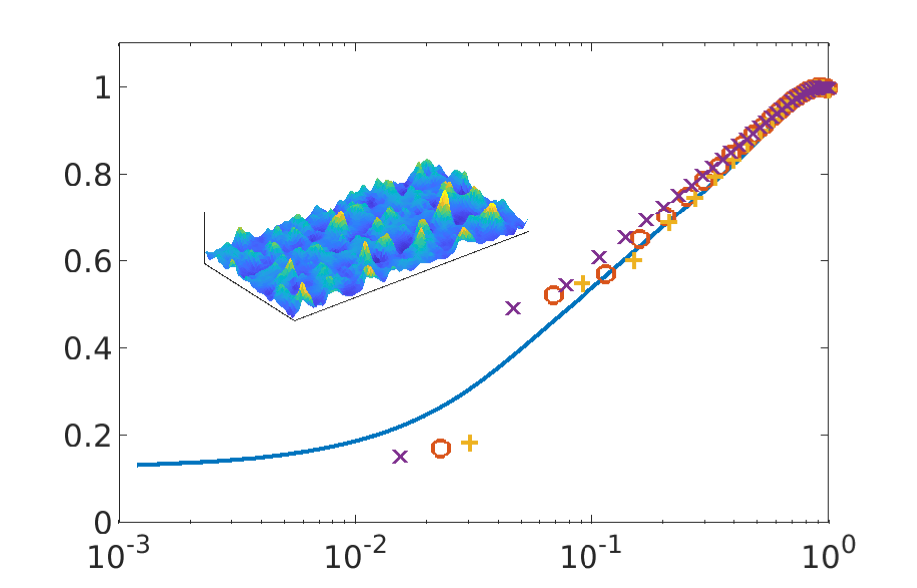}
\put(-107,-6){$(y-d)/\delta$}
\put(-187,50){\rotatebox{90}{$U/U_c$}}
\put(-140,88){WB13}
\put(-185,106){(a)}
\includegraphics[width=65mm]{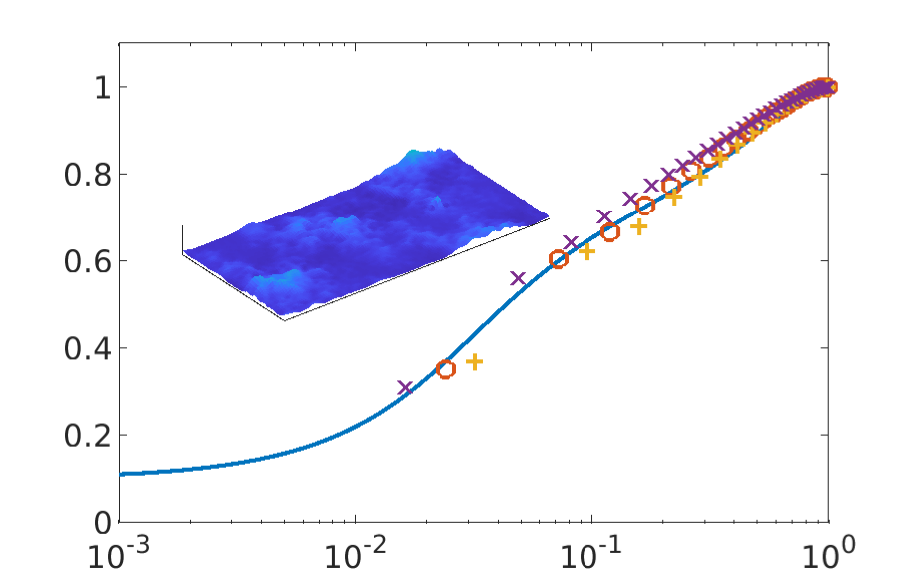}
\put(-107,-6){$(y-d)/\delta$}
\put(-187,50){\rotatebox{90}{$U/U_c$}}
\put(-140,88){WB06}
\put(-185,106){(b)}
\hspace{1mm}
\includegraphics[width=65mm]{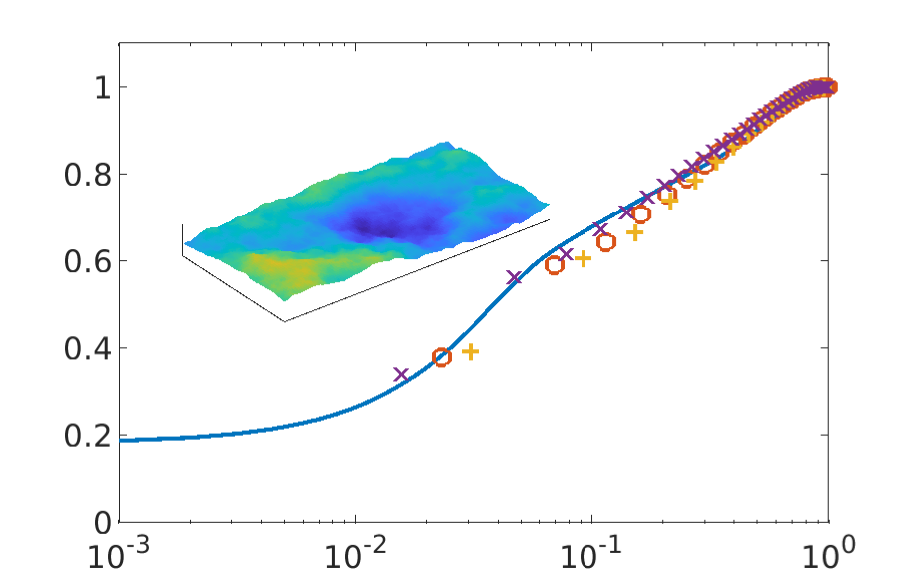}
\put(-107,-6){$(y-d)/\delta$}
\put(-187,50){\rotatebox{90}{$U/U_c$}}
\put(-140,88){GS03}
\put(-185,106){(c)}
\includegraphics[width=65mm]{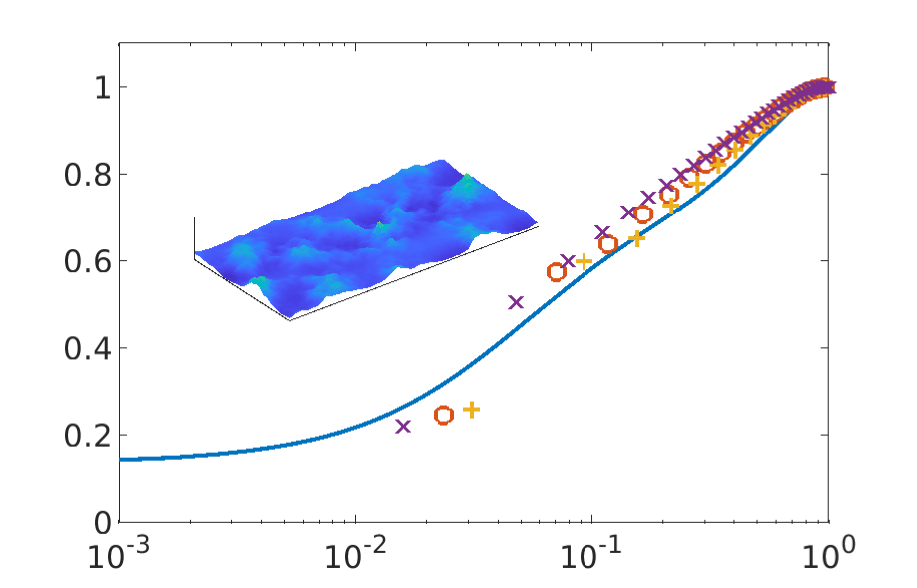}
\put(-107,-6){$(y-d)/\delta$}
\put(-187,50){\rotatebox{90}{$U/U_c$}}
\put(-140,88){WB05}
\put(-185,106){(d)}
\hspace{1mm}
\includegraphics[width=65mm]{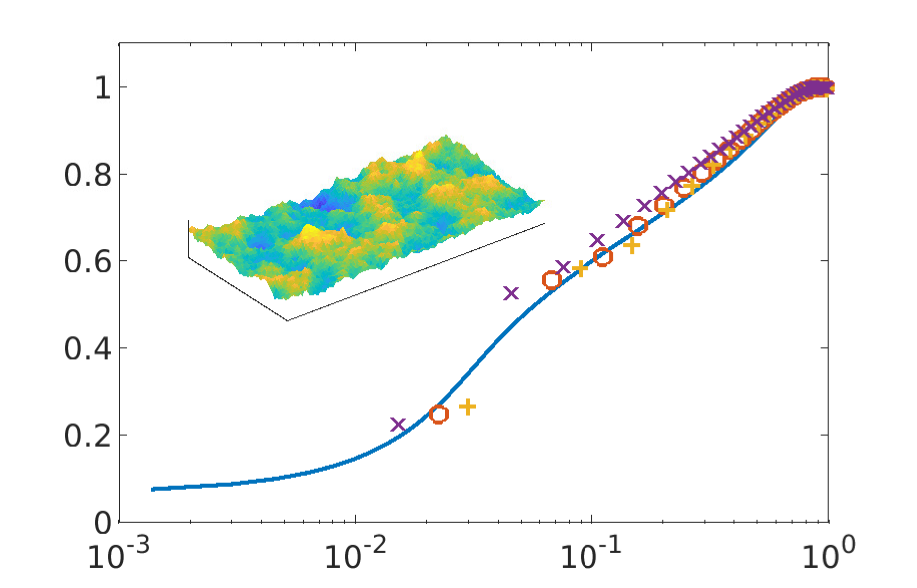}
\put(-107,-6){$(y-d)/\delta$}
\put(-187,50){\rotatebox{90}{$U/U_c$}}
\put(-140,88){GS14}
\put(-185,106){(e)}
\includegraphics[width=65mm]{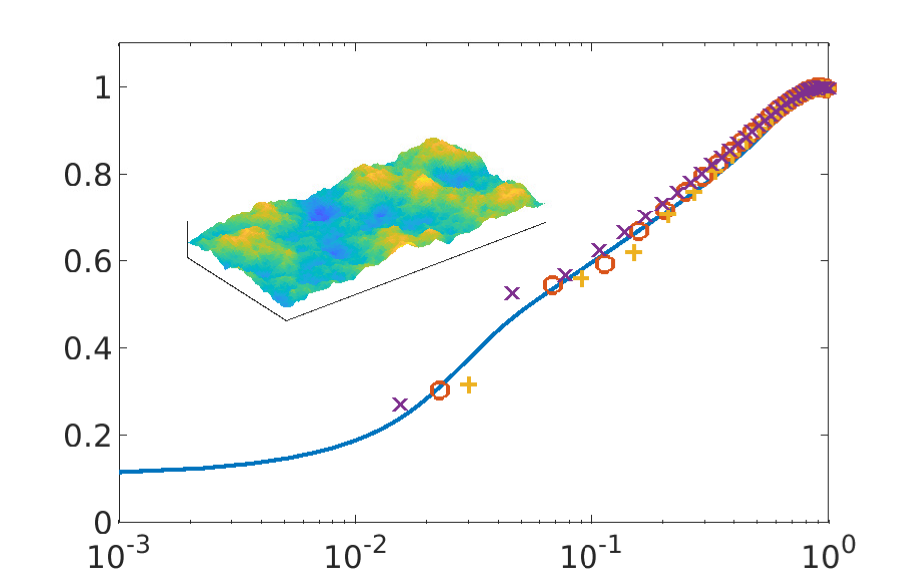}
\put(-107,-6){$(y-d)/\delta$}
\put(-187,50){\rotatebox{90}{$U/U_c$}}
\put(-140,88){GS18}
\put(-185,106){(f)}
\caption{Mean velocity profiles for DNS (line) and WMLES (symbols) of
  turbulent channel flows for selected test cases: (a) WB13 at
  $Re_\tau=720$, (b) WB06 at $Re_\tau=360$, (c) GS03 at $Re_\tau=720$,
  (d) WB05 at $Re_\tau=360$, (e) GS14 at $Re_\tau=540$, and (f) GS18
  at $Re_\tau=1000$. The roughness geometries are visualized with a
  section of $\delta$ in $x$ and $0.5\delta$ in $z$. Three grid
  resolutions are visualized: $\Delta/\delta=1/20$ ($\circ$),
  $\Delta/\delta=1/15$ ($+$), and $\Delta/\delta=1/30$ ($\times$).}
\label{fig:umean_test}
\end{figure}

Six cases from the testing datasets in \S\ref{sec:train}, spanning
both fully- and transitionally-rough regimes, are selected for
evaluation: WB13-$Re_{\tau}=720$, WB06-$Re_{\tau}=360$,
GS03-$Re_{\tau}=720$, WB05-$Re_{\tau}=360$, GS14-$Re_{\tau}=540$, and
GS18-$Re_{\tau}=1000$. Note that the six rough surfaces are seen
during the training process, however, they are evaluated at unseen
rough Reynolds numbers.  The cases are also examined for three
training grid sizes ($\Delta/\delta=1/20, 1/10, 1/5$) and three
testing grid sizes ($\Delta/\delta=1/30, 1/15, 1/8$). The grid sizes
$\Delta/\delta=1/15$ and $\Delta/\delta=1/8$ fall within the range of
the training grids, while $\Delta/\delta=1/30$ is outside this range.

The \emph{a-priori} and \emph{a-posteriori} errors for the predicted
wall shear stress are listed in Table
\ref{tab:test_datasets_error}. The \emph{a-priori} mean error is 6\%
with a standard deviation of 7\%. The errors range from 0\% to 30\%,
with the latter occurring for rough surfaces with Weibull
distributions at the finest grid resolution considered
($\Delta/\delta=1/30$). This discrepancy could be attributed to two
factors. First, the characterization of Weibull surfaces requires more
parameters than Gaussian roughness, complicating predictions due to
the increased dimensionality of the input space. Second, the finest
grid resolution falls outside the bounds of the training set. These
factors combined make Weibull surfaces at fine grid resolutions more
susceptible to inaccurate results due to model extrapolation. The mean
\emph{a-posteriori} error is 9\%, with a standard deviation of 5\%. In
some instances, \emph{a-posteriori} errors are lower than
\emph{a-priori} errors. However, this apparent improvement is due to
error cancellation arising from external errors in the SGS model.

The streamwise mean velocity profiles from WMLES are plotted and
compared with DNS results in Figure \ref{fig:umean_test}. To
facilitate the visualization of near-wall errors from WMLES, the mean
velocity $U$ is normalized by $U_c$ instead of $u_\tau$. The
wall-normal distance is non-dimensionalized by $(y-d)/\delta$. Three
grid resolutions are shown: $\Delta/\delta=1/20$,
$\Delta/\delta=1/15$, and $\Delta/\delta=1/30$. The agreement between
DNS and WMLES is within 5\%, demonstrating the capability of
BFWM-rough in predicting the mean velocity profiles for both fully-
and transitionally-rough cases. It is important to note that the SGS
model plays a crucial role in predicting the mean velocity
profile. Therefore, most of the errors observed in Figure
\ref{fig:umean_test} are likely dominated by deficiencies in the SGS
model rather than internal errors from BFWM-rough.

\subsubsection{Unseen Gaussian/Weibull rough surfaces}\label{sec:validation-2}
%
\begin{figure}
\centering
\includegraphics[width=130mm,trim={0.0cm 1.0cm 0.0cm 4.0cm},clip]{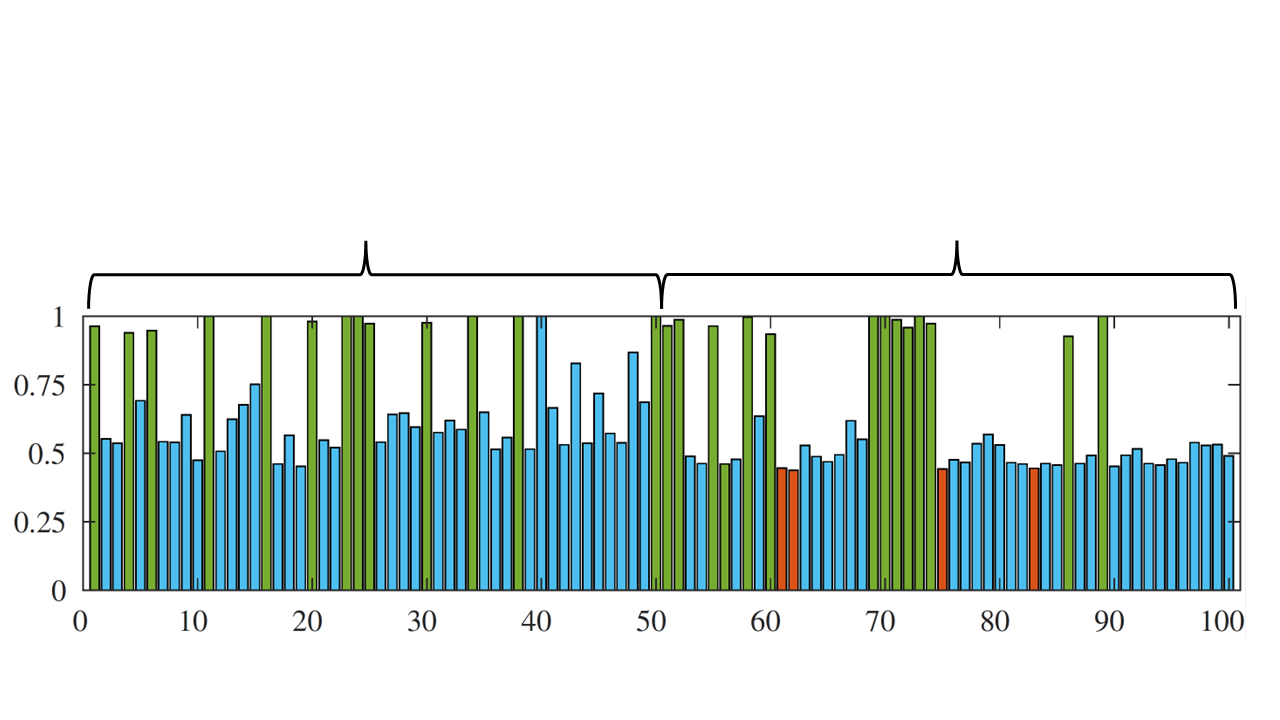}
\put(-300,132){Gaussian roughness}
\put(-135,132){Weibull roughness}
\put(-380,65){{$C$}}
\put(-215,5){Index of roughness}
\caption{Confidence score for each surface in the roughness
  repository. Surface indices 1 to 50 are Gaussian roughness, and
  surface indices 51 to 100 are Weibull roughness. The training cases
  are colored in green. The four cases with the smallest $C$ (colored
  in red) are selected for evaluation of BFWM-rough.  }
\label{fig:variance_test}
\end{figure}
\begin{figure}
\centering
\includegraphics[width=65mm]{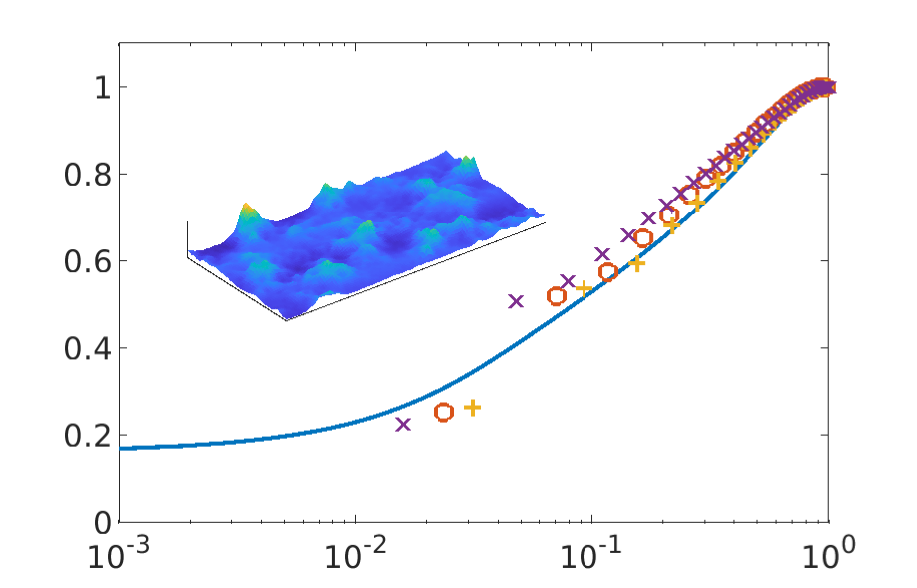}
\put(-107,-6){$(y-d)/\delta$}
\put(-187,50){\rotatebox{90}{$U/U_c$}}
\put(-140,88){WB14}
\put(-185,106){(a)}
\includegraphics[width=65mm]{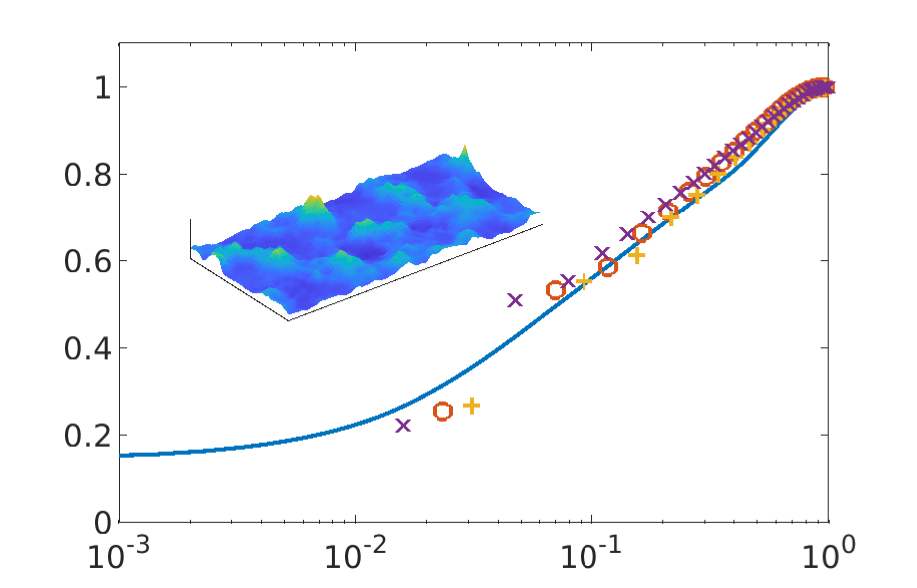}
\put(-107,-6){$(y-d)/\delta$}
\put(-187,50){\rotatebox{90}{$U/U_c$}}
\put(-140,88){WB15}
\put(-185,106){(b)}
\hspace{1mm}
\includegraphics[width=65mm]{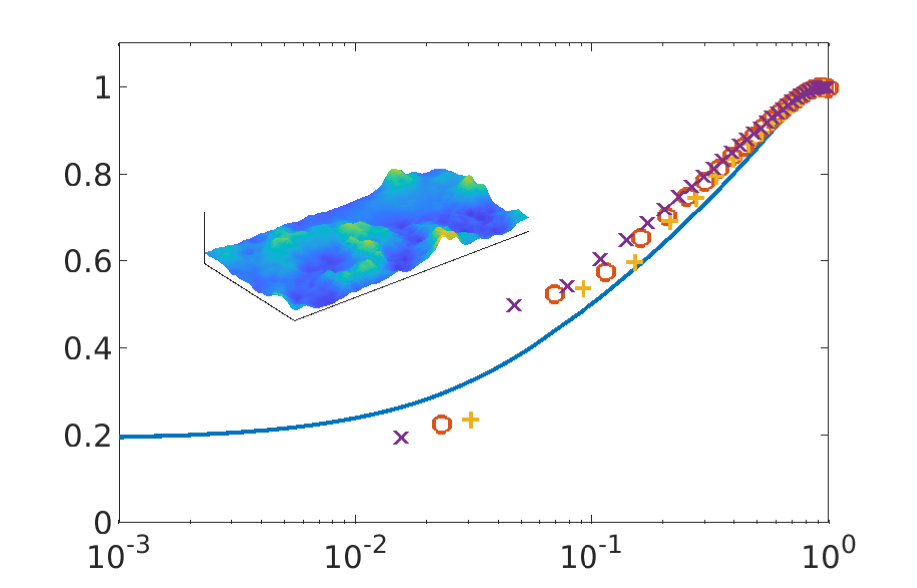}
\put(-107,-6){$(y-d)/\delta$}
\put(-187,50){\rotatebox{90}{$U/U_c$}}
\put(-140,88){WB16}
\put(-185,106){(c)}
\includegraphics[width=65mm]{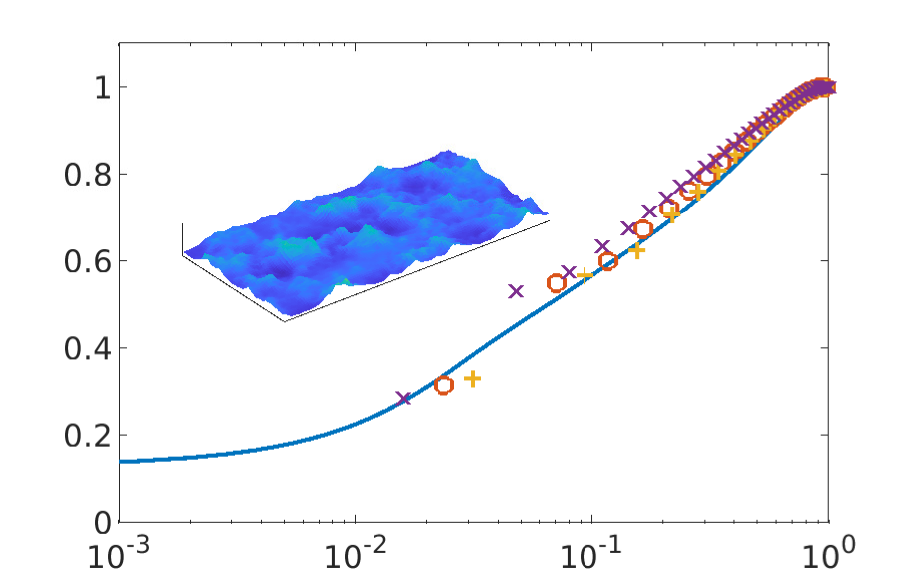}
\put(-107,-6){$(y-d)/\delta$}
\put(-187,50){\rotatebox{90}{$U/U_c$}}
\put(-140,88){WB17}
\put(-185,106){(d)}
\caption{Mean velocity profiles for DNS (line) and WMLES (symbols) of
  turbulent channel flows at $Re_{\tau}=1000$ with test Weibull
  surfaces: (a) WB14, (b) WB15, (c) WB16 and (d) WB17. The roughness
  geometries are visualized with a section of $\delta$ in $x$ and
  $0.5\delta$ in $z$. Three grid resolutions are visualized:
  $\Delta/\delta=1/20$ ($\circ$), $\Delta/\delta=1/15$ ($+$), and
  $\Delta/\delta=1/30$ ($\times$).}
\label{fig:umean_wb}
\end{figure}

We evaluate BFWM-rough on surfaces from the roughness repository whose
geometrical features are not included in the training
process. Additionally, these cases contain unseen $k_s$ and are
assessed in unseen grid resolutions. To identify the most challenging
cases, we use the confidence score $C$ introduced in
\S\ref{sec:confidence} to select rough surfaces with the lowest
confidence for evaluation. Figure \ref{fig:variance_test} displays the
confidence scores for each surface in the roughness repository. As
expected, the confidence levels for all training cases (colored in
green) are around 100\%. The confidence scores for the unseen rough
surfaces range from 45\% to 90\%. The four cases with the lowest $C$
scores (WB14, WB15, WB16, and WB17, colored in red) are selected for
evaluation of BFWM-rough. The geometrical parameters of these four
surfaces are summarized in Table \ref{tab:roughness_test}.

The errors of ${\tau}_w$ are also included in Table
\ref{tab:test_datasets_error} for both \emph{a-priori} and
\emph{a-posteriori} testing at different grid resolutions. The
streamwise mean velocity profiles from WMLES are plotted in Figure
\ref{fig:umean_wb} and compared with those from DNS. Both
\emph{a-priori} and \emph{a-posteriori} errors are similar, ranging
from 0\% to 20\%, with a mean value of 7\% and a standard deviation of
6\%. The best predictions are achieved for WB17, with errors remaining
below 10\% for all grid resolutions. The main difference between WB17
and the other three surfaces is its smaller $\hat{k}_s^+$ value
($\hat{k}_s^+=80$) compared to the larger values ($\hat{k}_s^+>130$)
for WB14, WB15, and WB16. The larger $\hat{k}_s^+$ values might
increase the difficulty of accurate model predictions, especially for
untrained grids. As noted previously, the largest errors are observed
for WB14, WB15, and WB16 at the finest (unseen) grid resolution.

\begin{table}
\begin{center}
\def~{\hphantom{0}}
\setlength\extrarowheight{-5pt}
\renewcommand{\arraystretch}{2}
    \begin{tabular}{cccccccccccccccc}
    \hline \\
      Case &  $k_{avg}$ &  $k_c$ &  $k_t$ &  $k_{rms}$ &  $R_a$ &  $S_k$ &  $K_u$ &  $ES$  &  $I$  &  $P_o$ &  $\lambda_f$ &  $L_{cor}$  \\[3pt]
      WB14 &  0.026 &  0.128 &  0.122 &  0.013 &  0.010 &  1.510 &  6.593 &  0.308   &  -0.012 &   0.798 &  0.172 &  0.064   \\
      WB15 &  0.031 &  0.140 &  0.119 &  0.013 &  0.010 &  1.227 &  5.402 &  0.336   &  0.005  &   0.779 &  0.144 &  0.060    \\
      WB16 &  0.040 &  0.184 &  0.162 &  0.019 &  0.014 &  1.384 &  6.008 &  0.289   &  0.063  &   0.786 &  0.201 &  0.105      \\
      WB17 &  0.028 &  0.109 &  0.089 &  0.011 &  0.009 &  0.952 &  4.495 &  0.327   &  0.012  &   0.748 &  0.177 &  0.061      \\
      BM01 &  0.047 &  0.151 &  0.139 &  0.022 &  0.018 &  0.597 &  3.007 &  0.574   &  -0.001 &   0.693 &  0.264 &  0.064     \\
      BM02 &  0.046 &  0.136 &  0.132 &  0.020 &  0.016 &  0.468 &  2.804 &  0.547   &  0.008  &   0.666 &  0.264 &  0.067     \\
      BM03 &  0.039 &  0.109 &  0.105 &  0.017 &  0.014 &  0.516 &  2.810 &  0.617   &  -0.014 &   0.646 &  0.298 &  0.058     \\
      BM04 &  0.084 &  0.123 &  0.122 &  0.018 &  0.014 &  -0.530 &  2.814 &  0.635  &  -0.028 &   0.315 &  0.276 &  0.055   \\
     \hline
    \end{tabular}
    \caption{\label{tab:roughness_test} Roughness parameters of Weibull and Bimodal rough surfaces for model evaluation. }
    \end{center}
\end{table}

\subsubsection{Unseen bimodal Gaussian rough surfaces}\label{sec:validation-3}

\begin{figure}
\centering
\includegraphics[width=100mm,trim={0.0cm 0.0cm 0.0cm 0.0cm},clip]{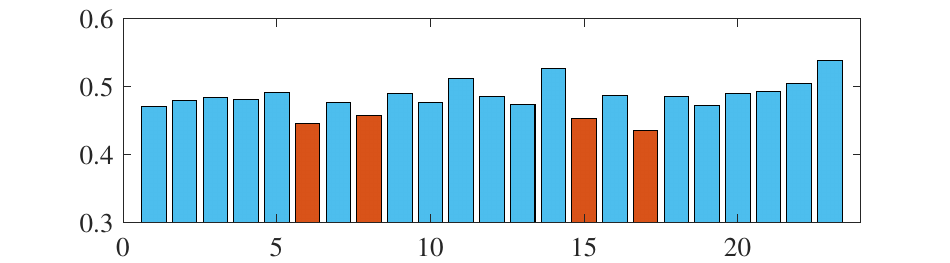}
\put(-280,40){{$C$}}
\put(-200,-10){Index of bimodal roughness}
\caption{Confidence scores for 23 bimodal rough surfaces not included
  in the roughness repository. The four cases with the smallest $C$
  (colored in red) are selected for evaluation of BFWM-rough. }
\label{fig:variance_BM}
\end{figure}

In this section, BFWM-rough is evaluated on rough surfaces with a
bimodal Gaussian distribution. The goal is to assess the performance
of BFWM-rough on a new roughness type that shares some similarities
with the surfaces included in the roughness repository but does not
follow the same generation process. The selection of bimodal Gaussian
roughness is motivated by the presence of engineering surfaces, which
are frequently generated by successive processes or multiple
factors. This introduces two or more different modal roughness
distributions into the final surface~\citep{peng2000modelling}. One
example of a bimodal roughness distribution is observed in ice
accretion on airfoils~\citep{bornhoft2022wall}. The bimodal Gaussian
distributions are constructed by combining two Gaussian distributions
according to \citep{peng2000modelling}:
\begin{equation}
    \text{PDF}_B(\Tilde{k}) = \text{PDF}_G(\Tilde{k};0,1) + \text{PDF}_G(\Tilde{k};\mu_G,\sigma^2_G) - \text{PDF}_G(\Tilde{k}; 0,1) \text{PDF}_G(\Tilde{k};\mu_G,\sigma^2_G),
\end{equation}
where
$\text{PDF}_B(\Tilde{k})$ is the PDF of the bimodal distribution,
$\text{PDF}_G(\Tilde{k};0,1)$ is a normal distribution, and
$\text{PDF}_G(\Tilde{k};\mu_G,\sigma^2_G)$ is the Gaussian distribution with
randomized mean $0 < \mu_G < 0.5$ and randomized variance $0 <
\sigma^2_G < 0.5$. The PS is specified in the same way as described in \S \ref{sec:repository}. The resulting surface map is then scaled from 0 to the root mean square height normalized by the channel half-height $k_{rms}/\delta$, whose values are randomly chosen in the range of $0.005<k_{rms}/\delta<0.030$. A total of 23 bimodal rough surfaces are generated.
Their confidence scores, shown in Figure \ref{fig:variance_BM}, are
about 50\%. The most challenging surfaces for evaluation are
identified as those with the lowest confidence score: BM01, BM02,
BM03, and BM04 (where BM refers to bimodal roughness). Their roughness
parameters are listed in Table \ref{tab:roughness_test}.

%
\begin{figure}
\centering
\includegraphics[width=65mm]{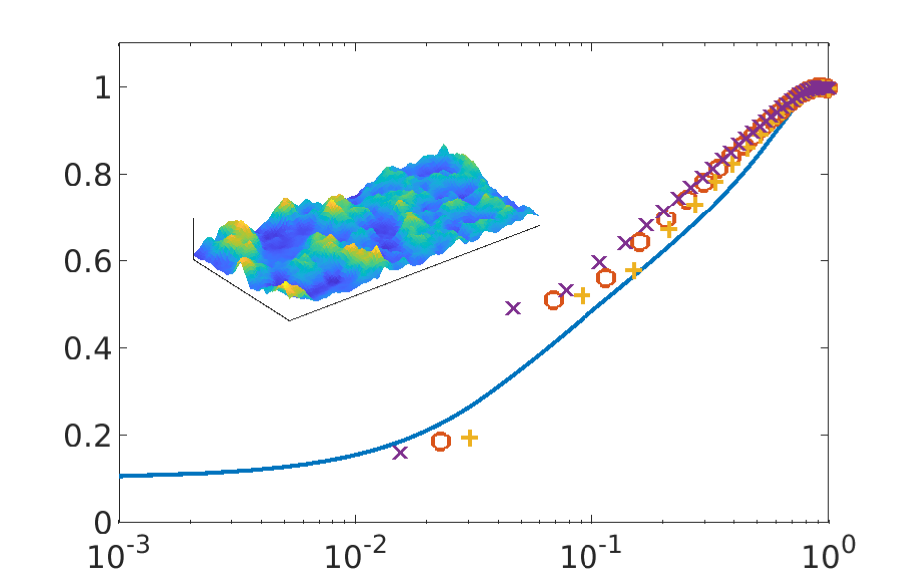}
\put(-107,-6){$(y-d)/\delta$}
\put(-187,50){\rotatebox{90}{$U/U_c$}}
\put(-140,88){BM01}
\put(-185,106){$(a)$}
\includegraphics[width=65mm]{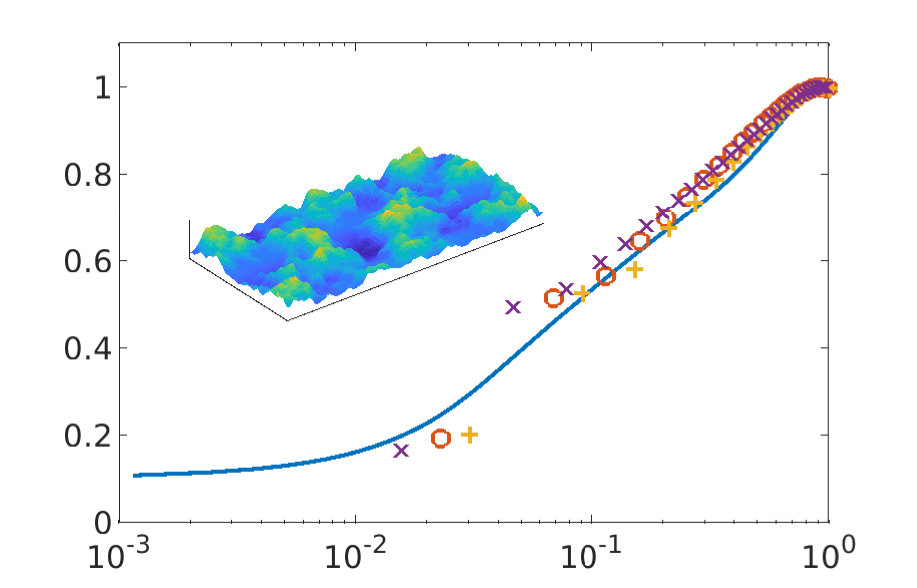}
\put(-107,-6){$(y-d)/\delta$}
\put(-187,50){\rotatebox{90}{$U/U_c$}}
\put(-140,88){BM02}
\put(-185,106){$(b)$}
\hspace{1mm}
\includegraphics[width=65mm]{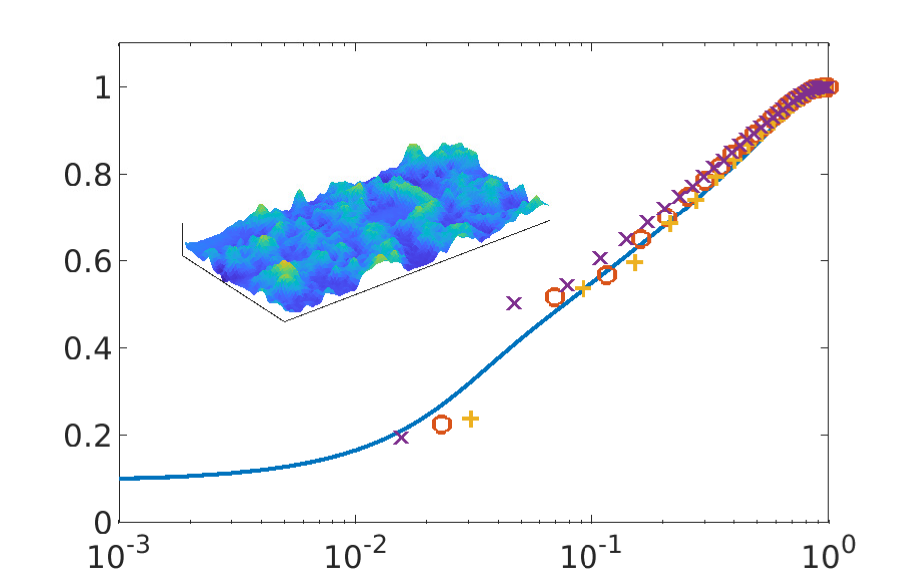}
\put(-107,-6){$(y-d)/\delta$}
\put(-187,50){\rotatebox{90}{$U/U_c$}}
\put(-140,88){BM03}
\put(-185,106){$(c)$}
\includegraphics[width=65mm]{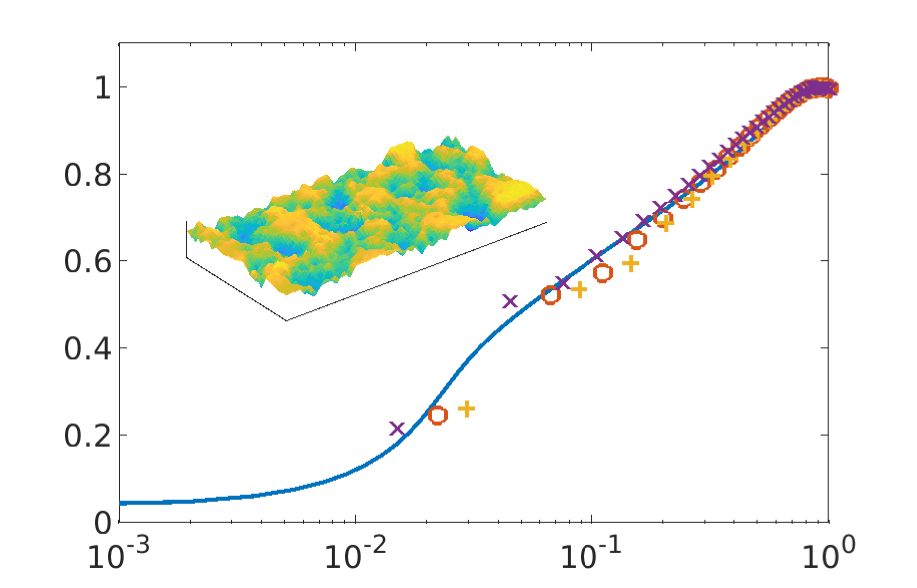}
\put(-107,-6){$(y-d)/\delta$}
\put(-187,50){\rotatebox{90}{$U/U_c$}}
\put(-140,88){BM04}
\put(-185,106){$(d)$}
\caption{Mean velocity profiles for DNS (line) and WMLES (symbols) of
  turbulent channel flows at $Re_{\tau}=1000$ with bimodal surfaces:
  $(a)$ BM01, $(b)$ BM02, $(c)$ BM03 and $(d)$ BM04. The roughness
  geometries are visualized with a section of $\delta$ in $x$ and
  $0.5\delta$ in $z$. Three grid resolutions are visualized:
  $\Delta/\delta=1/20$ ($\circ$), $\Delta/\delta=1/15$ (+), and
  $\Delta/\delta=1/30$ (x).}
\label{fig:umean_bm}
\end{figure}

The performance of BFWM-rough for BM01, BM02, BM03 and BM04 is shown in
Table \ref{tab:test_datasets_error}. Both \emph{a-priori} and
\emph{a-posteriori} errors are similar, with mean values close to 11\%
and a standard deviation of 8\%. Despite the errors are larger than in
the previous sections, the results still demonstrate that BFWM-rough
can offer reasonable predictions for new roughness types as long as
these follow similar distributions to those it was trained for. The
largest errors are obtained for BM04. The primary distinction between
BM04 and the other three surfaces, as well as surfaces in the training
database, is the combination of large negative $S_k$ and large
effective slopes for BM04. These roughness features can lead to less
drag; however, since the training database lacks this information, the
wall model results in an overestimation of the wall shear stress for
BM04. The mean velocity profiles of WMLES compared to DNS are shown in
Figure \ref{fig:umean_bm}. Note, however, that the improved agreement
with the DNS results for BM04 at the grid resolutions of
$\Delta/\delta=1/20$ and $\Delta/\delta=1/30$ is due to internal and
external error cancellation in \emph{a-posteriori} evaluation.

\subsubsection{Unseen rough surfaces from \cite{jouybari2021data}}\label{sec:validation-4}

\begin{figure}
\centering
\includegraphics[width=140mm,trim={0.5cm 0.0cm 0.0cm 0.0cm},clip]{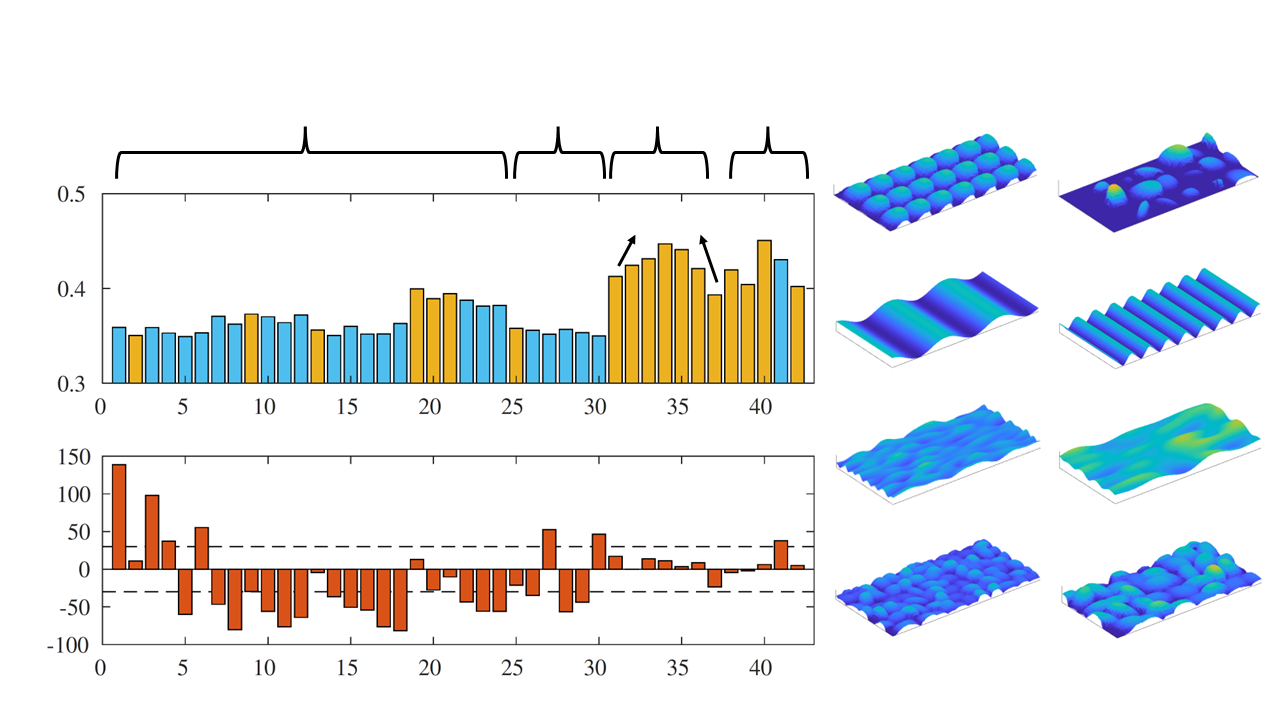}
\put(-400,170){(a)}
\put(-400,95){(b)}
\put(-400,133){$C$}
\put(-400,42){\rotatebox{90}{Error}}
\put(-345,190){Ellipsoidal roughness}
\put(-252,190){Sinusoidal}
\put(-205,190){Fourier-mode}
\put(-210,156){Sandgrain}
\put(-350,2){Index of roughness in \cite{jouybari2021data}}
\put(-130,182){\scriptsize{C04}}
\put(-60,182){\scriptsize{C21}}
\put(-130,140){\scriptsize{C29}}
\put(-60,140){\scriptsize{C30}}
\put(-130,98){\scriptsize{C34}}
\put(-60,98){\scriptsize{C39}}
\put(-130,54){\scriptsize{C31}}
\put(-60,54){\scriptsize{C37}}
\hspace{3mm}
\caption{(a) Confidence score for each rough surface and (b)
  \emph{a-priori} relative error (in \%) of predicted $\Tilde{\tau}_w$
  using the BFWM-rough and actual $\Tilde{\tau}_w$ from DNS of
  \cite{jouybari2021data}. The cases with the model error within $\pm
  30\%$ are highlighted in yellow in (a). The dashed lines in (b)
  denote the error values at $\pm 30\%$. Roughness index 1-42
  corresponds to Cases C01-C42 in \cite{jouybari2021data}. The
  topology of some of the surfaces is visualized on the right:
  ellipsoidal roughness: C04 and C21, surfaces with sinusoidal waves:
  C29 and C30, roughness generated by Fourier modes: C34 and C39,
  sandgrain roughness: C31 and C37.}
\label{fig:variance_JB}
\end{figure}
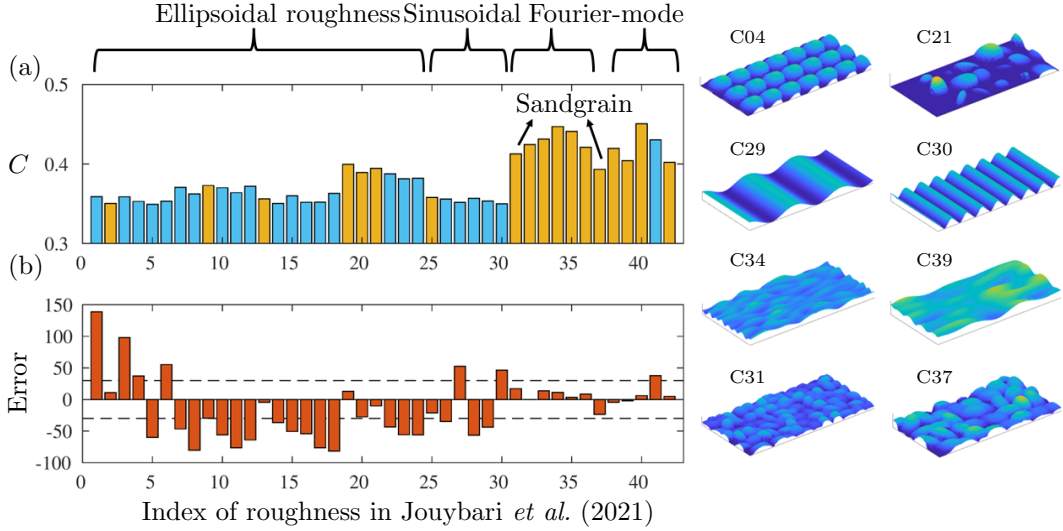

The final cases analyzed are the roughness types from
\cite{jouybari2021data}. The goal is to assess the performance of
BFWM-rough on surfaces with topologies significantly different from
the roughness repository used for training. A total of 42 rough
surfaces are evaluated (labeled from C01 to C42), including 27
fully-rough and 15 transitionally-rough cases. The cases comprise
ellipsoidal, sinusoidal, Fourier-mode, and sandgrain roughnesses. The
topology of some of the surfaces is shown in Figure
\ref{fig:variance_JB}. The exact $\Tilde{\tau}_w$ for all 42 rough
cases is obtained from the DNS of turbulent channel flows at
$Re_{\tau}=1000$ as reported by \cite{jouybari2021data}.

The confidence score for each surface is presented in Figure
\ref{fig:variance_JB}(a). Overall, the results demonstrate the
capability of the confidence score to identify unseen surfaces. The
confidence levels range from 0.35 to 0.45, which are lower than those
for the previous test cases involving unseen Weibull and Bimodal
roughnesses. The highest confidence scores are found for cases C31 to
C42, which correspond to sandgrain roughness and surfaces generated by
low-order Fourier modes. These cases share more similarities with the
isotropic roughness in the training dataset, explaining their higher
confidence scores. The lowest confidence is observed for ellipsoidal
roughness (C01 to C24) and roughness with streamwise sinusoidal waves
(C25 to C30). This is expected, as the ellipsoidal elements are
generated with varying orientations and semiaxis lengths, and the
sinusoidal rough surfaces feature only streamwise waves. Both of these
roughness types represent strongly anisotropic roughness, differing
significantly from the surfaces in the training database.

\emph{A-priori} relative errors for $\Tilde{\tau}_w$ are plotted in
Figure \ref{fig:variance_JB}(b). The grid resolution
  considered is $\Delta/\delta = 1/20$. Consistent with the evaluation
by the confidence score, BFWM-rough tends to perform best for cases
C31 to C42. However, the predictions can exhibit errors of up to 30\%,
which is significantly higher than the errors reported for previous
test cases. The majority of cases from C01 to C30 show errors above
30\%, correlating with their low confidence scores. There are
instances where the confidence score is low, yet the errors remain
below 30\%. This might be coincidental, and caution should always be
exercised for predictions accompanied by low values of the confidence
score.
%

\subsection{High-pressure turbine blade with roughness}\label{sec:blade}

We assess the performance of BFWM-rough in a high-pressure turbine
(HPT) with wall roughness. Surface roughness on HPT blades can result
from the manufacturing process or in-service degradation,
significantly affecting the aero-thermal performance of the blade
\citep{nardini2023direct}. The case selected for evaluation is the VKI
LS-89 HPT blade with surface roughness, and the results are compared
with previous numerical studies by \cite{jelly2023high} and
\cite{nardini2023direct}. This case involves laminar-turbulent
transition, strong pressure gradient effects, shock waves, and
vortical wakes. Therefore, it presents a challenging scenario to
evaluate the predictive capabilities of WMLES in practical flow
conditions involving complex geometries and flow physics.

Our WMLES follows the WRLES setup from \cite{jelly2023high}. The exit
Reynolds number is $Re_{ex} = \rho_{ex}U_{ex}C_{ax}/\mu_{\infty} =
590,000$, and the exit Mach number is $Ma_{ex}=U_{ex}/c_{\infty}=0.9$,
where $\rho_{ex}$ and $U_{ex}$ are the mean exit density and velocity,
respectively. $C_{ax}$ is the axial chord length, $\mu_{\infty}$ is
the dynamic viscosity of the reference state, and $c_{\infty}$ is the
acoustic velocity for the reference state. The inflow freestream
turbulence is generated by a spanwise array of parallel bars. The
turbulence intensity ($TI$) is defined as
$TI=1/3({u'_{rms}}^2+{v'_{rms}}^2+{w'_{rms}}^2)$. The value of $TI$ is
set to $8\%$ of the axial mean inlet velocity $U_{in}$ by adjusting
the distance of the bars upstream of the blade. The total pressure and
temperature are specified at the inlet with the characteristic
boundary conditions.  At the exit, the non-reflective Navier-Stokes
characteristic boundary conditions are enforced. The static pressure
is prescribed at the exit based on the isentropic Mach number
relationships. Periodic boundary conditions are prescribed at the
upper, lower, and spanwise boundaries. The spanwise extent of the
domain is set to $0.4C_{ax}$, as suggested by \cite{pichler2017high},
to ensure the correct development of the largest inflow turbulence
structures.

\begin{table}
\begin{center}
\def~{\hphantom{0}}
\setlength\extrarowheight{-5pt}
\renewcommand{\arraystretch}{2}
    \begin{tabular}{ccccccccc}
    \hline \\
     & $k_{s}/C_{ax}$ & $k_{rms}/C_{ax}$ & $S_k$ & $K_u$ & $ES$  & $L_{cor}/C_{ax}$ & $C$ \\[3pt]
     BS01 & $2.0 \times 10^{-3}$ & $0.4 \times 10^{-3}$ & 0.0 & 3.0 & 0.16 &  $3.6 \times 10^{-3}$ & 0.73 \\
     BS02 & $3.0 \times 10^{-3}$ & $0.6 \times 10^{-3}$ & 0.0 & 3.0 & 0.18 &  $9.5 \times 10^{-3}$ & 0.67 \\
     \hline
    \end{tabular}
    \caption{\label{tab:blade_roughness} The key roughness parameters
      of the blade surface roughness BS01 and BS02 from
      \cite{jelly2023high}. The last column contains the confidence
      score of BFWM-rough for BS01 and BS02.}
    \end{center}
\end{table}
\begin{figure}
\centering
\includegraphics[width=130mm]{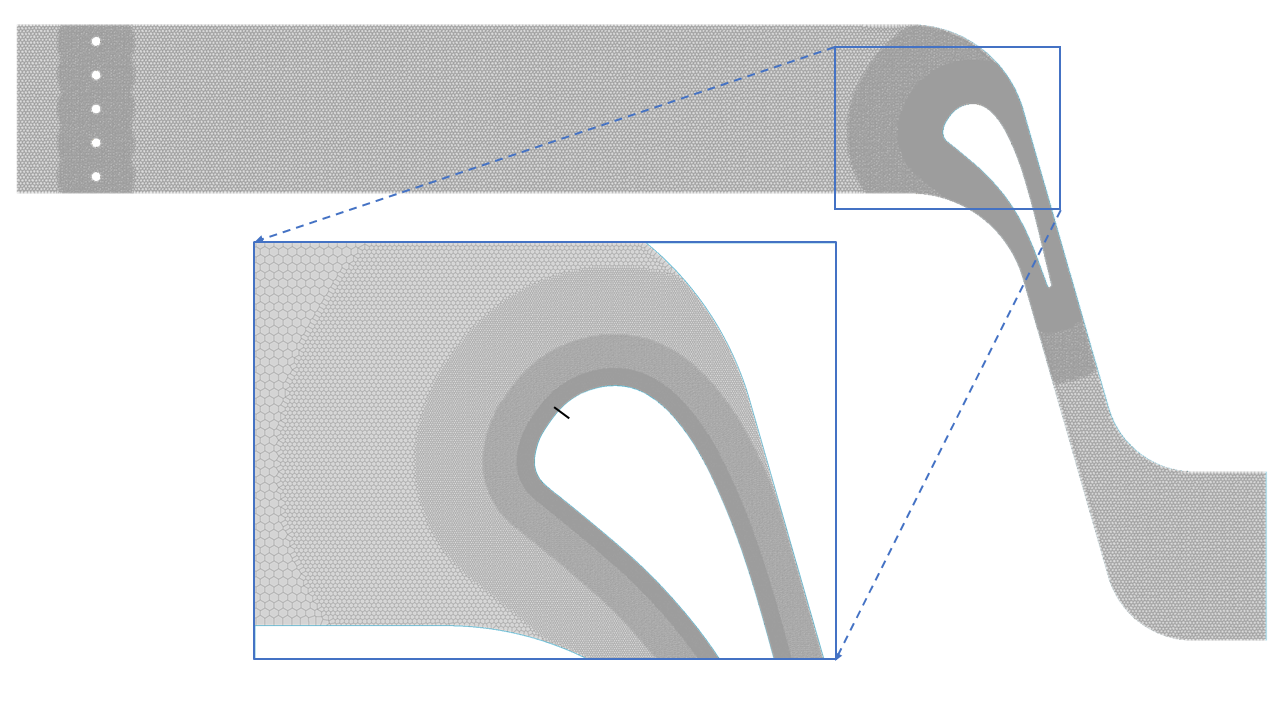}
\put(-293,120){\scriptsize{$\Delta_b$}}
\put(-268,110){\scriptsize{$\frac{1}{2}\Delta_b$}}
\put(-240,100){\scriptsize{$\frac{1}{4}\Delta_b$}}
\put(-225,93){\scriptsize{$\frac{1}{8}\Delta_b$}}
\put(-205,83){\scriptsize{$\frac{1}{16}\Delta_b$}}
\caption{Visualization of Voronoi control volumes for WMLES of the HPT
  blade: the whole computational domain with a zoom-in view near the
  leading edge of the blade.}
\label{fig:mesh}
\end{figure}
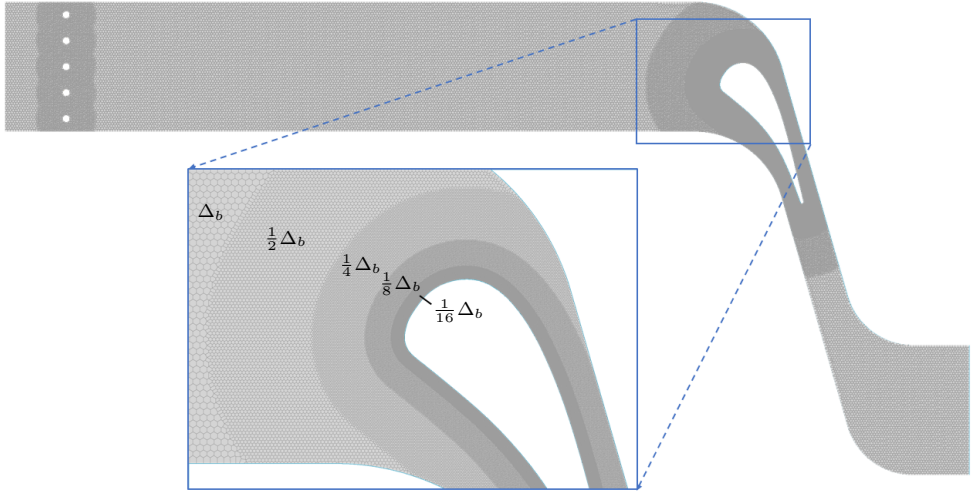
%

The mesh generation is based on the Voronoi tessellation of a
collection of points. The Voronoi control volumes are visualized in
Figure \ref{fig:mesh}. The size of the background grid is
$\Delta_b=0.0311C_{ax}$. The grid size near the upstream bars is
refined to $\frac{1}{2}\Delta_b$ with 20 layers, and the flow field
near the bars is wall-resolved. As shown in the zoom-in view of Figure
\ref{fig:mesh}, the grid size near the blade is refined by 4 levels,
and for each level, the grid size is reduced by 50\% with 30
layers. As a result, the minimum grid size near the blade surface is
$\frac{1}{16}\Delta_b=0.00195C_{ax}$, and the number of control
volumes per boundary layer thickness ranges from 0 (at the blade
leading edge) to 30 (at the blade trailing edge). The total number of
control volumes is 26 million.
\begin{figure}
\centering
\includegraphics[width=130mm]{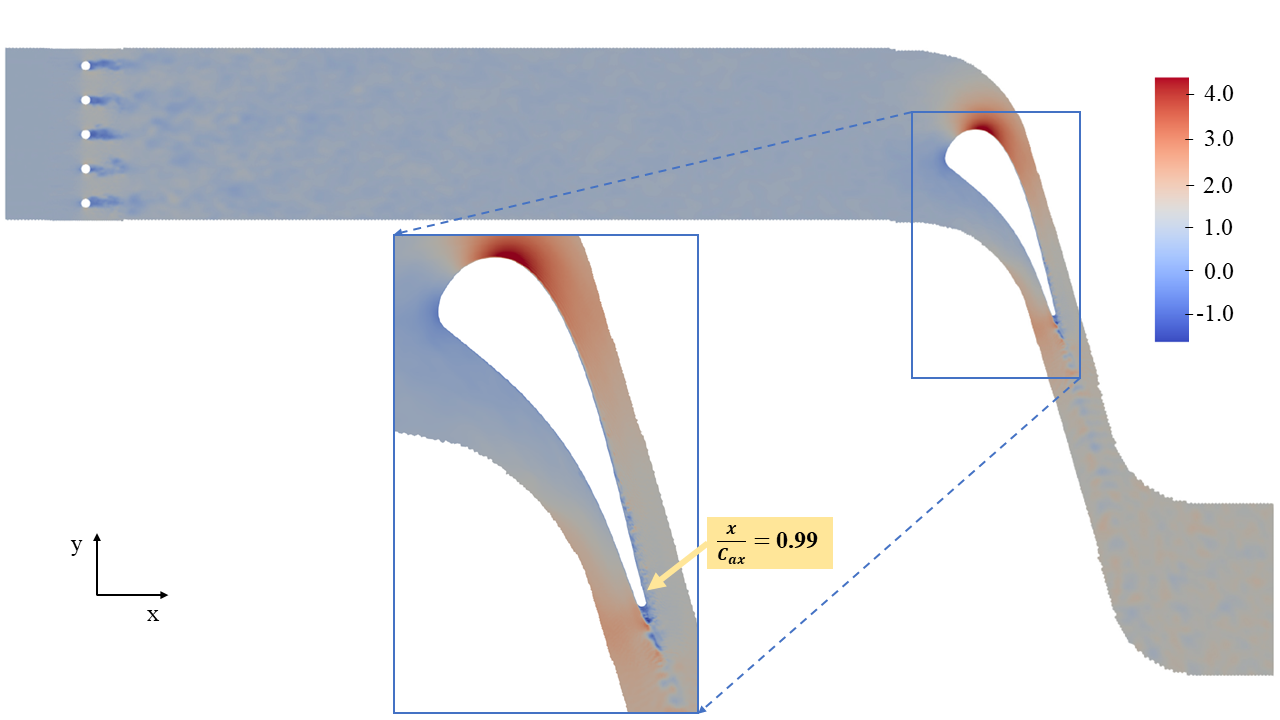}
\put(-40,195){$u_{ax}/U_{in}$}
\caption{Visualization of the instantaneous axial velocity field
  normalized by the axial mean inlet velocity $u_{ax}/U_{in}$ for
  WMLES of the HPT blade. The arrow indicates the location at $99\%$
  of the axial chord length for probing the mean tangential velocity.}
\label{fig:blade_velocity}
\end{figure}

Two different surface roughness profiles are considered for the
blade. The cases, denoted as BS01 and BS02, feature three-dimensional,
irregular Gaussian roughness. The key geometrical parameters of the
roughness are summarized in Table \ref{tab:blade_roughness}, which
also includes the confidence scores for both rough surfaces. The
confidence scores are $C=0.73$ for BS01 and
$C=0.67$ for BS02, indicating that the roughness
topologies can potentially be well-predicted by BFWM-rough. WMLES with
BFWM-rough is conducted for both blade surfaces, using the Vreman
model~\citep{Vreman2004} as the SGS model. The instantaneous axial
velocity field from WMLES with BFWM-rough for roughness BS01 is
visualized in Figure \ref{fig:blade_velocity}, with the boundary layer
transition observable in the zoom-in view. The results are compared to
the WRLES results of \cite{jelly2023high} and DNS data of
\cite{nardini2023direct} for the same rough surfaces.

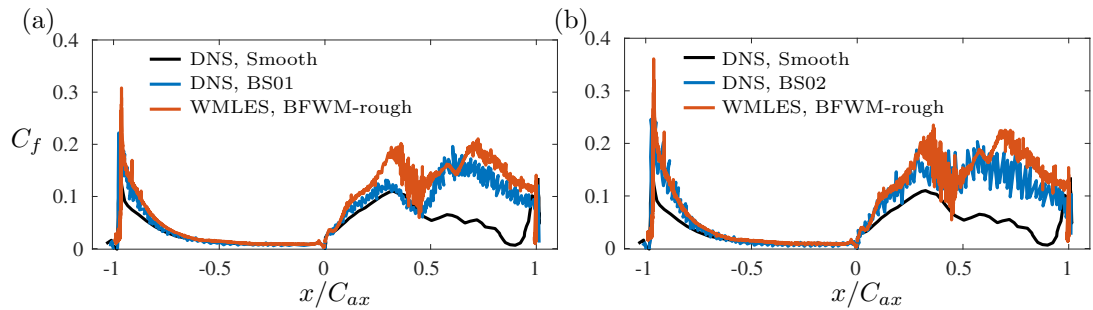
\begin{figure}
\centering
\includegraphics[width=70mm,trim={1.3cm 0.0cm 1.0cm 0.0cm},clip]{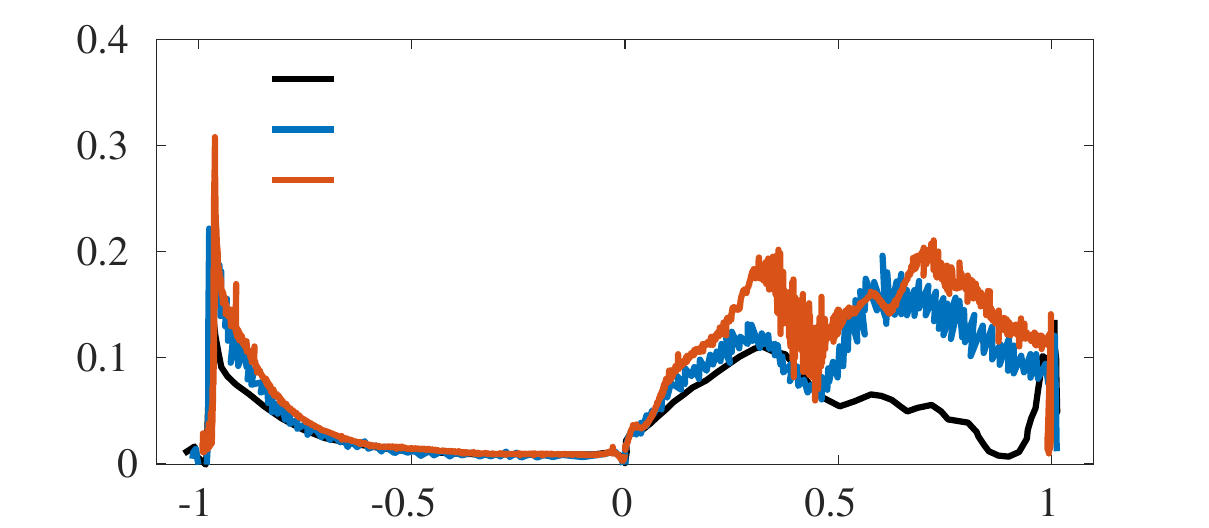}
\put(-107,-8){$x/C_{ax}$}
\put(-215,50){$C_f$}
\put(-148,81){\scriptsize{DNS, Smooth}}
\put(-148,72){\scriptsize{DNS, BS01}}
\put(-148,63){\scriptsize{WMLES, BFWM-rough}}
\put(-211,95){(a)}
\includegraphics[width=70mm,trim={1.3cm 0.0cm 1.0cm 0.0cm},clip]{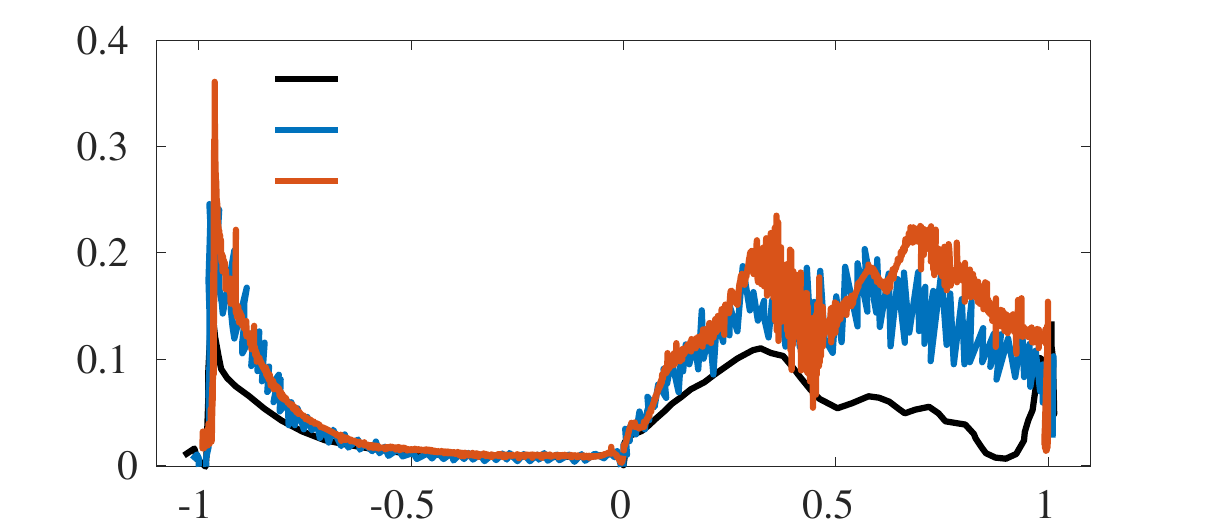}
\put(-107,-8){$x/C_{ax}$}
\put(-148,81){\scriptsize{DNS, Smooth}}
\put(-148,72){\scriptsize{DNS, BS02}}
\put(-148,63){\scriptsize{WMLES, BFWM-rough}}
\put(-211,95){(b)}
\caption{Time and spanwise-averaged distribution of skin friction
  coefficient $C_f$ along the axial position of the blade normalized
  by axial chord length $x/C_{ax}$. (a) Roughness BS01 and (b)
  roughness BS02. The DNS results of smooth and rough surfaces are
  from \cite{nardini2023direct}. The blade roughness BS01 and BS02
  correspond to cases $k_2^s$ and $k_3^s$ in
  \cite{nardini2023direct}. Note that $x/C_{ax} > 0$ and $x/C_{ax} <
  0$ correspond to the SS and PS, respectively, with $x/C_{ax} = 0$
  locate at the leading edge of the blade.}
\label{fig:cf}
\end{figure}

Figure \ref{fig:cf} shows the time- and spanwise-averaged
skin-friction coefficient $C_f$ for WMLES and DNS. The $C_f$ for a
smooth blade obtained from DNS is also included in Figure \ref{fig:cf}
as a reference to demonstrate the effects of roughness. On the
pressure side (PS) of the blade, the increase in $C_f$ is due to the
boundary layer transition, which occurs in the trailing-edge
region. The increase in roughness on the PS of the blade is accurately
captured by BFWM-rough for the two rough surfaces.  The roughness
effects are more significant on the suction side (SS) of the
blade. The sharp increase in skin friction corresponds to the
laminar-to-turbulent transition of the boundary layer, with larger
roughness elements (BS02) triggering an earlier transition compared to
the smaller roughness (BS01). WMLES with BFWM-rough captures the key
trends of the rough wall blade that are absent in the smooth wall
case, such as the faster and larger increase in $C_f$ and the second
peak after $x/C_{ax}>0.5$. However, the value of $C_f$ is
overpredicted by WMLES from $x/C_{ax}=0.7$ to the trailing edge for
the two rough cases. This might be related to the interaction of the
turbulent boundary layer over the blade with the shock waves within
the region $0.7<x/C_{ax}<0.9$.  This effect was described by
\cite{nardini2023direct}, who noted that a normal shock wave is
induced by the roughness, with larger roughness magnitude increasing
the intensity of normal shock patterns. The presence of shock waves
may lead to the formation of shock-induced vortices, and these
vortices can interact with the boundary layer, influencing its
stability and altering the skin friction pattern. This intricate
effect might not be correctly captured by WMLES, although multiple
factors beyond the wall model (e.g., SGS models and grid resolution)
are probably at play.

\begin{figure}
\centering
\includegraphics[width=70mm]{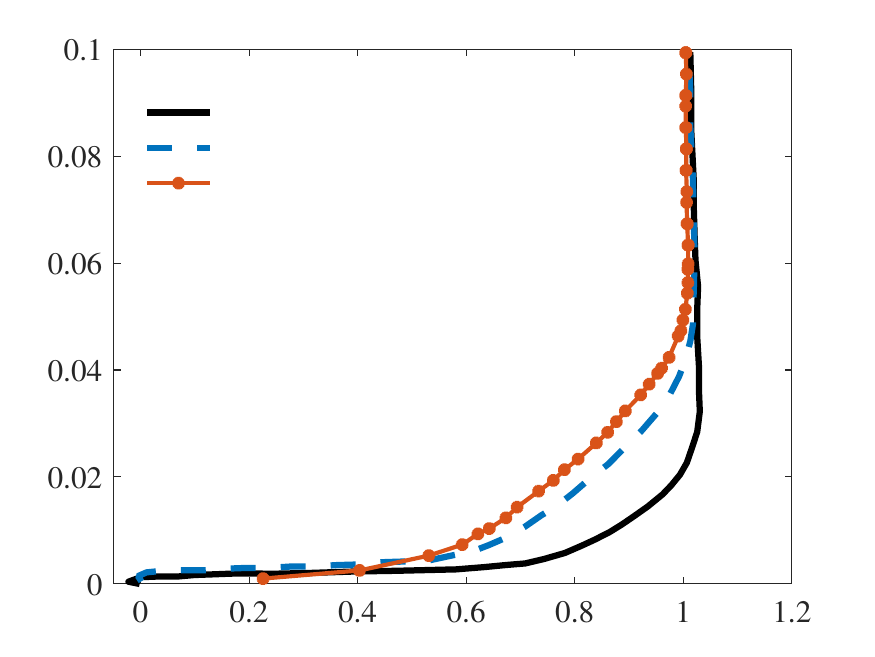}
\put(-107,-8){$U_t/U_{\infty}$}
\put(-209,65){\rotatebox{90}{$y_n/C_{ax}$}}
\put(-148,121){\scriptsize{WRLES, Smooth}}
\put(-148,112){\scriptsize{WRLES, BS01}}
\put(-148,103){\scriptsize{WMLES, BFWM-rough}}
\put(-211,135){(a)}
\includegraphics[width=70mm]{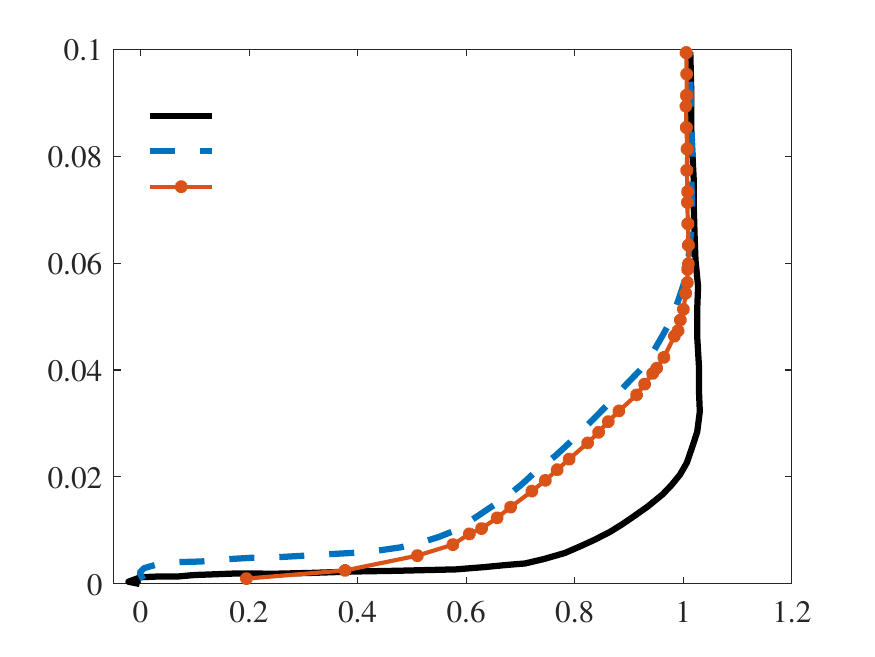}
\put(-107,-8){$U_t/U_{\infty}$}
\put(-209,65){\rotatebox{90}{$y_n/C_{ax}$}}
\put(-148,121){\scriptsize{WRLES, Smooth}}
\put(-148,112){\scriptsize{WRLES, BS02}}
\put(-148,103){\scriptsize{WMLES, BFWM-rough}}
\put(-211,135){(b)}
\caption{Mean tangential velocity at $x/C_{ax}=0.99$ normalized by the
  local freestream velocity $U_t/U_{\infty}$ along the blade-normal
  direction normalized by axial chord length $y_n/C_{ax}$.  (a)
  Roughness BS01 and (b) roughness BS02. The WRLES results of smooth
  and rough surfaces are from \cite{jelly2023high}.}
\label{fig:ut}
\end{figure}
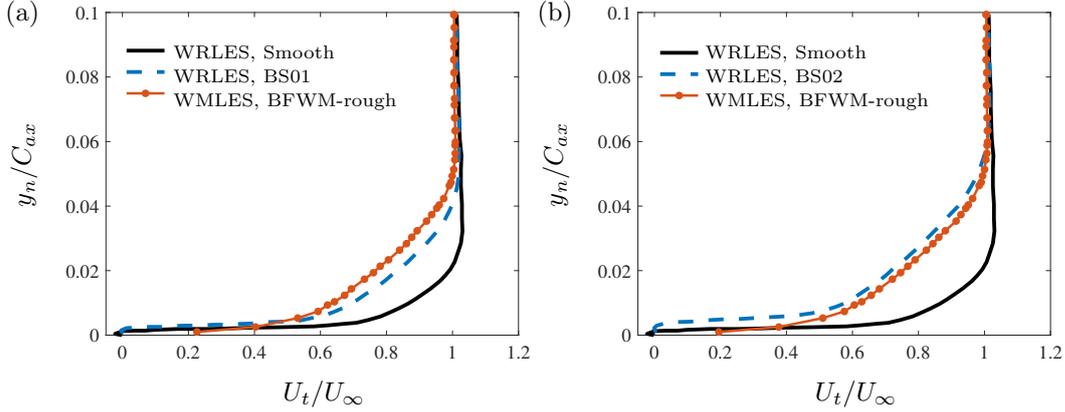

Figure \ref{fig:ut} shows the mean tangential velocity profiles close
to the trailing edge at the axial location $x/C_{ax}=0.99$. The
results from WMLES are compared with WRLES from
\cite{jelly2023high}. Although both the smooth- and rough-blade
boundary layers have transitioned to turbulence in the trailing-edge
region of the SS surface, the flow has yet to reach fully rough
conditions \citep{jelly2023high}. Figure \ref{fig:ut} demonstrates
that the velocity deficit is captured by the WMLES for the two rough
surfaces considered. These results indicate that BFWM-rough performs
well in modeling the integrated momentum deficit along the blade,
which is ultimately responsible for the lower-speed mean velocity profiles
at the trailing edge.
%



\section{Conclusions}\label{sec:conclusion}

We introduce a wall model for WMLES applicable to rough surfaces with
Gaussian and Weibull distributions for both the transitionally- and
fully-rough regimes. The model can be applied to arbitrary complex
geometries where the surface roughnesses are assumed to be underresolved. The wall model is implemented using a
multi-hidden-layer feedforward neural network (FNN), with the statistical
geometric parameters of the surface roughness and near-wall flow
quantities serving as input. The wall model, referred to as
BFWM-rough, extends the building-block flow wall model introduced by
\cite{lozano2023machine} to rough walls.

A roughness repository containing 100 random rough surfaces was
created, encompassing a wide range of Gaussian and Weibull roughness
features. Active learning was then employed to optimally construct the
DNS training database by selecting rough surfaces from the repository
with maximum uncertainty. This approach effectively improved the wall
model prediction capabilities while minimizing the number of DNS cases
required for training. A total of 19 Gaussian and 13 Weibull rough
surfaces were selected through active learning and used to conduct the
DNS of turbulent channel flows. The DNS cases were performed in a
minimal-span channel flow at six different $Re_{\tau}$ ranging from
180 to 1000. The final DNS database comprises approximately 200 cases.

The optimal set of non-dimensional inputs to the model was selected
using information theory. The approach  identifies a
collection of non-dimensional inputs with minimum redundant information among them and
maximum information about the output. Over 30 input candidates were
ranked in order of importance to predict the wall shear stress. The
most informative inputs were found to be based on flow
state variables such as local Reynolds numbers at the first and second
points off the wall ($u_1y_1/\nu$ and $u_2y_2/\nu$)
and mean roughness features related to roughness height
fluctuations and effective slope ($k_{rms}/R_a$ and $ES^2$).

The wall model also incorporates a confidence score to detect
potential low performance in the presence of unseen rough
surfaces. The score is computed using Gaussian process model to
evaluate the uncertainty of the roughness topology compared to the
roughness repository used for training. The confidence score was
calculated for different roughness types, such as Gaussian, Weibull,
Bimodal, sandgrain, and Fourier-mode roughness. The results demonstrated the ability of the
confidence score to highlight potential model deficiencies for rough
surfaces with strong anisotropic characteristics. In such cases, the
confidence score can be used for \emph{a-priori} detection of low
performance scenarios for BFWM-rough. This information can also be
leveraged to inform future extensions of the model by incorporating
roughness types with low confidence scores into the training dataset.

The BFWM-rough model has been tested \emph{a priori} and \emph{a
posteriori} in more than 120 turbulent channel flows across various
rough surfaces and flow conditions. These include cases with untrained
rough Reynolds numbers and grid resolutions over unseen surface roughness
with Gaussian and Weibull distributions both at transitionally- and
fully-rough flow conditions. The model was also evaluated in over 40
rough surfaces whose geometrical features were not incorporated in the
training process, such as bimodal rough surfaces and rough surfaces
from \cite{jouybari2021data} that contain ellipsoidal, sinusoidal,
sandgrain roughness, and rough surfaces generated with low-order
Fourier modes. The results show that the rough-wall model typically
predicts the wall shear stress within a 1\% to 10\% accuracy range for
roughness types resembling Gaussian and Weibull distributions. While
the performance of BFWM-rough degrades for bimodal distributions, the
accuracy remains comparable. Model errors increase to around
30\%-50\% for roughness types exhibiting strong anisotropy, as these
geometries differ considerably from the training set. However, 
the low performance in these cases can be anticipated by the low confidence score. 

The BFWM-rough has also been evaluated in a complex flow involving a
high-pressure turbine blade with two different rough surfaces. This
case includes laminar-turbulent transition, strong pressure gradient
effects, shock waves, and vortical wakes, making it a challenging
scenario for assessing the predictive capabilities of WMLES in
practical flow conditions. The results show that BFWM-rough captures
key trends of the rough wall blade that are absent in the smooth wall
case, such as the faster and larger increase in friction coefficient
and the second peak after the shock wave, with errors ranging between
1\% and 10\%. BFWM-rough also accurately predicts the integrated
momentum deficit along the blade, which is ultimately responsible for
the slower mean velocity profiles at the trailing edge.

The current version of BFWM-rough is designed for equilibrium
turbulence over isotropic rough surfaces. Future developments aim to
extend its applicability to more general flow conditions, including
mean pressure gradient effects and separation, following the approach
from \cite{lozano2023machine}. Additionally, the wall model is
expected to be expanded to cover a broader range of roughness types,
including those with anisotropic geometries.

\section*{Acknowledgements}

This work was supported by the MIT Research Support Committee under
the Chang Foundation and the National Science Foundation under grant
number \#2317254 for the project titled ``Building-Block-Flow Model
for Large-Eddy Simulation''. We gratefully acknowledge Prof. Junlin
Yuan and her student Sai Chaitanya Mangavelli for their help and
guidance on using the data and the DNS solver they provided.  We thank
Dr. Mostafa Aghaei Jouybari for providing the roughness database to
test our wall model. The authors acknowledge MIT SuperCloud and
Extreme Science and Engineering Discovery Environment (XSEDE) for
providing computing resources that have contributed to the research
results reported in this paper.

\section*{Declaration of Interests}
The authors report no conflict of interest.


\bibliographystyle{jfm}
\bibliography{jfm-instructions}

\end{document}